\documentclass[aps,prx,superscriptaddress,twocolumn,longbibliography]{revtex4-2}

\usepackage[normalem]{ulem} 

\usepackage[hidelinks]{hyperref}
\hypersetup{
	colorlinks,
	linkcolor={red},
	citecolor={blue},
	urlcolor={blue}
}
\usepackage{physics}
\usepackage{balance}
\usepackage{amssymb}
\usepackage[table]{xcolor}

\definecolor{highlightcolor}{rgb}{1, 1, 1} 
\definecolor{highlightcolor1}{rgb}{1, 1, 1} 
\definecolor{diagcolor}{gray}{0.88}
\definecolor{diagcolor1}{gray}{0.65}
\usepackage{suffix}
\usepackage{mathtools}
\usepackage[utf8]{inputenc}
\usepackage{booktabs}
\usepackage{cases}
\usepackage[multiple]{footmisc}
\usepackage{dcolumn}
\usepackage{color,soul}
\usepackage{rotating}
\usepackage{perpage}
\usepackage{siunitx}
\usepackage{xcolor}
\usepackage{soul}
\usepackage{amsmath}
\usepackage{tikz}
\usepackage[T1]{fontenc}
\usepackage{etoolbox}
\usepackage{graphics}
\usepackage{siunitx}
\usepackage{float}	
\usepackage{hhline}
\usepackage{collref}
\usepackage{multirow}
\usepackage{mathtools}
\usepackage{bm}
\usepackage{url}

\usepackage{tikz}
\usepackage{tikz-3dplot}
\usepackage{accents}

\makeatletter
\newcommand{\doublewidetilde}[1]{{%
		\mathpalette\double@widetilde{#1}%
}}
\newcommand{\double@widetilde}[2]{%
	\sbox\z@{$\m@th#1\widetilde{#2}$}%
	\ht\z@=.9\ht\z@
	\widetilde{\box\z@}%
}
\makeatother

\usepackage{etoolbox,lipsum}

\newcommand{\bs}[1]{\boldsymbol{#1}}

\begin{document}
	
	\title{Bichromatic light-induced electronic torques on the N\'eel vector in Rashba \\magnetic multipoles from $p$-wave magnets to $i$-wave altermagnets}
	\author{Mohsen Yarmohammadi}
	\email{mohsen.yarmohammadi@georgetown.edu}
	\address{Department of Physics, Georgetown University, Washington DC 20057, USA}
	
	\begin{abstract}
		Light can control magnetic order on ultrafast timescales, but for a two-dimensional magnet with an out-of-plane order parameter $\boldsymbol N\parallel\hat z$, the inverse Faraday field is collinear with $\boldsymbol N$ and gives no transverse torque. Here we show that bichromatic $(\omega,b\omega)$ light generates electronic torques perpendicular to $\boldsymbol N$ in Rashba-coupled magnetic multipoles, from $p$-wave ($n=1$) magnets to $i$-wave ($n=6$) altermagnets. Beyond high-frequency effective fields, we evaluate the exact Floquet response and the generalized torque per occupied electron, for prescribed occupations. The momentum-dependent exchange terms dominate this Fermi-sea torque and remove the leading-order zeros (amplitude-ratio blindspots) of the momentum-independent field in the channels tested. For out-of-plane $\boldsymbol N$, rotational symmetry permits a transverse torque only if $\gcd(n,b+1)=1$ for counter-rotating beams or $\gcd(n,b-1)=1$ for co-rotating beams. For in-plane $\boldsymbol N$, the Fermi sea also gives a torque component along $\hat z$ that the momentum-independent field cannot produce. We classify the active channels, with $n=1,\dots,6$ and $b=1,\dots,7$, by frequency scaling and origin (cycle-averaged versus oscillating exchange). As a control, a Rashba ferromagnet ($n=0$) has one symmetry-allowed out-of-plane channel and a weak in-plane response without a selection rule. Finite-momentum occupied-state physics is therefore essential for light-driven torques in unconventional magnets, and the exact Floquet calculation validates and extends the high-frequency description.
	\end{abstract}
	
	\maketitle
	{\allowdisplaybreaks
		\section{Introduction}
		\label{sec:intro}
		
		Magnetic materials include a broad family of multipolar orders characterized by momentum-dependent spin textures~\cite{smejkal2022beyond,smejkal2022emerging,doi:10.1126/sciadv.aaz8809}. These include odd-parity $p$-, $f$-, and $h$-wave magnets~\cite{shen2026roomtemperatureferroelectricpwavemagnet,Chakraborty2025,hellenes2024pwavemagnets,liu2026oddparity,yamada2025pwave,song2025electrical}, as well as even-parity $d$-, $g$-, and $i$-wave altermagnets~\cite{	doi:10.1126/sciadv.adj4883,smejkal2022beyond,smejkal2022emerging,doi:10.1126/sciadv.aaz8809,Feng2022,PhysRevLett.131.256703,PhysRevLett.132.056701,Reichlova2024,Adamantopoulos2024,cheong2025classification,xiao2024spin}. Such systems combine features of antiferromagnets, including vanishing net magnetization and fast magnetic dynamics, with the spin-polarized electronic transport associated with magnetic order~\cite{jungwirth2016antiferromagnetic,jungwirth2018review,zelezny2018spintransport,song2025altermagnets,guo2025spinpolarized}. Utilizing them in spintronic devices calls for fast and energy-efficient methods to manipulate their magnetic order, and Floquet engineering with laser pulses offers a natural route to such control~\cite{kimel2019writing,nemec2018opto,kirilyuk2010ultrafast,oka2019floquet,bukov2015universal}.
		
		Circularly polarized light can induce an inverse Faraday effect (IFE), producing an effective magnetic field along the light propagation direction~\cite{pershan1966inverse,kirilyuk2010ultrafast,battiato2014quantum,xt23-9pnv,k3xb-8pts,zp5c-vpd9,mslj-zz6t,shah2026floquetengineeringtopologicalphases,liu2026linearlypolarizedlightinducedanomalous,li2026dynamicalunmaskingpwavemagnetism,cheraghchi2026floquettopologicalphasesanomalous,mori2023floquetopen}. For a two-dimensional (2D) film with an out-of-plane N\'eel vector $\boldsymbol{N}\parallel\hat{z}$ under normal incidence, this field is parallel to $\boldsymbol{N}$ and therefore produces no transverse torque, since $\boldsymbol{\tau}\propto\boldsymbol{N}\times\boldsymbol{h}_{\rm eff}=0$. Generating a finite transverse response requires an effective field with a component perpendicular to the N\'eel vector~\cite{dannegger2021switching,wang2020ultrafast}.
		
		Bichromatic light~\cite{trevisan2022bicircular,yarmohammadi2026giant,yarmohammadi2026amr,6jv2-nmnn}, consisting of a fundamental frequency $\omega$ and an integer harmonic $b\omega$, provides an additional degree of freedom for generating such a transverse torque. Interference between the two frequency components produces a time-dependent optical pattern whose form depends on their relative phase and helicities. In 2D interfaces and monolayers, Rashba spin-orbit coupling (RSOC, $\lambda$), arising from broken structural inversion symmetry, couples the driven orbital motion to the electron spin~\cite{amin2020interfacial,hellman2017interface,wang2025orbitronics,song2021spintorques,manchon2019current,shao2021roadmap,gao2020spintorques,ji2024controlling,bychkov1984oscillatory,manchon2015new,zollner2025soc}. Here, torque refers to the electronic generalized torque obtained
		from the occupied Floquet states; it is not a direct calculation
		of the ensuing local-moment dynamics. Specifically, we define $\boldsymbol{\tau}=\boldsymbol{N}\times\boldsymbol{h}_{\rm eff}$, where $\boldsymbol{h}_{\rm eff}$ is the energy-valued generalized field acting on the electron spins per occupied state. This is a field-driven electronic response and does not rely on charge-current-induced spin transfer.
		
		To identify general principles that are not tied to a specific material, we use a generalized low-energy continuum model for 2D magnetic multipoles. In the absence of RSOC, the spin of each Floquet state remains collinear with $\boldsymbol{N}$, and the transverse torque is identically zero. Within the dipole approximation, the optical vector potential is spatially uniform and shifts the electronic momentum identically across the Brillouin zone. The resulting drive introduces two distinct types of terms to the exchange interaction: a spatially uniform, momentum-independent effective field and momentum-dependent terms arising from the light-induced mixing of lower multipoles into the exchange texture.
		
		We first treat the momentum-independent effective field using a high-frequency expansion and photon-counting arguments. Crucially, however, the field evaluated at a single momentum, such as the zone center ($\Gamma$ point), does not represent the true macroscopic torque of a metallic Fermi sea. In the interior of the Fermi sea where both spin-split bands are occupied, their torque contributions cancel identically, leaving only the thin shell between the two spin-split Fermi surfaces. To capture the full response, we move beyond high-frequency effective field approximations and compute the torque by integrating the exact, nonperturbative Floquet states at each momentum over the occupied Fermi sea. Our calculation uses a minimal model with occupations connected to the undriven bands, neglecting relaxation and laser heating, establishing rigorous selection rules and scaling laws. We obtain the following main results for a magnetic multipole with static order $\mathcal{O}(\boldsymbol{k}^n)$ under $(\omega,b\omega)$ drive with amplitudes $(\mathcal{A}_1,\mathcal{A}_2)$ and helicities $(\eta_1,\eta_2)$.
		
		The $\boldsymbol{k}$-independent effective field follows photon-counting selection rules. Its leading resonant contribution occurs at $b=n$ and scales as $M\epsilon\lambda\mathcal{A}_1^n\mathcal{A}_2/(\hbar\omega)$ with $M$ being the exchange amplitude of the multipole. Its direction depends on relative phase $\phi$ of beams, and destructive interference between competing virtual photon-pair pathways ($b=n-2$) produces leading-order zeros of the effective field (``optical blindspots'') at specific amplitude ratios $\mathcal{A}_2/\mathcal{A}_1$. For $b<n$, this transverse field requires counter-rotating beams, whereas for $b>n\ge2$ all multipole-induced transverse terms vanish, leaving only the longitudinal IFE field. 
		
		In a metallic Fermi sea, occupied-band cancellation restricts the contributing states and, together with spin-matrix-element and momentum-integration effects that we do not separate, suppresses the resonant torque by $10^2$--$10^4$ relative to the bare $\boldsymbol{k}$-independent field. Instead, $\boldsymbol{k}$-dependent exchange terms dominate the Fermi-sea response across several nonresonant channels. These torques exhibit no amplitude-ratio blindspots in the seven channels (defined by multipole order $n=1$--$6$, harmonic $b=1$--$7$, and helicity pair) tested, are approximately independent of $\omega$, and scale approximately as $\lambda^2$ (exponents 1.1--2.0). 
		
		For an out-of-plane N\'eel vector ($\boldsymbol{N}\parallel\hat{z}$), rotational symmetry imposes an exact selection rule on the Fermi-sea torque: a transverse response is symmetry-allowed if and only if $\gcd(n,b+1)=1$ for counter-rotating beams ($\eta_1\eta_2=-1$) or $\gcd(n,b-1)=1$ for co-rotating beams ($\eta_1\eta_2=+1$). Across the 84 channels, this rule excludes 34 channels, for which the calculated torque vanishes to numerical precision. The rule gives the allowed channels; it does not guarantee that an allowed channel is nonzero. The allowed channels fall into two categories: Class~I (six counter-rotating channels with $b+1\le n$), which originate from the cycle-averaged exchange texture and are frequency-independent; and Class~II, which arise from oscillating exchange terms and fall off as $1/\hbar\omega$ or $1/(\hbar\omega)^2$. The relative phase $\phi$ controls the torque direction, rotating it with a discrete winding number $v$ per $2\pi$ cycle, while the amplitudes set its magnitude. Several of the largest responses occur at $b=1$, where both beams have the same frequency and the drive is a single-frequency elliptically polarized field (linearly polarized for equal amplitudes); the relative phase then sets its polarization axis.
		
		For an in-plane N\'eel vector ($\boldsymbol{N}\parallel\hat{x}$ or $\hat{y}$), the out-of-plane rotation symmetry is absent and the $\gcd$ restriction is lifted. Ten counter-rotating channels with $b+1<n$ (Class~I$'$) produce an in-plane field perpendicular to $\boldsymbol N$; the resulting torque is along $\hat z$ and rotates $\boldsymbol N$ within the plane. It follows the cycle-averaged exchange and is frequency-independent, reaching amplitudes up to 2.5 times larger than the maximum out-of-plane response. A weaker response occurs at $b=n-1$ (Class~II$'$); its field is along $\hat z$, so its torque tilts $\boldsymbol N$ out of the plane, and it decreases as $1/\hbar\omega$.
		
		Comparing the full response with calculations isolating individual parts of the drive allows us to disentangle their microscopic origins. Because the torque per magnetic moment scales down by the carrier-to-spin ratio, the resulting torque functions as an ultrafast electronic steering torque that acts on the order parameter $\boldsymbol N$ (local-moment dynamics are not calculated), rather than a deterministic mechanism for $180^\circ$ switching of local moments.
		
		As a control, we repeat the calculation for a Rashba ferromagnet ($n=0$): only one out-of-plane channel is allowed, and the in-plane torque is small and has no selection rule. Thus, momentum-dependent spin splitting is essential for torques studied here.
		
		\section{Model and Floquet Formalism}
		\label{sec:model}
		\begin{figure*}[t]
			\centering
			\includegraphics[width=0.75\linewidth]{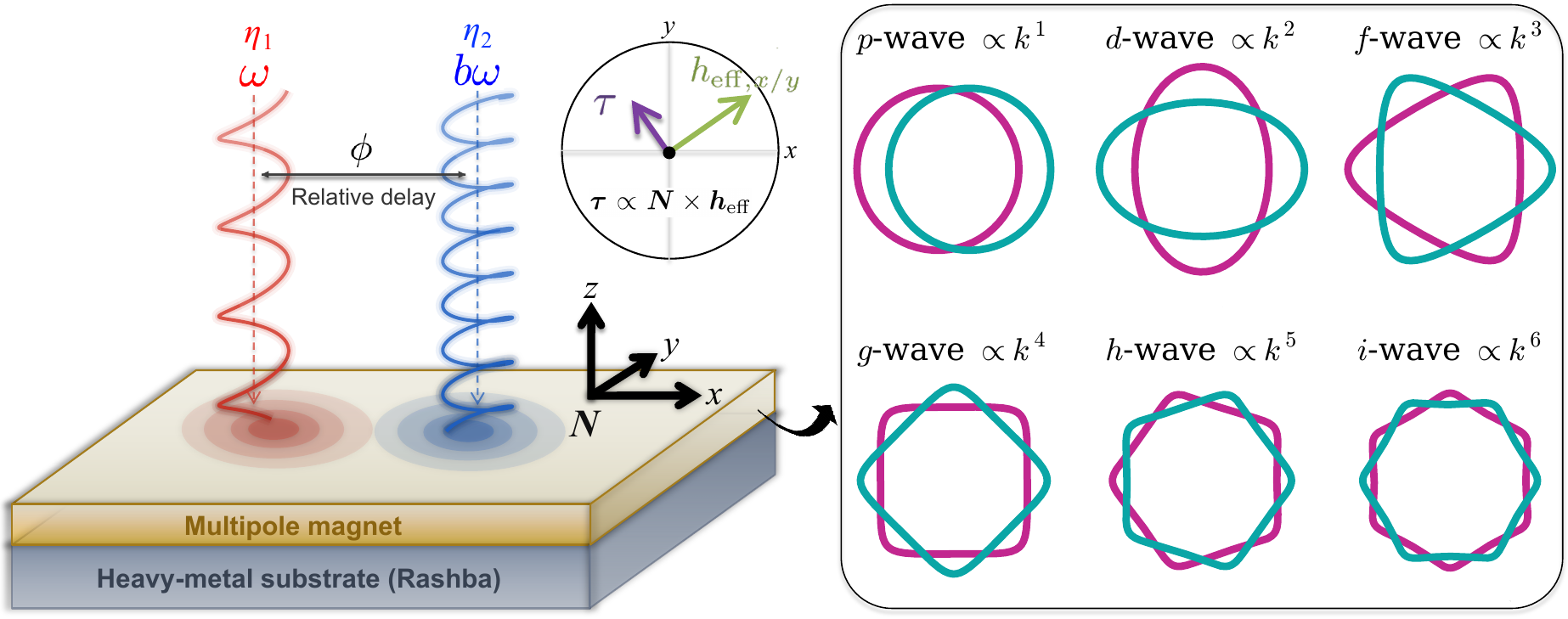}
			\caption{Schematic of the optical driving setup and magnetic multipole family.
				Left: Two collinear beams with frequencies $\omega$ and $b\omega$,
				helicities $\eta_1$ and $\eta_2$ ($+1$ and $-1$ denote counterclockwise
				and clockwise polarization, respectively), and relative phase $\phi$
				illuminate a thin magnetic film on a heavy-metal substrate. The substrate
				provides the interfacial RSOC.
				The inset shows the top view for $\boldsymbol{N}\parallel\hat{z}$,
				illustrating the in-plane light-induced effective field
				$\boldsymbol{h}_{{\rm eff}} = (h_{{\rm eff},x},h_{{\rm eff},y})$ and the associated electronic
				generalized torque $\boldsymbol{\tau}\propto
				\boldsymbol{N}\times\boldsymbol{h}_{\rm eff}$.
				Right: Fermi contours illustrating the momentum-dependent exchange
				textures $\mathcal{M}_n(\boldsymbol{k})\propto k^n$ defined in
				Eq.~\eqref{eq:multipoles}, from $p$-wave ($n=1$) to $i$-wave ($n=6$).
				Magenta and cyan indicate spin polarization parallel and antiparallel
				to $\boldsymbol{N}$, respectively. The $p$-, $f$-, and $h$-wave
				textures have odd parity, whereas the $d$-, $g$-, and $i$-wave
				textures have even parity and describe altermagnetic order.
				The exchange amplitude is enhanced for clarity, and RSOC is omitted
				from the schematic Fermi contours.}
			\label{f0}
		\end{figure*}
		We consider a 2D low-energy model with a quadratic kinetic energy, RSOC~[Fig.~\ref{f0}] arising from broken structural inversion symmetry~\cite{bychkov1984oscillatory,manchon2015new,zollner2025soc}, and a momentum-dependent exchange field~\cite{hayami2019momentum,smejkal2022beyond,smejkal2022emerging,bai2024altermagnetism,song2025altermagnets}. The N\'eel vector $\boldsymbol{N}$ couples to the electron spin through the Pauli matrices $\boldsymbol{\sigma}=(\sigma_x,\sigma_y,\sigma_z)$. For the three principal orientations $\boldsymbol{N}\parallel\hat{\boldsymbol{j}}$, with $j\in\{x,y,z\}$, write the static Hamiltonian as
		{\small\begin{equation}
				\label{eq:hamiltonian}
				H_{\text{s}}^{(j)}(\boldsymbol{k})
				=\epsilon(k_x^2+k_y^2)\sigma_0
				+\lambda(k_x\sigma_y-k_y\sigma_x)
				+\epsilon\mathcal{M}_n(\boldsymbol{k}) \bs N\cdot\bs \sigma\,.
		\end{equation}}Here, $\epsilon=\hbar^2k_{\rm F}^2/2m^*$ and $\lambda=\alpha_Rk_{\rm F}$, where $\alpha_R$ is the Rashba coefficient. Momentum is expressed in units of $k_{\rm F}$, so $\epsilon$ and $\lambda$ have units of energy. 
		
		The function $\mathcal{M}_n(\boldsymbol{k})$ describes the momentum dependence of the exchange texture and has order $\mathcal{O}(\bs k^n)$. We use the real part of $(k_x+ik_y)^n$; choosing its imaginary part gives the same texture after an in-plane rotation of the coordinate axes~\cite{xl2d-88z7}. With dimensionless exchange amplitude $M$, the multipoles considered here are~\cite{5ys6-pq68,zxcv-xcyc,zt4l-y18j,p21c-rtv7,PhysRevLett.133.236703,PhysRevB.111.L220403,shen2026roomtemperatureferroelectricpwavemagnet}
		\begin{subequations}
			\label{eq:multipoles}
			\begin{align}
				p\text{-wave} &: \quad
				\mathcal{M}_1(\boldsymbol{k})=M k_x,\\
				d\text{-wave} &: \quad
				\mathcal{M}_2(\boldsymbol{k})=M(k_x^2-k_y^2),\\
				f\text{-wave} &: \quad
				\mathcal{M}_3(\boldsymbol{k})=M(k_x^3-3k_xk_y^2),\\
				g\text{-wave} &: \quad
				\mathcal{M}_4(\boldsymbol{k})
				=M(k_x^4-6k_x^2k_y^2+k_y^4),\label{eq_2d}\\
				h\text{-wave} &: \quad
				\mathcal{M}_5(\boldsymbol{k})
				=M(k_x^5-10k_x^3k_y^2+5k_xk_y^4),\\
				i\text{-wave} &: \quad
				\mathcal{M}_6(\boldsymbol{k})
				=M(k_x^6-15k_x^4k_y^2+15k_x^2k_y^4-k_y^6).
			\end{align}
		\end{subequations}
		The $d$-, $g$-, and $i$-wave exchange textures ($n=2,4,6$) are even under $\boldsymbol{k}\rightarrow-\boldsymbol{k}$ and describe even-parity altermagnetic order. The $p$-, $f$-, and $h$-wave textures ($n=1,3,5$) are odd under this transformation. These parity labels refer to the exchange texture, not to the full Hamiltonian, since the Rashba term is also odd in momentum. In the odd-parity cases, the exchange field reverses under $\boldsymbol{k}\rightarrow-\boldsymbol{k}$, giving a momentum-dependent spin texture with zero net magnetization in the model. The case $n=0$ is a uniform exchange field, a Rashba ferromagnet.
		
		Real odd-parity magnets generally contain several magnetic sublattices. For the $p$-wave case, we assume that these sublattices can be integrated out, leaving an effective description with the exchange term in Eq.~\eqref{eq:hamiltonian}. Appendix~\ref{app:pwave} derives this reduction, including Rashba coupling, for a four-band model~\cite{p21c-rtv7}. Accordingly, the Hamiltonian is intended
		as a minimal effective description of the symmetry and momentum
		dependence of the magnetic exchange texture, rather than as a
		microscopic model derived for a specific material. For odd $n$, $\boldsymbol N$ is not a conventional time-reversal-odd N\'eel vector. In the $p$-wave case it is the normal of the plane of coplanar moments, which is even under time reversal, and the reduced Hamiltonian has ordinary spinful time-reversal symmetry (Appendix~\ref{app:pwave}). The torque $\boldsymbol N\times\boldsymbol h_{\rm eff}$ is then the generalized torque on the plane normal, without conventional N\'eel-vector dynamics. We have not established an analogous microscopic reduction for the $f$- and $h$-wave cases. Their extension to these higher odd-parity orders is therefore
		a modeling assumption. 
		
		We drive the system with two frequency components, $\omega$ and its integer harmonic $b\omega$ [Fig.~\ref{f0}]~\cite{oka2019floquet,trevisan2022bicircular,yarmohammadi2026giant,yarmohammadi2026amr}. Within the dipole approximation, the vector potential is spatially uniform and takes the form
		\begin{subequations}
			\label{eq:vector_potential}
			\begin{align}
				\mathcal{A}_x(t)
				&=\mathcal{A}_1\cos(\omega t)
				+\mathcal{A}_2\cos(b\omega t+\phi),\\
				\mathcal{A}_y(t)
				&=\eta_1\mathcal{A}_1\sin(\omega t)
				+\eta_2\mathcal{A}_2\sin(b\omega t+\phi).
			\end{align}
		\end{subequations}
		The dimensionless amplitudes are
		$\mathcal{A}_1=eE_1/(\hbar\omega k_{\rm F})$ and
		$\mathcal{A}_2=eE_2/(\hbar b\omega k_{\rm F})$, where $E_1$ and $E_2$ are the electric-field amplitudes. The helicity $\eta=+1$ ($-1$) denotes counterclockwise (clockwise) rotation of the vector potential. Thus, the two components are co-rotating for $\eta_1=\eta_2$ and counter-rotating for $\eta_1=-\eta_2$~\cite{trevisan2022bicircular,ogawa2024bicircular}. The relative phase $\phi$ controls the phase offset between the two components and provides a means of changing the direction of the induced response.
		
		The light couples to the electrons through the Peierls substitution
		$\boldsymbol{k}\rightarrow\boldsymbol{k}+\boldsymbol{\mathcal{A}}(t)$, giving $H(\boldsymbol{k},t)
		=H_{\text{s}}\big(\boldsymbol{k}+\boldsymbol{\mathcal{A}}(t)\big).$
		This substitution acts on both the Rashba and exchange terms. In particular, expanding
		$\mathcal{M}_n(\boldsymbol{k}+\boldsymbol{\mathcal{A}}(t))$
		generates terms with powers of the vector-potential amplitude up to order $n$. Thus, the multipole order sets the maximum power of the optical drive entering the exchange term: the $p$-wave texture contains terms up to first order in $\mathcal{A}$, whereas the $i$-wave texture contains terms up to sixth order.
		
		For integer $b$, the driven Hamiltonian is periodic with period
		$T=2\pi/\omega$. At each momentum $\boldsymbol{k}$, we write its Floquet
		states as
		$|\psi_s(t)\rangle=e^{-i\varepsilon_st/\hbar}|u_s(t)\rangle$,
		where $|u_s(t+T)\rangle=|u_s(t)\rangle$ and
		$\varepsilon_s(\boldsymbol{k})$ are the quasienergies, with $s=\pm$.
		We obtain the Floquet states numerically from the one-period evolution
		operator, without using a high-frequency expansion~\cite{oka2019floquet,bukov2015universal}.
		
		Since $\boldsymbol{N}$ enters the Hamiltonian only through the exchange
		term, the derivative with respect to the N\'eel vector is
		$\partial H/\partial\boldsymbol{N}
		=\epsilon\mathcal{M}_n(\boldsymbol{k}+\boldsymbol{\mathcal{A}}(t))
		\boldsymbol{\sigma}$.
		The Hellmann--Feynman theorem~\cite{hellmann1937einfuhrung,PhysRev.56.340} then gives the energy-valued generalized
		field exerted by the occupied electronic states,
		\begin{equation}
			\label{eq:heff}
			\boldsymbol{h}_{\rm eff}
			=-\frac{1}{N_{\rm el}}
			\sum_{\boldsymbol{k}}\sum_{s=\pm}
			n_s(\boldsymbol{k})\,
			\overline{
				\epsilon\mathcal{M}_n\big(\boldsymbol{k}+\boldsymbol{\mathcal{A}}(t)\big)
				\langle\boldsymbol{\sigma}\rangle_s(t)
			}\,,
		\end{equation}
		where
		$\langle\boldsymbol{\sigma}\rangle_s(t)
		=\langle\psi_s(t)|\boldsymbol{\sigma}|\psi_s(t)\rangle$,
		the overline denotes an average over one period,
		$n_s(\boldsymbol{k})$ is the occupation of Floquet band $s$, and
		$N_{\rm el}=\sum_{\boldsymbol{k},s}n_s(\boldsymbol{k})$ is the total
		number of occupied states. We subtract the corresponding undriven
		contribution and retain only the light-induced part of
		$\boldsymbol{h}_{\rm eff}$. The associated electronic generalized
		torque is
		$\boldsymbol{\tau}=\boldsymbol{N}\times\boldsymbol{h}_{\rm eff}$.
		Only the component perpendicular to $\boldsymbol{N}$ contributes. We call this component the transverse torque $\tau_\perp=\boldsymbol N\times\boldsymbol h_{\rm eff}$; it is perpendicular to $\boldsymbol N$ by construction, so $|\tau_\perp|=|\boldsymbol\tau|$. For a unit N\'eel vector,
		$|\boldsymbol{\tau}|=|\boldsymbol{h}_{{\rm eff},\perp}|$. For $\boldsymbol N\parallel\hat z$ its components are $(\tau_x,\tau_y)$, and for $\boldsymbol N\parallel\hat x$ they are $(\tau_y,\tau_z)$.
		Both $\boldsymbol{h}_{\rm eff}$ and $\boldsymbol{\tau}$ are energy-valued quantities per occupied electron. Because $H$ is linear in $\boldsymbol N$, $\partial H/\partial\boldsymbol N$ does not depend on $\boldsymbol N$ and Eq.~\eqref{eq:heff} is exact within the occupation prescription. It is the force conjugate to $\boldsymbol N$, weighted by the spin expectation values of the occupied Floquet states. It differs from the $\boldsymbol k$-independent effective spin field of Sec.~\ref{sec:kindep}, which is the coefficient of $\boldsymbol\sigma$; we call the latter an effective field, and we compare the two only in Sec.~\ref{sec:both}. The conversion of this electronic torque into a dynamical
		equation for the local Néel vector requires additional
		microscopic information about the magnetic moments and their
		coupling to the electronic system, which is beyond the scope
		of the present work.
		
		The occupations are assigned from the undriven bands connected to the
		Floquet states during the drive ramp:
		$n_s(\boldsymbol{k})=f[E_s(\boldsymbol{k})]$, where $E_s$ are the
		static band energies of Eq.~\eqref{eq:hamiltonian} and $f$ is the Fermi
		function at chemical potential $\mu=400$~meV and temperature $T$. Numerically,
		we increase the drive amplitudes in eight steps (128 for in-plane $\boldsymbol N$, Appendix~\ref{app:num}), each lasting one period, and track the Floquet states by maximum overlap
		(Appendix~\ref{app:num}). This prescription specifies how the
		occupations are assigned in our calculation; it does not include
		relaxation, heating, or redistribution between Floquet states.
		Equation~\eqref{eq:heff} therefore gives the electronic response of
		the driven model using its Floquet states, rather than an effective
		field obtained from a truncated expansion in $1/\omega$. Thus, the calculated response corresponds to this prescribed
		occupation scheme and should not be interpreted as the
		steady-state response of a system with relaxation.
		
		Expanding the exchange texture as
		$
		\mathcal{M}_n(\boldsymbol{k}+\boldsymbol{\mathcal{A}})
		=M\sum_{j=0}^{n}\binom{n}{j}
		\operatorname{Re}\left[k_+^{\,n-j}\mathcal{A}_+^{\,j}\right],
		$
		where $k_\pm=k_x\pm ik_y$ and
		$\mathcal{A}_\pm=\mathcal{A}_x\pm i\mathcal{A}_y$, separates the
		drive-induced exchange terms into a $\boldsymbol{k}$-independent
		contribution ($j=n$) and $\boldsymbol{k}$-dependent contributions
		($j<n$). We consider three cases: the $\boldsymbol{k}$-independent
		contribution alone (Sec.~\ref{sec:kindep}), the
		$\boldsymbol{k}$-dependent contributions alone (Sec.~\ref{sec:kdep}),
		and the full driven exchange texture, which gives the complete
		response of the model (Sec.~\ref{sec:both}). In all three calculations,
		the Rashba term is driven through the same Peierls substitution, and
		the Floquet states are obtained from the full time-dependent
		Hamiltonian. The separate contributions are therefore useful for
		identifying the roles of different terms in the exchange texture.
		
		Without RSOC, the Hamiltonian contains only $\sigma_0$ and $\sigma_j$. Its Floquet eigenstates can then be chosen as eigenstates of $\sigma_j$, and their spin expectation values remain parallel or antiparallel to $\boldsymbol{N}$. Consequently, no transverse electronic torque is generated. RSOC introduces spin components perpendicular to $\boldsymbol{N}$ and allows the driven spin expectation values to acquire a transverse component. Thus, RSOC is necessary for the torque here.
		
		Throughout this work, $\tau_\perp$ denotes the magnitude of the torque component perpendicular to $\boldsymbol N$. Since $\boldsymbol\tau=\boldsymbol N\times\boldsymbol h_{\rm eff}$ is perpendicular to $\boldsymbol N$, and
		\begin{equation}
			|\tau_\perp|=
			\begin{cases}
				(\tau_y^2+\tau_z^2)^{1/2}, & \boldsymbol N\parallel\hat x,\\[2pt]
				(\tau_x^2+\tau_z^2)^{1/2}, & \boldsymbol N\parallel\hat y,\\[2pt]
				(\tau_x^2+\tau_y^2)^{1/2}, & \boldsymbol N\parallel\hat z.
			\end{cases}
			\label{eq:tauperp}
		\end{equation}
		
		\section{Torque from the $\textbf{\textit k}$-independent light-induced part}
		\label{sec:kindep}\label{sec:analytical_fields}
		
		We first consider the $\boldsymbol{k}$-independent contribution ($j=n$) to the driven exchange texture. This term depends only on the vector potential and gives the momentum-independent part of the first-order high-frequency effective Hamiltonian, $H_{\mathrm{eff}}=h_0\sigma_0+\boldsymbol{h}_{\mathrm{eff}}\cdot\boldsymbol{\sigma}$. Here, $\boldsymbol{h}_{\mathrm{eff}}$ is an energy-valued effective field acting on the electron spin, not the macroscopic magnetic field used directly in a conventional Landau--Lifshitz--Gilbert equation. $\boldsymbol{\tau}=\boldsymbol{N}\times\boldsymbol{h}_{\mathrm{eff}}$ is a field-driven electronic response, rather than a conventional current-induced spin-transfer torque; the calculation does not assume or require a charge-current-induced spin accumulation. Throughout this section we use $M\epsilon=10$~meV, $\lambda=0.5$~meV, and $\hbar\omega=0.124$~eV (30~THz), with optical amplitudes up to $0.5$. We express the fields and generalized torques in units of $u_0=M\epsilon\lambda_0/\hbar\omega=40~\mu$eV, with $\lambda_0=0.5$~meV kept fixed when $\lambda$ is varied. These parameters are representative and are not intended to describe a specific material. For these values, $\max_t|\boldsymbol{h}(t)|/\hbar\omega\lesssim0.08$. The first-order high-frequency results agree with the numerical Floquet calculation of Appendix~\ref{app:floquet} to within a few percent.
		
		Table~\ref{tab:effective_fields} summarizes the momentum-independent effective fields for the three principal N\'eel-vector orientations. Apart from the IFE contribution, $h_{\mathrm{IFE}}\propto\lambda^2$, the terms containing $\lambda$ are transverse to $\boldsymbol{N}$, whereas the terms without $\lambda$ are parallel to $\boldsymbol{N}$. Their presence depends on the relation between the harmonic number $b$ and the multipole order $n$, as well as on the beam helicities. In particular, for $n\geq2$ and $b>n$, only the IFE contribution remains in this momentum-independent field. These selection rules characterize this contribution; they should not be identified with selection rules for the complete Fermi-sea torque, which also contains momentum-dependent exchange terms.\begin{table*}[t!]
			\centering
			\caption{Momentum-independent components of the light-induced effective field, $h_{\mathrm{eff},x}$, $h_{\mathrm{eff},y}$, and $h_{\mathrm{eff},z}$, for multipole orders $n=1$--$6$ under bichromatic $(\omega,b\omega)$ driving. The Kronecker deltas $\delta_{b,m}$ identify the harmonic numbers for which a static contribution occurs. Terms without $\lambda$ are parallel to $\boldsymbol{N}$ and do not contribute to $\boldsymbol{\tau}$. The IFE contribution, $h_{\mathrm{IFE}}=-\lambda^2\Omega_b$ [Eq.~\eqref{eq:h_IFE_def}], points along $\hat{\boldsymbol z}$. The remaining terms containing $\lambda$ are transverse to $\boldsymbol{N}$. For $n\geq2$ and $b>n$, only the IFE contribution remains; the $p$-wave terms proportional to $\Omega_b$ occur for every $b$. The table gives effective fields, not the complete Fermi-sea torque.}
			\label{tab:effective_fields}
			\renewcommand{\arraystretch}{2.5}
			\resizebox{0.745\width}{!}{%
				\begin{tabular}{c|c|c|c|c}
					\hline\hline
					Multipole& N\'eel state & $h_{\text{eff},x}^{(\boldsymbol{N})}$ & $h_{\text{eff},y}^{(\boldsymbol{N})}$ & $h_{\text{eff},z}^{(\boldsymbol{N})}$ \\
					\hline
					
					\rowcolor{highlightcolor1}[\tabcolsep][\tabcolsep]
					& $\boldsymbol{N} \parallel \hat{x}$ & $0$ & $0$ & $h_{\text{IFE}}$ \\
					\hhline{~|-|-|-|-}
					
					\rowcolor{highlightcolor1}[\tabcolsep][\tabcolsep]
					& $\boldsymbol{N} \parallel \hat{y}$ & $0$ & $0$ & 
					$\begin{aligned}
						&h_{\text{IFE}}
						-\lambda M\epsilon \, \Omega_b
					\end{aligned}$ \\
					\hhline{~|-|-|-|-}
					
					\rowcolor{highlightcolor1}[\tabcolsep][\tabcolsep]
					\multirow{-3}{*}{\shortstack{$p$-wave \\ ($n=1$) \\ Odd-parity}} 
					& $\boldsymbol{N} \parallel \hat{z}$ 
					& $0$ & $M\epsilon\lambda  \, \Omega_b$ & $h_{\text{IFE}}$ \\
					\hline
					
					\multirow{3}{*}{\shortstack{$d$-wave \\ ($n=2$) \\ Even-parity}} 
					& $\boldsymbol{N} \parallel \hat{x}$ 
					& $-\delta_{b,1} \mathcal{A}_1\mathcal{A}_2M\epsilon (\eta_1\eta_2-1) \cos\phi$ & 
					$0$ & 
					$\begin{aligned}
						&h_{\text{IFE}} - \delta_{b,2} \frac{\mathcal{A}_1^2\mathcal{A}_2M\epsilon\lambda }{2\hbar\omega}(1-2\eta_1\eta_2)\sin\phi
					\end{aligned}$ \\\hhline{~|-|-|-|-}
					& $\boldsymbol{N} \parallel \hat{y}$ 
					& $0$ & 
					$-\delta_{b,1} \mathcal{A}_1\mathcal{A}_2M\epsilon (\eta_1\eta_2-1) \cos\phi$ & 
					$\begin{aligned}
						&h_{\text{IFE}} - \delta_{b,2} \frac{\mathcal{A}_1^2\mathcal{A}_2M\epsilon\lambda }{2\hbar\omega}(2\eta_1-\eta_2)\cos\phi
					\end{aligned}$ \\\hhline{~|-|-|-|-}
					& $\boldsymbol{N} \parallel \hat{z}$ 
					& $\delta_{b,2} \frac{\mathcal{A}_1^2\mathcal{A}_2M\epsilon\lambda }{2\hbar\omega}(1-2\eta_1\eta_2)\sin\phi$ & 
					$\delta_{b,2} \frac{\mathcal{A}_1^2\mathcal{A}_2M\epsilon\lambda }{2\hbar\omega}(2\eta_1-\eta_2)\cos\phi$ & 
					$\begin{aligned}
						&h_{\text{IFE}}- \delta_{b,1}\mathcal{A}_1 \mathcal{A}_2 M\epsilon(\eta_1\eta_2-1) \cos\phi
					\end{aligned}$ \\
					\hline
					
					\rowcolor{highlightcolor}[\tabcolsep][\tabcolsep]
					& $\boldsymbol{N} \parallel \hat{x}$ & $-\delta_{b,2} \frac{3}{2}\mathcal{A}_1^2\mathcal{A}_2M\epsilon(\eta_1\eta_2-1)\cos\phi$ & 
					$0$ & 
					\shortstack{$h_{\text{IFE}} - \delta_{b,1} \frac{3\mathcal{A}_1\mathcal{A}_2M\epsilon\lambda }{2\hbar\omega}(\mathcal{A}_2^2-\mathcal{A}_1^2)(1-\eta_1\eta_2)\sin\phi$ \\ 
						$- \delta_{b,3} \frac{\mathcal{A}_1^3\mathcal{A}_2M\epsilon\lambda }{6\hbar\omega}(7-9\eta_1\eta_2)\sin\phi$} \\\hhline{~|-|-|-|-}
					\rowcolor{highlightcolor}[\tabcolsep][\tabcolsep]
					& $\boldsymbol{N} \parallel \hat{y}$ & $0$ & 
					$-\delta_{b,2} \frac{3}{2}\mathcal{A}_1^2\mathcal{A}_2M\epsilon(\eta_1\eta_2-1)\cos\phi$ & 
					\shortstack{$h_{\text{IFE}} - \delta_{b,1} \frac{3M\epsilon\lambda  \mathcal{A}_1\mathcal{A}_2}{2\hbar\omega}(\eta_1-\eta_2)(\mathcal{A}_1^2-\mathcal{A}_2^2)\cos\phi$ \\ 
						$- \delta_{b,3} \frac{M\epsilon\lambda  \mathcal{A}_1^3\mathcal{A}_2}{6\hbar\omega}(9\eta_1-7\eta_2)\cos\phi$}  \\\hhline{~|-|-|-|-}
					\rowcolor{highlightcolor}[\tabcolsep][\tabcolsep]
					\multirow{-3}{*}[1em]{\shortstack{$f$-wave \\ ($n=3$) \\ Odd-parity}} 
					& $\boldsymbol{N} \parallel \hat{z}$ 
					& \shortstack{$\delta_{b,1} \frac{3\mathcal{A}_1\mathcal{A}_2M\epsilon\lambda}{2\hbar\omega}(\mathcal{A}_2^2-\mathcal{A}_1^2)(1-\eta_1\eta_2)\sin\phi$ \\ 
						$+ \delta_{b,3} \frac{\mathcal{A}_1^3\mathcal{A}_2M\epsilon\lambda}{6\hbar\omega}(7-9\eta_1\eta_2)\sin\phi$} & 
					\shortstack{$\delta_{b,1} \frac{3\mathcal{A}_1\mathcal{A}_2M\epsilon\lambda}{2\hbar\omega}(\eta_1-\eta_2)(\mathcal{A}_1^2-\mathcal{A}_2^2)\cos\phi$ \\ 
						$+ \delta_{b,3} \frac{\mathcal{A}_1^3\mathcal{A}_2M\epsilon\lambda}{6\hbar\omega}(9\eta_1-7\eta_2)\cos\phi$} & 
					\shortstack{$h_{\text{IFE}}$ $- \delta_{b,2} \frac{3}{2}\mathcal{A}_1^2\mathcal{A}_2M\epsilon(\eta_1\eta_2-1)\cos\phi$} \\
					\hline
					
					\multirow{3}{*}[-2em]{\shortstack{$g$-wave \\ ($n=4$) \\ Even-parity}} 
					& $\boldsymbol{N} \parallel \hat{x}$ 
					& $\begin{aligned}
						&\delta_{b,1} 3\mathcal{A}_1^2\mathcal{A}_2^2M\epsilon(1-\eta_1\eta_2)\cos(2\phi) \\ 
						&+ \delta_{b,3} 2\mathcal{A}_1^3\mathcal{A}_2M\epsilon(1-\eta_1\eta_2)\cos\phi
					\end{aligned}$ & 
					$0$ & 
					$\begin{aligned}
						&h_{\text{IFE}} + \delta_{b,2} \frac{\lambda M\epsilon}{2\hbar\omega}\mathcal{A}_1^2\mathcal{A}_2(4\mathcal{A}_1^2-3\mathcal{A}_2^2)(1-\eta_1\eta_2)\sin\phi \\ 
						&- \delta_{b,4} \frac{\lambda M\epsilon}{4\hbar\omega}\mathcal{A}_1^4\mathcal{A}_2(7-8\eta_1\eta_2)\sin\phi
					\end{aligned}$ \\\hhline{~|-|-|-|-}
					& $\boldsymbol{N} \parallel \hat{y}$ 
					& $0$ & 
					$\begin{aligned}
						&\delta_{b,1} 3\mathcal{A}_1^2\mathcal{A}_2^2M\epsilon(1-\eta_1\eta_2)\cos(2\phi) \\ 
						&+ \delta_{b,3} 2\mathcal{A}_1^3\mathcal{A}_2M\epsilon(1-\eta_1\eta_2)\cos\phi
					\end{aligned}$ & 
					$\begin{aligned}
						&h_{\text{IFE}} - \delta_{b,2} \frac{\lambda M\epsilon}{2\hbar\omega}\mathcal{A}_1^2\mathcal{A}_2(4\mathcal{A}_1^2-3\mathcal{A}_2^2)(\eta_1-\eta_2)\cos\phi \\ 
						&- \delta_{b,4} \frac{\lambda M\epsilon}{4\hbar\omega}\mathcal{A}_1^4\mathcal{A}_2(8\eta_1-7\eta_2)\cos\phi
					\end{aligned}$ \\\hhline{~|-|-|-|-}
					& $\boldsymbol{N} \parallel \hat{z}$ 
					& $\begin{aligned}
						&-\delta_{b,2} \frac{\lambda M\epsilon}{2\hbar\omega}\mathcal{A}_1^2\mathcal{A}_2(4\mathcal{A}_1^2-3\mathcal{A}_2^2)(1-\eta_1\eta_2)\sin\phi \\ 
						&+ \delta_{b,4} \frac{\lambda M\epsilon}{4\hbar\omega}\mathcal{A}_1^4\mathcal{A}_2(7-8\eta_1\eta_2)\sin\phi
					\end{aligned}$ & 
					$\begin{aligned}
						&\delta_{b,2} \frac{\lambda M\epsilon}{2\hbar\omega}\mathcal{A}_1^2\mathcal{A}_2(4\mathcal{A}_1^2-3\mathcal{A}_2^2)(\eta_1-\eta_2)\cos\phi \\ 
						&+ \delta_{b,4} \frac{\lambda M\epsilon}{4\hbar\omega}\mathcal{A}_1^4\mathcal{A}_2(8\eta_1-7\eta_2)\cos\phi
					\end{aligned}$ & 
					$\begin{aligned}
						&h_{\text{IFE}}+ \delta_{b,1} 3\mathcal{A}_1^2\mathcal{A}_2^2M\epsilon(1-\eta_1\eta_2)\cos(2\phi) \\ 
						&+ \delta_{b,3} 2\mathcal{A}_1^3\mathcal{A}_2M\epsilon(1-\eta_1\eta_2)\cos\phi
					\end{aligned}$ \\
					\hline
					
					\multirow{3}{*}[-5.5em]{\shortstack{$h$-wave \\ ($n=5$) \\ Odd-parity}} 
					& $\boldsymbol{N} \parallel \hat{x}$ 
					& $\delta_{b,4}\frac{5}{2}\mathcal{A}_1^4\mathcal{A}_2M\epsilon(1-\eta_1\eta_2)\cos\phi$ & 
					$0$ & 
					$\begin{aligned}
						&h_{\text{IFE}}
						+\delta_{b,1}\frac{5M\epsilon\lambda}{\hbar\omega}\mathcal{A}_1^2\mathcal{A}_2^2(\mathcal{A}_1^2-\mathcal{A}_2^2)(1-\eta_1\eta_2)\sin(2\phi)\\
						&-\delta_{b,2}\frac{15M\epsilon\lambda}{4\hbar\omega}\mathcal{A}_1^4\mathcal{A}_2^2(1-\eta_1\eta_2)\sin(2\phi)\\
						&+\delta_{b,3}\frac{5M\epsilon\lambda}{6\hbar\omega}\mathcal{A}_1^3\mathcal{A}_2(3\mathcal{A}_1^2-2\mathcal{A}_2^2)(1-\eta_1\eta_2)\sin\phi\\
						&+\delta_{b,5}\frac{M\epsilon\lambda}{10\hbar\omega}\mathcal{A}_1^5\mathcal{A}_2(25\eta_1\eta_2-23)\sin\phi
					\end{aligned}$ \\\hhline{~|-|-|-|-}
					& $\boldsymbol{N} \parallel \hat{y}$ 
					& $0$ 
					& $\delta_{b,4}\frac{5}{2}\mathcal{A}_1^4\mathcal{A}_2M\epsilon(1-\eta_1\eta_2)\cos\phi$ 
					& $\begin{aligned}
						&h_{\text{IFE}}
						-\delta_{b,1}\frac{5M\epsilon\lambda}{\hbar\omega}\mathcal{A}_1^2\mathcal{A}_2^2(\mathcal{A}_1^2-\mathcal{A}_2^2)(\eta_1-\eta_2)\cos(2\phi)\\
						&-\delta_{b,2}\frac{15M\epsilon\lambda}{4\hbar\omega}\mathcal{A}_1^4\mathcal{A}_2^2(\eta_1-\eta_2)\cos(2\phi)\\
						&-\delta_{b,3}\frac{5M\epsilon\lambda}{6\hbar\omega}\mathcal{A}_1^3\mathcal{A}_2(3\mathcal{A}_1^2-2\mathcal{A}_2^2)(\eta_1-\eta_2)\cos\phi\\
						&-\delta_{b,5}\frac{M\epsilon\lambda}{10\hbar\omega}\mathcal{A}_1^5\mathcal{A}_2(25\eta_1-23\eta_2)\cos\phi
					\end{aligned}$ \\\hhline{~|-|-|-|-}
					& $\boldsymbol{N} \parallel \hat{z}$ 
					& $\begin{aligned}
						&-\delta_{b,1}\frac{5M\epsilon\lambda}{\hbar\omega}\mathcal{A}_1^2\mathcal{A}_2^2(\mathcal{A}_1^2-\mathcal{A}_2^2)(1-\eta_1\eta_2)\sin(2\phi)\\
						&+\delta_{b,2}\frac{15M\epsilon\lambda}{4\hbar\omega}\mathcal{A}_1^4\mathcal{A}_2^2(1-\eta_1\eta_2)\sin(2\phi)\\
						&-\delta_{b,3}\frac{5M\epsilon\lambda}{6\hbar\omega}\mathcal{A}_1^3\mathcal{A}_2(3\mathcal{A}_1^2-2\mathcal{A}_2^2)(1-\eta_1\eta_2)\sin\phi\\
						&-\delta_{b,5}\frac{M\epsilon\lambda}{10\hbar\omega}\mathcal{A}_1^5\mathcal{A}_2(25\eta_1\eta_2-23)\sin\phi
					\end{aligned}$ & 
					$\begin{aligned}
						&\delta_{b,1}\frac{5M\epsilon\lambda}{\hbar\omega}\mathcal{A}_1^2\mathcal{A}_2^2(\mathcal{A}_1^2-\mathcal{A}_2^2)(\eta_1-\eta_2)\cos(2\phi)\\
						&+\delta_{b,2}\frac{15M\epsilon\lambda}{4\hbar\omega}\mathcal{A}_1^4\mathcal{A}_2^2(\eta_1-\eta_2)\cos(2\phi)\\
						&+\delta_{b,3}\frac{5M\epsilon\lambda}{6\hbar\omega}\mathcal{A}_1^3\mathcal{A}_2(3\mathcal{A}_1^2-2\mathcal{A}_2^2)(\eta_1-\eta_2)\cos\phi\\
						&+\delta_{b,5}\frac{M\epsilon\lambda}{10\hbar\omega}\mathcal{A}_1^5\mathcal{A}_2(25\eta_1-23\eta_2)\cos\phi
					\end{aligned}$ & 
					$\begin{aligned}
						&h_{\text{IFE}}+\delta_{b,4}\tfrac{5}{2} \mathcal{A}_1^4\mathcal{A}_2M\epsilon(1-\eta_1\eta_2)\cos\phi
					\end{aligned}$ \\
					\hline
					
					\rowcolor{highlightcolor}[\tabcolsep][\tabcolsep]
					& $\boldsymbol{N} \parallel \hat{x}$ & 
					$\begin{aligned}
						&\delta_{b,1} 10\mathcal{A}_1^3\mathcal{A}_2^3M\epsilon(1-\eta_1\eta_2)\cos(3\phi) \\ 
						&+ \delta_{b,2} \frac{15}{2}\mathcal{A}_1^4\mathcal{A}_2^2M\epsilon(1-\eta_1\eta_2)\cos(2\phi) \\
						&+ \delta_{b,5} 3\mathcal{A}_1^5\mathcal{A}_2M\epsilon(1-\eta_1\eta_2)\cos\phi
					\end{aligned}$ & 
					$0$ & 
					$\begin{aligned}
						&h_{\text{IFE}}
						+ \delta_{b,4} \frac{3\lambda M\epsilon}{8\hbar\omega}\mathcal{A}_1^4\mathcal{A}_2(8\mathcal{A}_1^2-5\mathcal{A}_2^2)(1-\eta_1\eta_2)\sin\phi \\ 
						&- \delta_{b,6} \frac{\lambda M\epsilon}{6\hbar\omega}\mathcal{A}_1^6\mathcal{A}_2(17-18\eta_1\eta_2)\sin\phi
					\end{aligned}$ \\
					\hhline{~|-|-|-|-}
					
					\rowcolor{highlightcolor}[\tabcolsep][\tabcolsep]
					& $\boldsymbol{N} \parallel \hat{y}$ & $0$ & 
					$\begin{aligned}
						&\delta_{b,1} 10\mathcal{A}_1^3\mathcal{A}_2^3M\epsilon(1-\eta_1\eta_2)\cos(3\phi) \\ 
						&+ \delta_{b,2} \frac{15}{2}\mathcal{A}_1^4\mathcal{A}_2^2M\epsilon(1-\eta_1\eta_2)\cos(2\phi) \\
						&+ \delta_{b,5} 3\mathcal{A}_1^5\mathcal{A}_2M\epsilon(1-\eta_1\eta_2)\cos\phi
					\end{aligned}$ & 
					$\begin{aligned}
						&h_{\text{IFE}}
						- 				\delta_{b,4} \frac{3\lambda M\epsilon}{8\hbar\omega}\mathcal{A}_1^4\mathcal{A}_2(8\mathcal{A}_1^2-5\mathcal{A}_2^2)(\eta_1-\eta_2)\cos\phi \\ 
						&- \delta_{b,6} \frac{\lambda M\epsilon}{6\hbar\omega}\mathcal{A}_1^6\mathcal{A}_2(18\eta_1-17\eta_2)\cos\phi
					\end{aligned}$ \\
					\hhline{~|-|-|-|-}
					
					\rowcolor{highlightcolor}[\tabcolsep][\tabcolsep]
					\multirow{-3}{*}[2em]{\shortstack{$i$-wave \\ ($n=6$) \\ Even-parity}} 
					& $\boldsymbol{N} \parallel \hat{z}$ 
					& $\begin{aligned}
						&-\delta_{b,4} \frac{3\lambda M\epsilon}{8\hbar\omega}\mathcal{A}_1^4\mathcal{A}_2(8\mathcal{A}_1^2-5\mathcal{A}_2^2)(1-\eta_1\eta_2)\sin\phi \\ 
						&+ \delta_{b,6} \frac{\lambda M\epsilon}{6\hbar\omega}\mathcal{A}_1^6\mathcal{A}_2(17-18\eta_1\eta_2)\sin\phi
					\end{aligned}$ & 
					$\begin{aligned}
						&\delta_{b,4} \frac{3\lambda M\epsilon}{8\hbar\omega}\mathcal{A}_1^4\mathcal{A}_2(8\mathcal{A}_1^2-5\mathcal{A}_2^2)(\eta_1-\eta_2)\cos\phi \\ 
						&+ \delta_{b,6} \frac{\lambda M\epsilon}{6\hbar\omega}\mathcal{A}_1^6\mathcal{A}_2(18\eta_1-17\eta_2)\cos\phi
					\end{aligned}$ & 
					$\begin{aligned}
						&h_{\text{IFE}}+ \delta_{b,1} 10\mathcal{A}_1^3\mathcal{A}_2^3M\epsilon(1-\eta_1\eta_2)\cos(3\phi) \\ 
						&+ \delta_{b,2} \frac{15}{2}\mathcal{A}_1^4\mathcal{A}_2^2M\epsilon(1-\eta_1\eta_2)\cos(2\phi) \\
						&+ \delta_{b,5} 3\mathcal{A}_1^5\mathcal{A}_2M\epsilon(1-\eta_1\eta_2)\cos\phi
					\end{aligned}$ \\ 
					\hline\hline
				\end{tabular}
			}
		\end{table*}
		
		\subsection{Photon-counting selection rules}
		\label{sec:photon_counting}
		
		The selection rules in Tab.~\ref{tab:effective_fields} follow from frequency matching between the driven exchange term and the {\color{white} Rashba term. For the momentum-independent part, $\mathcal{M}_n(\boldsymbol{\mathcal{A}}(t))=$} \clearpage\hspace*{-0.3cm}Rashba term. For the momentum-independent part, $\mathcal{M}_n(\boldsymbol{\mathcal{A}}(t))=$ $M\,\mathrm{Re}\,[z(t)^n]$ with
		\begin{subequations}
			\begin{align}
				z={}&\mathcal{A}_x+i\mathcal{A}_y=\mathcal{A}_1e^{i\eta_1\omega t}+\mathcal{A}_2e^{i\eta_2(b\omega t+\phi)}\,,\\
				z^n={}&\sum_{q=0}^{n}\binom{n}{q}\mathcal{A}_1^{n-q}\mathcal{A}_2^{q}\,e^{i[(n-q)\eta_1+qb\eta_2]\omega t+iq\eta_2\phi}\,.
			\end{align}
		\end{subequations}
		Each term contains $n-q$ fundamental photons and $q$ harmonic photons. It oscillates at $[(n-q)\eta_1+qb\eta_2]\omega$ and carries the phase $q\eta_2\phi$. The Rashba term has harmonics at $\pm\omega$ and $\pm b\omega$. A static contribution to the first-order effective field can therefore arise in two ways: 
		\begin{itemize}
			\item[(i)] A cycle-averaged exchange term, $(n-q)\eta_1+qb\eta_2=0$, which is independent of $\lambda$ and parallel to $\boldsymbol{N}$
			\item[(ii)] A commutator between an exchange harmonic and a Rashba harmonic, for which $|(n-q)\eta_1+qb\eta_2|\in\{1,b\}$ and the resulting term contains $\lambda$. 
		\end{itemize}These are necessary frequency-matching conditions; the actual coefficients, including possible cancellations, are given in Tab.~\ref{tab:effective_fields} and Appendix~\ref{app:derivations}. Three consequences follow. First, for co-rotating light ($\eta_1=\eta_2$), the cycle-average condition (i) cannot be satisfied for $n>0$. Condition (ii) is met only for $q=0$ with $n=b$ (the photon-number-matching condition) and for the $p$-wave~($n=1$). Thus, for $n\geq2$, only $b=n$ gives a multipole-induced transverse term. Here, ``resonance'' refers to photon-number matching, not to an electronic or magnetic resonance. Second, for counter-rotating light ($\eta_1=-\eta_2$), condition (i) requires $(b+1)q=n$. Thus, terms without $\lambda$ occur only when $b+1$ divides $n$: $b=1$ for $d$; $b=2$ for $f$; $b=1,3$ for $g$; $b=4$ for $h$; and $b=1,2,5$ for $i$. These terms are longitudinal and do not by themselves contribute to torque. Third, for $b>n$ and $n\geq2$, neither condition can be met, so only the IFE contribution remains. For the $i$-wave at $b=3$, $|n-4q|=|6-4q|$ is always even, so condition (ii) is not satisfied.
		
		\subsection{Baseline relativistic IFE}
		
		The interaction between the vector potential and the Rashba term creates a non-magnetic out-of-plane field, the IFE,
		\begin{equation}\label{eq:h_IFE_def}
			h_{\text{IFE}}\equiv-\lambda^2\Omega_b\,,
		\end{equation}where
		\begin{equation}\label{eq:Omega_b_def}
			\Omega_b\equiv\frac{1}{\hbar\omega}\left[\eta_1\mathcal{A}_1^2+\frac{\eta_2\mathcal{A}_2^2}{b}+\delta_{b,1}\mathcal{A}_1\mathcal{A}_2(\eta_1+\eta_2)\cos\phi\right].
		\end{equation}
		The first two terms of $\Omega_b$ arise from the two beams separately; the factor $1/b$ in the second term comes from its frequency denominator. The last term describes their interference and occurs only for $b=1$. This relativistic IFE background is present for every multipole and is kept separate from the multipole-induced contributions. For $b=1$, the two components have the same frequency and
		combine into a single-frequency field with, in general,
		elliptical polarization; this case is therefore not a
		two-frequency drive in the strict sense.\begin{figure}[t!]
			\centering
			\includegraphics[width=1\linewidth]{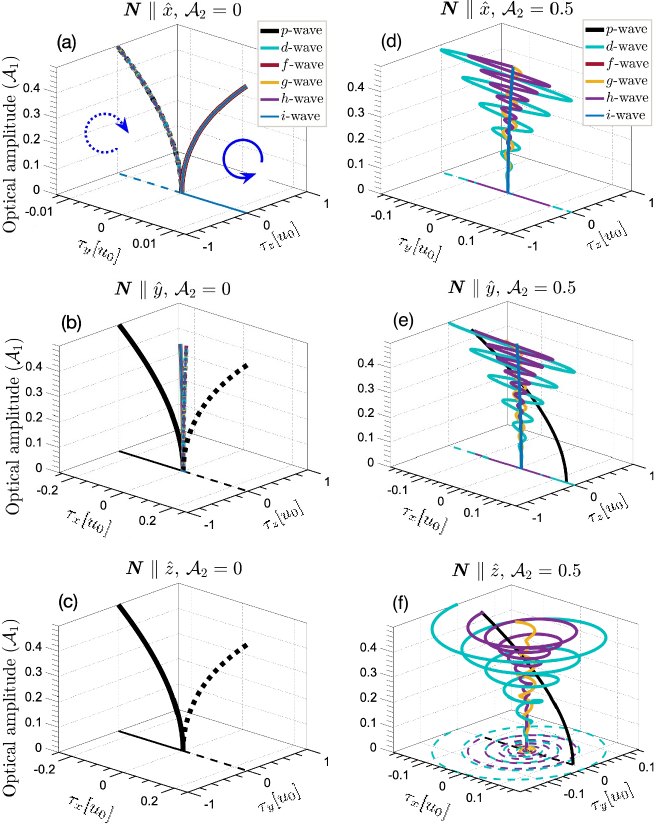}
			\caption{Torque trajectories for $\mathcal{A}_1$ from 0 to 0.5, in units of $u_0=M\epsilon\lambda/\hbar\omega$ ($=40~\mu$eV), for $\boldsymbol{N}\parallel\hat{x}$ (a, d), $\hat{y}$ (b, e) and $\hat{z}$ (c, f). Left (a--c): $\mathcal{A}_2=0$; solid (dotted) lines are $\eta_1=+1$ ($-1$). (a) All multipoles follow one trajectory along $\tau_y$. (b) The $d$- to $i$-wave coincide along $\tau_x$, and the $p$-wave splits off. (c) Only the $p$-wave gives a torque, along $\tau_x$. Right (d--f): $\mathcal{A}_2=0.5$, $b=2$, $\eta_1=+1$, $\eta_2=-1$, with $\phi$ swept over $12\pi$; dashed lines are projections onto the base plane. The torque forms ribbons in (d, e) and spirals in (f), largest for the $d$- and $h$-wave. The $p$-wave torque does not depend on $\phi$.}
			\label{fo2}
		\end{figure}
		
		\subsection{Single-frequency baseline ($\mathcal{A}_2=0$)}
		
		Without the harmonic beam only the IFE field $h_{\text{IFE}}=-\lambda^2\eta_1\mathcal{A}_1^2/\hbar\omega$ and the $p$-wave terms $\propto\Omega_b$ remain [Fig.~\ref{fo2}]. For $\boldsymbol{N}\parallel\hat{x}$ the torque is $\boldsymbol{\tau}=-h_{\text{IFE}}\hat{y}$ for all multipoles. It is not specific to the multipole and equals $(\lambda/M\epsilon)\mathcal{A}_1^2$ in units of $u_0$, only $0.0125$ at $\mathcal{A}_1=0.5$ [Fig.~\ref{fo2}(a)]. For $\boldsymbol{N}\parallel\hat{y}$, $\tau_x=h_{\text{eff},z}$, equal to $h_{\text{IFE}}$ for the $d$- to $i$-wave and to $h_{\text{IFE}}-M\epsilon\lambda\,\Omega_b$ for the $p$-wave [Fig.~\ref{fo2}(b)]. For $\boldsymbol{N}\parallel\hat{z}$ the $d$- to $i$-wave have no transverse field at $\mathcal{A}_2=0$, and the $p$-wave gives $\boldsymbol{\tau}^{(p)}=-M\epsilon\lambda\,\Omega_b\,\hat{x}$ [Fig.~\ref{fo2}(c)]. Reversing $\eta_1$ reverses every torque, as shown by solid and dashed circles in Fig.~\ref{fo2}(a).
		
		\subsection{Two-frequency drive ($\mathcal{A}_2\neq0$)}
		
		The second frequency makes the fields depend on $\phi$. We use $b=2$ and plot the torque as a trajectory in the right column of Fig.~\ref{fo2}, with $\mathcal{A}_1$ ramped from 0 to 0.5 at fixed $\mathcal{A}_2=0.5$ while $\phi$ goes through six $2\pi$ turns. For the $p$-wave the field $\propto\Omega_b$ does not depend on $\phi$ for $b\ge2$, so its torque cannot be steered. For the transverse channels with $b<n$ the phase-dependent fields vanish for co-rotating light, so we use counter-rotating light from here on. At $b=n$ the fields are finite for co-rotating light and larger for counter-rotating light. The orientation of $\boldsymbol{N}$ sets the shape of the trajectory.
		
		\subsubsection{Sub-harmonic drives ($1<b<n$)} Table~\ref{tab:effective_fields} contains three kinds of terms. (a) Longitudinal terms without $\lambda$ at $b=n-1$ ($f$ at $b=2$, $g$ at $b=3$, $h$ at $b=4$, $i$ at $b=5$) and, for the $i$-wave, also at $b=2$; they scale as $M\mathcal{A}_1^{n-1}\mathcal{A}_2\cos\phi$, except $i$-wave at $b = 2$, which scales as $M\mathcal{A}_1^{4}\mathcal{A}^2_2\cos (2\phi)$.  (b) Virtual-pair splitting at $b=n-2$: two processes contribute with opposite signs, $n-1$ fundamental photons and one harmonic photon matching the Rashba harmonic at $\omega$, and $n-2$ fundamental photons and two harmonic photons matching the one at $b\omega$. This gives the intensity-difference  polynomial $\lambda M\mathcal{A}_1^{n-2}\mathcal{A}_2(c_1\mathcal{A}_1^2-c_2\mathcal{A}_2^2)$ with $(c_1,c_2)=(4,3)$ for the $g$-wave, $(3,2)$ for the $h$-wave and $(8,5)$ for the $i$-wave; the $f$-wave at $b=1$ belongs to the same family with $(1,1)$. (c) Higher phase harmonics: the $h$-wave at $b=2$ has a transverse term $\lambda M\mathcal{A}_1^4\mathcal{A}_2^2\{\cos2\phi,\sin2\phi\}$ (degree 6), and the $h$-wave at $b=1$ has a similar $2\phi$ term. The transverse terms vanish for co-rotating light. The $i$-wave at $b=3$ has no term, so only $h_{\rm IFE}$ remains. For $n\ge2$ the sub-harmonic ($b<n$) fields contain the factors $(1-\eta_1\eta_2)$ or $(\eta_1-\eta_2)$, which vanish for co-rotating light, so counter-rotating light is required. 
		
		\subsubsection{Resonance (harmonic matching, $b=n$)} Two processes contribute. In the first, $n$ fundamental photons give an exchange harmonic at $n\omega=b\omega$, which matches the Rashba harmonic  at $b\omega$. In the second, $n-1$ fundamental photons and one harmonic photon give a harmonic at $\omega$ for counter-rotating light, which matches the Rashba harmonic at $\omega$. Both give $\mathcal{A}_1^n\mathcal{A}_2$, so the degree is $n+1$ and the scaling $\lambda M\mathcal{A}_1^n\mathcal{A}_2$ is the same for both parities. Each process has one harmonic photon, so the phase enters only as $\{\sin\phi,\cos\phi\}$. With $\Gamma_n=\mathcal{A}_1^n\mathcal{A}_2\epsilon\lambda/\hbar\omega$, the fields for $\boldsymbol{N}\parallel\hat{z}$ and $2\le n\le6$ are
		\begin{subequations}\label{eq:resonant_general}
			\begin{align}
				h_{\text{eff},x}^{b=n}&=\Gamma_nM\left[\frac{n}{2}(1-\eta_1\eta_2)-\frac{1}{n}\right]\sin\phi\,,\\
				h_{\text{eff},y}^{b=n}&=\Gamma_nM\left[\frac{n}{2}(\eta_1-\eta_2)+\frac{\eta_2}{n}\right]\cos\phi\,.
			\end{align}
		\end{subequations}
		The terms $\mp1/n$ come from the first process, which does not depend on the helicities, and the terms $\propto n$ from the second, which needs counter-rotating light. The field size is $1/n$ for co-rotating light and $(n^2-1)/n$ for counter-rotating light. The $p$-wave is special: for $\boldsymbol N\parallel\hat z$, $h_{\text{eff},x}=0$ and $h_{\text{eff},y}=M\epsilon\lambda\,\Omega_b$ for every $b$, from the same two-photon processes that give the IFE. Because $\mathcal{A}_1^n\mathcal{A}_2$ falls with $n$ for $\mathcal{A}<1$, higher multipoles need stronger fields. At the resonance the fields are finite for co-rotating light, and counter-rotating light adds a second process [Eq.~\eqref{eq:resonant_general}].
		
		We take the $h$-wave at $b=2$, where an $\mathcal{A}_1^4\mathcal{A}_2^2$ process is locked to $2\phi$. For $\boldsymbol{N}\parallel\hat{x}$ and $\eta_1=+1$, $\eta_2=-1$ the torque lies along $\tau_y$,
		\begin{equation}
			\boldsymbol{\tau}^{(5)}_{\boldsymbol{N}\parallel\hat{x}}=-\left[h_{\text{IFE}}-\frac{15}{4}\frac{M\epsilon\lambda}{\hbar\omega}\mathcal{A}_1^4\mathcal{A}_2^2\sin2\phi\right]\hat{y}\,,
		\end{equation}
		so changing $\phi$ moves it back and forth along a line, and sweeping $\mathcal{A}_1$ gives the ribbon of Fig.~\ref{fo2}(d). The $d$- and $g$-wave give similar ribbons, and the $f$- and $i$-wave stay at the $\phi$-independent IFE value. For $\boldsymbol{N}\parallel\hat{y}$ the torque oscillates along $\tau_x$ [Fig.~\ref{fo2}(e)]. For $\boldsymbol{N}\parallel\hat{z}$ both transverse fields depend on the phase,
		\begin{subequations}
			\begin{align}
				h_{\text{eff},x}={} &\frac{15M\epsilon\lambda}{4\hbar\omega}\mathcal{A}_1^4\mathcal{A}_2^2\sin(2\phi)\,,\\
				h_{\text{eff},y}={} &\frac{15M\epsilon\lambda}{4\hbar\omega}\mathcal{A}_1^4\mathcal{A}_2^2\cos(2\phi)\,.
			\end{align}
		\end{subequations}
		They have equal size and a $\pi/2$ phase shift, so the torque moves on a circle that it traverses twice per $2\pi$ of $\phi$. The radius grows as $\mathcal{A}_1^4\mathcal{A}_2^2$ and gives the spiral of Fig.~\ref{fo2}(f); it is $0.12$ at $\mathcal{A}_1=\mathcal{A}_2=0.5$. The $d$-wave circle turns once per $2\pi$ and has the larger radius $\tfrac32\mathcal{A}_1^2\mathcal{A}_2=0.19$, and the $g$-wave circle has the smaller radius $0.03$. Swapping to a clockwise fundamental and a counterclockwise harmonic ($\eta_1=-1$, $\eta_2=+1$) keeps $\eta_1\eta_2$ and flips $\eta_1-\eta_2$, so $\tau_y$ stays and $\tau_x$ changes sign. This reverses the winding direction of the spirals without changing their size.
		
		\subsubsection{High harmonics ($b>n$)}\label{sec:decoupling} For $b>n\ge2$ all multipole-induced terms vanish and only $h_{\text{IFE}}$ is left,
		\begin{equation}\label{eq:decoupling_b_gt_n}
			\boldsymbol{h}_{\text{eff}}^{(b>n)}=-\lambda^2\Omega_b\,\hat{\boldsymbol{z}}=-\frac{\lambda^2}{\hbar\omega}\left(\eta_1\mathcal{A}_1^2+\frac{\eta_2\mathcal{A}_2^2}{b}\right)\hat{\boldsymbol{z}}\,.
		\end{equation}This follows from condition (ii) in Sec.~\ref{sec:photon_counting}. The exchange term has $n$ photons, $n-q$ from the fundamental and $q$ from the harmonic. For $q\ge1$ we have $n-q\le n-1$, so $|(n-q)\eta_1+qb\eta_2|\ge b-n+1\ge2$; it cannot equal $1$, and it equals $b$ only for $q=1$, $n=q$, i.e.\ $n=1$. For $q=0$ it equals $n<b$, which is $1$ only for $n=1$. The $p$-wave is the exception, because each beam acts on the single exchange photon separately. For $n\ge2$ the magnetic lattice is transparent to the harmonic beam, and for $\boldsymbol{N}\parallel\hat{z}$ the remaining $h_{\mathrm{IFE}}$ is parallel to $\boldsymbol{N}$.

		\subsubsection{Optical blindspots}
		
		Higher intensity does not always give a larger torque: for some channels the torque vanishes at a specific ratio $\mathcal{A}_2/\mathcal{A}_1$ (Fig.~\ref{fo4}, $\boldsymbol{N}\parallel\hat{z}$, $\phi=\pi/3$, $b=2$, $\eta_1=-1$, $\eta_2=+1$). The $p$-wave gives the simplest zero: its torque is $\tau_x=-M\epsilon\lambda\,\Omega_b\propto\mathcal{A}_1^2-\mathcal{A}_2^2/2$, which changes sign at $\mathcal{A}_2=\sqrt{2}\mathcal{A}_1$. Here $\Omega_b$ is the sum of the IFE contributions of the two beams, which have opposite sign for counter-rotating light, so the zero is their cancellation and not an interference effect. Since all three $p$-wave torques in Tab.~\ref{tab:effective_fields} are $\propto\Omega_b$, the zero at $\mathcal{A}_2=\sqrt{b}\,\mathcal{A}_1$ exists for every orientation of $\boldsymbol{N}$ and every $b$.\begin{figure}[t!]
			\centering
			\includegraphics[width=1\linewidth]{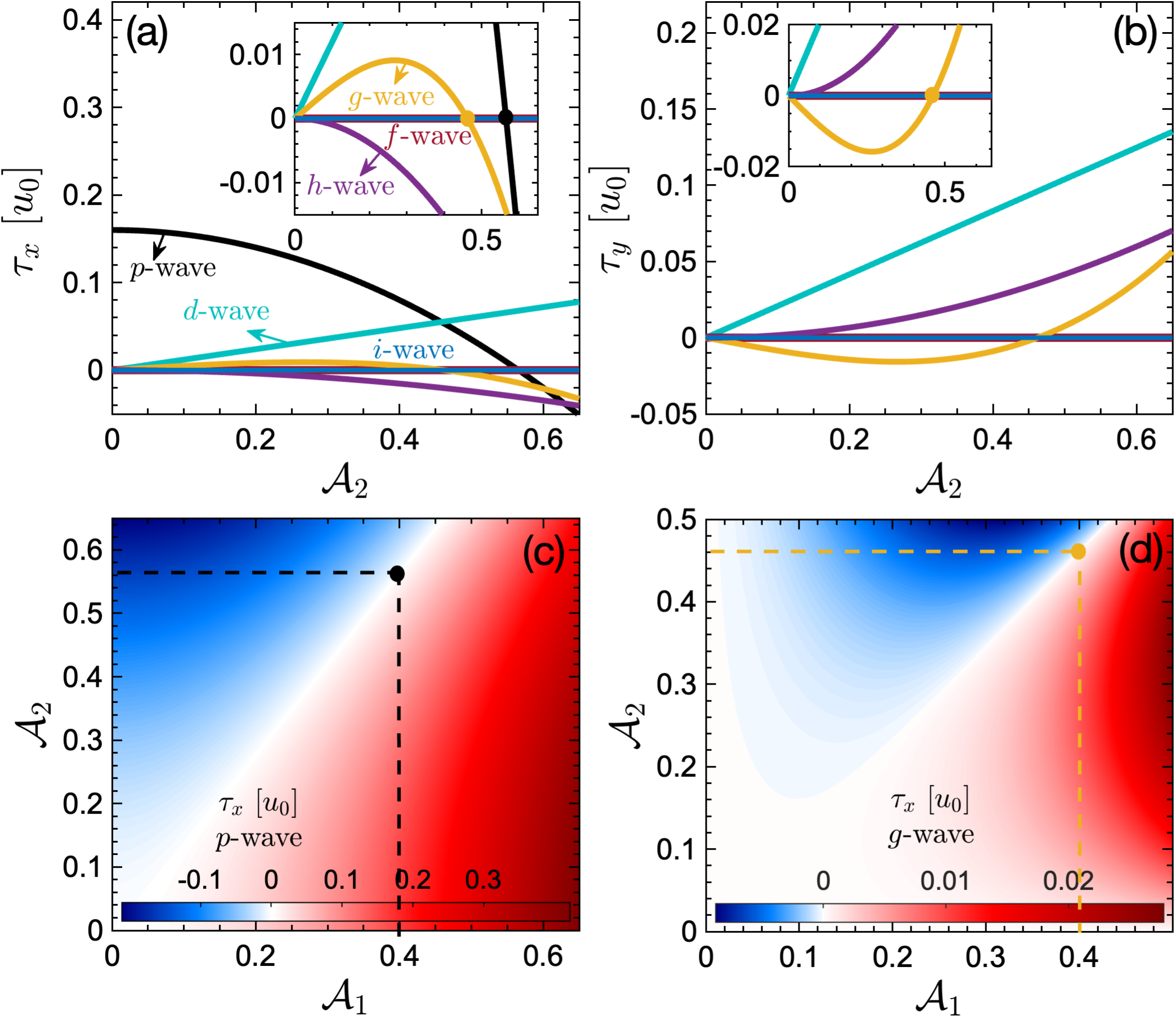}
			\caption{Optical blindspots for $\boldsymbol{N}\parallel\hat{z}$, counter-rotating beams ($\eta_1=-1$, $\eta_2=+1$), $b=2$, $\phi=\pi/3$. (a, b) $\tau_x$ and $\tau_y$ against $\mathcal{A}_2$ at fixed $\mathcal{A}_1=0.4$; the $f$- and $i$-wave vanish, and the insets show the low-torque region. The black dot marks the $p$-wave zero ($\mathcal{A}_2=\sqrt{2}\mathcal{A}_1\approx0.57$) and the orange dots the $g$-wave zero ($\mathcal{A}_2=\sqrt{4/3}\mathcal{A}_1\approx0.46$). (c, d) Maps of $\tau_x$ over $(\mathcal{A}_1,\mathcal{A}_2)$ for the $p$-wave (c) and $g$-wave (d), with the zero lines in white. Dashed lines mark the cut $\mathcal{A}_1=0.4$ to match (a,b) zeros.}
			\label{fo4}
		\end{figure}\begin{table}[b!]
			\vspace*{-0.6em}\centering
			\caption{Optical blindspots of the constant-field torque for $\boldsymbol{N}\parallel\hat{z}$ and counter-rotating light: amplitude ratio $\mathcal{A}_2/\mathcal{A}_1$ at which the transverse torque vanishes. The $p$-wave zero is $\Omega_b=0$ and holds for every $\boldsymbol{N}$ and $b$. For the $f$- and $h$-wave at $b=1$ the zero is a linearly polarized drive.}
			\label{tab:blindspots}
			\footnotesize
			\renewcommand{\arraystretch}{1.3}
			\resizebox{\columnwidth}{!}{%
				\begin{tabular}{c|c|c|c}
					\hline\hline
					Multipole & $b$ & Factor & $\mathcal{A}_2/\mathcal{A}_1$ \\
					\hline
					$p$-wave ($n=1$) & any & $\eta_1\mathcal{A}_1^2+\eta_2\mathcal{A}_2^2/b$ & $\sqrt{b}$ \\\hline
					$d$-wave ($n=2$) & -- & -- & none \\\hline
					$f$-wave ($n=3$) & 1 & $\mathcal{A}_1^2-\mathcal{A}_2^2$ & $1$ \\\hline
					$g$-wave ($n=4$) & 2 & $4\mathcal{A}_1^2-3\mathcal{A}_2^2$ & $\sqrt{4/3}\approx1.15$ \\\hline
					$h$-wave ($n=5$) & 1 & $\mathcal{A}_1^2-\mathcal{A}_2^2$ & $1$ \\\hline
					$h$-wave ($n=5$) & 3 & $3\mathcal{A}_1^2-2\mathcal{A}_2^2$ & $\sqrt{3/2}\approx1.22$ \\\hline
					$i$-wave ($n=6$) & 4 & $8\mathcal{A}_1^2-5\mathcal{A}_2^2$ & $\sqrt{8/5}\approx1.26$ \\
					\hline\hline
				\end{tabular}
			}
		\end{table}
		
		The $g$-wave at $b=2$ is the virtual-pair case $b=n-2$. For $\eta_1=-1$, $\eta_2=+1$,
		\begin{subequations}
			\begin{align}
				\tau_x={} &\frac{\lambda M\epsilon}{\hbar\omega}\mathcal{A}_1^2\mathcal{A}_2(4\mathcal{A}_1^2-3\mathcal{A}_2^2)\cos\phi\,,\\
				\tau_y={} &-\frac{\lambda M\epsilon}{\hbar\omega}\mathcal{A}_1^2\mathcal{A}_2(4\mathcal{A}_1^2-3\mathcal{A}_2^2)\sin\phi\,.
			\end{align}
			
		\end{subequations}
		
		Both contain the factor $(4\mathcal{A}_1^2-3\mathcal{A}_2^2)$ from two competing five-photon processes, so the torque vanishes at $\mathcal{A}_2=\sqrt{4/3}\,\mathcal{A}_1$ for every $\phi$ (orange dots in Fig.~\ref{fo4}(a,b)). The white lines in Figs.~\ref{fo4}(c,d) show that the blindspots are lines in the $(\mathcal{A}_1,\mathcal{A}_2)$ plane, so changing the amplitude ratio switches the torque on or off without changing polarization or phase. Channels without a subtraction polynomial, such as the resonant $d$-wave or the $h$-wave at $b=2$, have no such zero. Table~\ref{tab:blindspots} lists the zeros. For counter-rotating light and $\mathcal{A}_1=\mathcal{A}_2$ at $b=1$ the drive is linearly polarized, so that zero is a symmetry effect. For in-plane $\boldsymbol{N}$ the polynomial zeros do not zero the torque, because $h_{\text{IFE}}$ stays in $h_{\text{eff},z}$; the only zero valid for all orientations is the $p$-wave one.\begin{figure}[t!]
			\centering
			\includegraphics[width=1\linewidth]{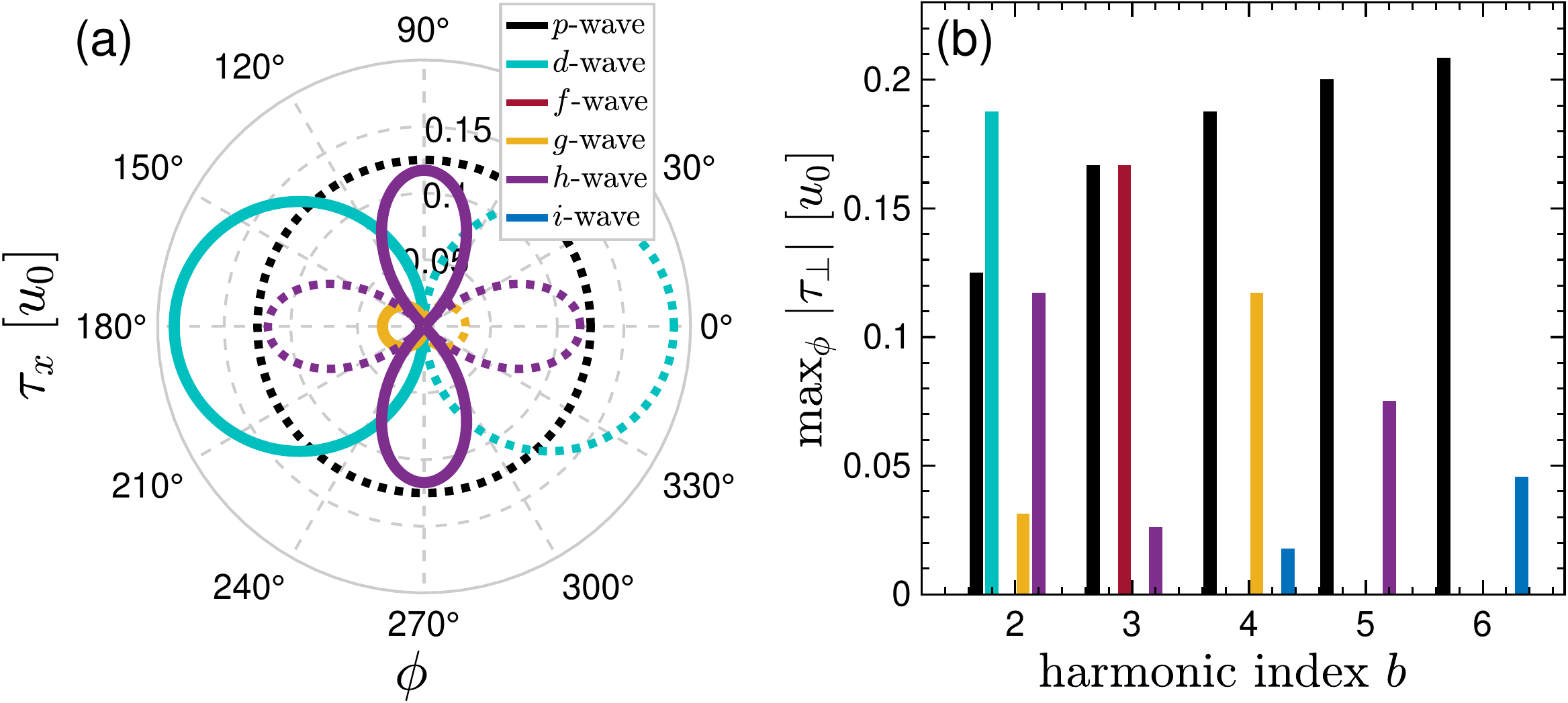}
			\caption{Torque for $\boldsymbol{N}\parallel\hat{z}$, counter-rotating beams ($\eta_1=+1$, $\eta_2=-1$), $\mathcal{A}_1=\mathcal{A}_2=0.5$. (a) $\tau_x$ against $\phi$ at $b=2$ (solid/dotted: positive/negative). (b) Largest transverse torque $|\boldsymbol{\tau}_\perp| = (\tau_x^2 + \tau_y^2)^{1/2}$ over $\phi$ against $b$ ($b=1$ gives zero for equal amplitudes). The $d$-, $f$-, $g$- and $i$-wave peak at $b=n$, the $h$-wave is largest at $b=2$, and the $p$-wave grows slowly with $b$.}
			\label{fo5}
		\end{figure}
		
		\subsubsection{Phase dependence and strength across harmonics}
		
		Figure~\ref{fo5}(a) shows $\tau_x$ for all multipoles for $\boldsymbol{N}\parallel\hat{z}$ under an $\omega$--$2\omega$ drive with $\mathcal{A}_1=\mathcal{A}_2=0.5$ and $\eta_1=+1$, $\eta_2=-1$. The $f$- and $i$-wave give only fields along $\boldsymbol{N}$ at $b=2$. The $p$-wave torque, $-0.125$, does not depend on $\phi$. The $d$-wave (resonant) and $g$-wave ($b=n-2$) follow $\phi$ with radii $0.19$ and $0.03$, and the $h$-wave follows $2\phi$ with radius $0.12$. At equal amplitudes $4\mathcal{A}_1^2-3\mathcal{A}_2^2$ equals $\mathcal{A}^2$, which reduces the $g$-wave torque, and the $h$-wave term carries six powers of $\mathcal{A}$, so it is smaller than the $d$-wave.\begin{table}[t!]
			\vspace*{-0.6em}\centering
			\caption{Activation of $(\tau_x,\tau_y)$ for $\boldsymbol{N}\parallel\hat{z}$ and counter-rotating light, for generic amplitudes. A check mark ($\checkmark$) is a nonzero component. A cross ($\times$) means that the transverse field is absent or that the only field is parallel to $\boldsymbol{N}$. $\tau_z=0$ in all cases. Light cells mark $b=n$, dark cells $b=n-2$.}
			\label{tab:torque_activation}
			\renewcommand{\arraystretch}{1.75}
			\resizebox{0.95\width}{!}{%
				\begin{tabular}{c|c|c|c|c|c|c}
					\hline\hline
					Multipole & $b=1$ & $b=2$ & $b=3$ & $b=4$ & $b=5$ & $b=6$ \\
					\hline
					$p$-wave ($n=1$) & \cellcolor{diagcolor}$(\checkmark, \, \times)$ & $(\checkmark, \, \times)$ & $(\checkmark, \, \times)$ & $(\checkmark, \, \times)$ & $(\checkmark, \, \times)$ & $(\checkmark, \, \times)$ \\\hline
					$d$-wave ($n=2$) & $(\times, \, \times)$ & \cellcolor{diagcolor}$(\checkmark, \, \checkmark)$ & $(\times, \, \times)$ & $(\times, \, \times)$ & $(\times, \, \times)$ & $(\times, \, \times)$ \\\hline
					$f$-wave ($n=3$) & \cellcolor{diagcolor1}$(\checkmark, \, \checkmark)$ & $(\times, \, \times)$ & \cellcolor{diagcolor}$(\checkmark, \, \checkmark)$ & $(\times, \, \times)$ & $(\times, \, \times)$ & $(\times, \, \times)$ \\\hline
					$g$-wave ($n=4$) & $(\times, \, \times)$ & \cellcolor{diagcolor1}$(\checkmark, \, \checkmark)$ & $(\times, \, \times)$ & \cellcolor{diagcolor}$(\checkmark, \, \checkmark)$ & $(\times, \, \times)$ & $(\times, \, \times)$ \\\hline
					$h$-wave ($n=5$) & $(\checkmark, \, \checkmark)$ & $(\checkmark, \, \checkmark)$ & \cellcolor{diagcolor1}$(\checkmark, \, \checkmark)$ & $(\times, \, \times)$ & \cellcolor{diagcolor}$(\checkmark, \, \checkmark)$ & $(\times, \, \times)$ \\\hline
					$i$-wave ($n=6$) & $(\times, \, \times)$ & $(\times, \, \times)$ & $(\times, \, \times)$ & \cellcolor{diagcolor1}$(\checkmark, \, \checkmark)$ & $(\times, \, \times)$ & \cellcolor{diagcolor}$(\checkmark, \, \checkmark)$ \\
					\hline\hline
				\end{tabular}
			}
		\end{table}
		
		Figure~\ref{fo5}(b) gives the largest transverse torque over $\phi$ for $b=1,\dots,6$. The $d$-, $f$-, $g$- and $i$-wave are largest at $b=n$, and their virtual-pair channels ($g$ at $b=2$, $i$ at $b=4$) are weaker. The $h$-wave is different: $b=2$ ($0.117$) is larger than the resonance $b=5$ ($0.075$). The $p$-wave vanishes at $b=1$ for equal amplitudes and grows from $0.125$ at $b=2$ to $0.208$ at $b=6$. Every multipole-induced transverse channel has degree $n+1$ in the amplitudes, so at equal amplitudes it scales as $\mathcal{A}^{n+1}$ (multiply the $d$- to $i$-wave rows by $(2\mathcal{A})^{n+1}$ and the $p$-wave row by $(2\mathcal{A})^2$). At $\mathcal{A}=0.5$ the resonant torque falls from $0.19$ ($d$) to $0.05$ ($i$), and the ranking of the multipoles depends on the amplitude.
		
		\subsubsection{Activation across harmonics}
		
		Table~\ref{tab:torque_activation} shows which components $(\tau_x,\tau_y)$ are nonzero for $\boldsymbol{N}\parallel\hat{z}$ and counter-rotating light. A transverse field needs condition (ii) in Sec.~\ref{sec:photon_counting}, which for $q\ge1$ harmonic photons reads $|n-(b+1)q|\in\{1,b\}$. For $n\ge2$ and $b>n$ all entries are $(\times,\times)$, and the $p$-wave row is $(\checkmark,\times)$ throughout. On the diagonal ($b=n$) both components are active, because the $(n+1)$-photon loop locks the torque to $\phi$. For $b<n$ the condition is met in the virtual-pair cases $b=n-2$ and for the $h$-wave at $b=1,2$; the channels at $b=n-1$ give only a field along $\boldsymbol{N}$.
		
		\section{Torque from the $\textbf{\textit{k}}$-dependent light-induced part}
		\label{sec:kdep}
		
		Unlike the $\boldsymbol{k}$-independent term discussed in Sec.~\ref{sec:kindep}, these contributions modify the momentum dependence of the exchange field. We first examine their cycle average, which is the zeroth-order term of the high-frequency effective Hamiltonian~\cite{bukov2015universal,mikami2016brillouin}. This provides a useful way to identify the lower-order magnetic textures generated by the drive. The full Fermi-sea torque, including the oscillating parts of the drive, is evaluated separately using the Floquet states in Sec.~\ref{sec:both}.
		
		The cycle-averaged Hamiltonian is
		\begin{equation}
			H_0(\boldsymbol{k})=\frac{1}{T}\int_0^T H(\boldsymbol{k},t)\,dt.
		\end{equation}
		Because $\langle\boldsymbol{\mathcal A}\rangle=0$, its Rashba term is unchanged. Its exchange term becomes
		\begin{equation}
			\label{eq:H0exch}
			\epsilon\langle\mathcal M_n\rangle\,\hat{\boldsymbol N}\cdot\boldsymbol\sigma
			=\epsilon M\sum_{j=0}^{n}\binom{n}{j}
			\operatorname{Re}\!\left[k_+^{\,n-j}\langle\mathcal A_+^{\,j}\rangle\right]
			\hat{\boldsymbol N}\cdot\boldsymbol\sigma.
		\end{equation}The $j=0$ term is the undriven exchange texture, while terms with $j\geq1$ describe drive-induced lower-order textures. For counter-rotating beams, which we take as $\eta_1=+1$ and $\eta_2=-1$, the time average is
		\begin{equation}
			\label{eq:Aj}
			\left\langle\mathcal A_+^{\,j}\right\rangle
			=\sum_{c\geq1,\,j=(b+1)c}
			\frac{j!}{(bc)!\,c!}
			\mathcal A_1^{\,bc}\mathcal A_2^{\,c}e^{-ic\phi}.
		\end{equation}
		Thus, a nonzero average requires $j=(b+1)c$. For co-rotating beams, $\langle\mathcal A_+^{\,j}\rangle=0$ for every $j\geq1$. The cycle-averaged drive therefore generates lower multipoles of order $n-(b+1)c$ only for counter-rotating beams. Their amplitudes depend on both optical amplitudes, while their orientation is controlled by the relative phase $\phi$. For example, for an $f$-wave texture ($n=3$) at $b=1$, the leading induced term is $6\mathcal A_1\mathcal A_2\operatorname{Re}\!\left[k_+e^{-i\phi}\right],$
		which is a $p$-wave contribution added to the original $f$-wave exchange texture.
		
		The spin-dependent parts of the undriven and cycle-averaged Hamiltonians can then be written as $\boldsymbol b\cdot\boldsymbol\sigma$ and $\boldsymbol b'\cdot\boldsymbol\sigma$, respectively, with
		\begin{subequations}
			\label{eq:h0all}
			\begin{align}
				\boldsymbol b(\boldsymbol{k})&=
				\lambda(-k_y,k_x,0)+\epsilon M\operatorname{Re}(k_+^n)\hat{\boldsymbol N},\\
				\boldsymbol b'(\boldsymbol{k})&=
				\lambda(-k_y,k_x,0)+\epsilon M c_n(\boldsymbol{k})\hat{\boldsymbol N},
			\end{align}
		\end{subequations}where\begin{equation}
			\label{eq:cn}
			c_n(\boldsymbol{k})=
			\sum_{j=0}^{n}\binom{n}{j}
			\operatorname{Re}\!\left[k_+^{\,n-j}\langle\mathcal A_+^{\,j}\rangle\right].
		\end{equation}
		The static bands of $H_0$ are $\epsilon k^2\pm|\boldsymbol b'|$, with spin expectation values $\pm\boldsymbol b'/|\boldsymbol b'|$. We assign their occupations from the corresponding undriven bands, denoted by $n_\pm(\boldsymbol{k})$. Subtracting the undriven contribution gives the cycle-averaged, light-induced effective field\begin{align}\label{eq:h0}
			\boldsymbol h_{\rm eff}^{(0)}
			={}&-\frac{\epsilon M}{N_{\rm el}}
			\sum_{\boldsymbol{k}}\big[n_+(\boldsymbol{k})-n_-(\boldsymbol{k})\big]\notag \\ {} &\times
			\left[
			\frac{c_n(\boldsymbol{k})}{|\boldsymbol b'(\boldsymbol{k})|}\boldsymbol b'(\boldsymbol{k})
			-\frac{\operatorname{Re}(k_+^n)}{|\boldsymbol b(\boldsymbol{k})|}\boldsymbol b(\boldsymbol{k})
			\right].
		\end{align}
		Here, $N_{\rm el}$ is the number of occupied states. Equation~\eqref{eq:h0} is the contribution obtained from the cycle-averaged Hamiltonian; it is not, in general, the full Floquet effective field, because the latter also contains contributions from the oscillating terms and micromotion.
		
		At fixed vector-potential amplitudes $\mathcal A_1$ and $\mathcal A_2$, Eq.~\eqref{eq:h0} has no explicit dependence on $\omega$. It vanishes for co-rotating beams, for $b+1>n$, and for the $p$-wave texture, for which there are no terms with $j<n$ that survive the cycle average. For an out-of-plane N\'eel vector, $\boldsymbol N\parallel\hat{\boldsymbol z}$, the cycle-averaged field is nonzero in six counter-rotating channels: the $f$-wave at $b=1$, the $g$-wave at $b=2$, the $h$-wave at $b=1,2,3$, and the $i$-wave at $b=4$. These are precisely the channels satisfying
		\begin{equation}
			b+1\leq n,\qquad \gcd(n,b+1)=1.
		\end{equation}
		We refer to them as class~I. For the leading phase harmonic with $|v|=1$, the amplitude dependence is $\mathcal A_1^{b}\mathcal A_2$, of total order $b+1$. More generally, the lowest-order torque scales as $\mathcal A_1^{b|v|}\mathcal A_2^{|v|}$, with $|v|$ determined by the symmetry condition discussed in Sec.~\ref{sec:gcdrule}. The remaining out-of-plane channels are referred to as class~II; their full Fermi-sea torque requires contributions from the oscillating part of the drive and, in some cases, from the $\boldsymbol{k}$-independent exchange term.
		
		For an in-plane N\'eel vector, the cycle-averaged field in Eq.~\eqref{eq:h0} is nonzero for a larger set of channels. In this case, it is finite for the ten counter-rotating channels satisfying $b+1<n$: the $f$-wave at $b=1$; the $g$-wave at $b=1,2$; the $h$-wave at $b=1,2,3$; and the $i$-wave at $b=1,2,3,4$. We call these channels class~I$'$. Unlike the out-of-plane case, no additional greatest-common-divisor condition is required. The six out-of-plane class~I channels are included in this in-plane set. For $\boldsymbol N\parallel\hat x$ ($\hat y$) the cycle-averaged field lies in the plane along $\hat y$ ($\hat x$), because $\boldsymbol b$ and $\boldsymbol b'$ in Eqs.~\eqref{eq:h0all} have no $z$ component. The torque $\boldsymbol N\times\boldsymbol h_{\rm eff}$ is therefore along $\hat z$ and rotates $\boldsymbol N$ within the plane; it does not tilt $\boldsymbol N$ out of the plane.
		
		The numerical Floquet calculation confirms that the cycle-averaged contribution captures the dominant response in the class~I channels for $\boldsymbol N\parallel\hat{\boldsymbol z}$: retaining only the $j<n$ terms reproduces the full Fermi-sea torque to within approximately $2\%$, while the $\boldsymbol{k}$-independent contribution is at most $10^{-4}$ of the full result in these channels. For an in-plane N\'eel vector, the same $j<n$ contribution reproduces the full torque to within approximately $2\%$ in class~I$'$ and in the class~II$'$ channels with $b=n-1\geq2$. In the latter comparison, the $\boldsymbol{k}$-independent contribution is at most $0.4\%$ of the full torque for the tested cases with $b=1$ and $n\geq3$. These comparisons show that the cycle-averaged exchange texture identifies the dominant mechanism in these channels, but it does not replace the full Floquet calculation for the remaining channels.
		
		For completeness, class~II$'$ denotes the in-plane channels with $b=n-1$. In these channels the cycle average in Eq.~\eqref{eq:h0} vanishes or is a small fraction of the full response, and the torque is mainly generated by the oscillating part of the drive. Their frequency dependence and detailed orientation are discussed with the full Floquet results in Sec.~\ref{sec:inplane}.
		
		\section{Full Fermi-sea torque from the $\textbf{\textit k}$-independent and $\textbf{\textit k}$-dependent light-induced parts}
		\label{sec:both}
		
		Here we retain the complete driven exchange texture, including all terms in the expansion of $\mathcal{M}_n(\boldsymbol{k}+\boldsymbol{\mathcal A})$, and calculate the Floquet states from the full time-dependent Hamiltonian. The results therefore give the full electronic response of the model, within the occupation prescription and numerical accuracy described in Sec.~\ref{sec:model} and Appendix~\ref{app:num}. We refer to this quantity as the full Fermi-sea torque to distinguish it from the momentum-independent effective field discussed in Sec.~\ref{sec:kindep}. Unless stated otherwise, we use $\lambda=1$~meV and $\mathcal A_1=\mathcal A_2=0.15$; the other parameters are as in Sec.~\ref{sec:kindep}, and torques are given in units of $u_0$.\begin{figure*}[t!]
			\centering
			\includegraphics[width=0.8\linewidth]{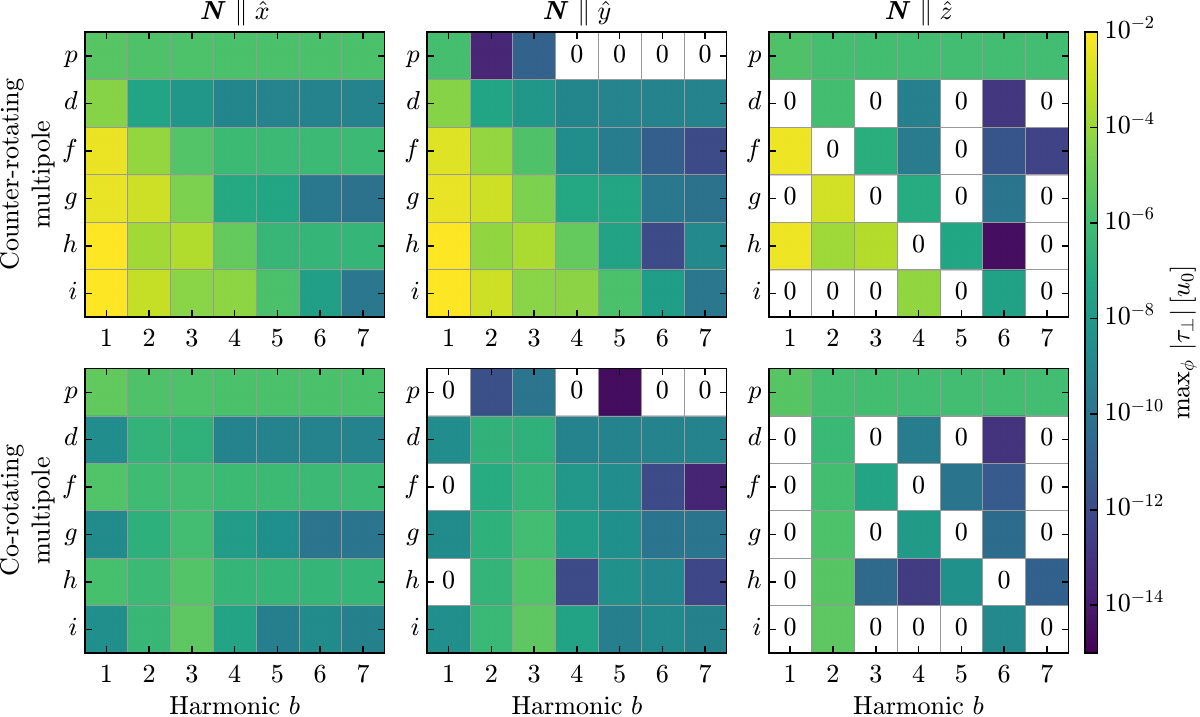}
			\caption{Channel map of the Fermi-sea torque: maximum over the sampled phases of $|\tau_\perp|$, for $p$ ($n=1$) to $i$ ($n=6$) and $b=1$ to 7, for counter-rotating (top) and co-rotating (bottom) beams and $\boldsymbol{N}$ along $\hat x$, $\hat y$, $\hat z$. Cells marked 0 are below $10^{-10}\,u_0$. For $\boldsymbol{N}\parallel\hat z$ the 34 channels excluded by Eq.~\eqref{eq:gcd} are among them. $\mathcal{A}_1=\mathcal{A}_2=0.15$, $\lambda=1$~meV, $\epsilon=400$~meV, $M\epsilon=10$~meV, $\hbar\omega=124$~meV, $k_{\rm B}T=1$~meV.}
			\label{f2}
		\end{figure*}
		
		\subsection{Symmetry rule for an out-of-plane N\'eel vector}
		\label{sec:gcdrule}
		
		For $\boldsymbol{N}\parallel\hat z$, the isotropic kinetic term, the Rashba term, and the exchange texture are invariant under a simultaneous rotation of momentum and spin about $\hat z$ by $\alpha=2\pi/n$. The occupation prescription preserves this symmetry, so the occupied-state response obeys the same covariance. Rotating the drive by $\alpha$ therefore rotates $\boldsymbol{h}_{\rm eff}$ by $\alpha$: for $h_\pm=h_{{\rm eff},x}\pm ih_{{\rm eff},y}$ we have $h_+\to e^{i\alpha}h_+$. The same rotation multiplies $\mathcal{A}_+$ by $e^{i\alpha}$, which equals a time shift combined with the phase shift $\phi\to\phi+\alpha(\eta_2-b\eta_1)$. The period average does not depend on a time shift, so
		\begin{equation}
			h_+\big(\phi+\alpha(\eta_2-b\eta_1)\big)=e^{i\alpha}h_+(\phi)\,.
		\end{equation}
		We next write $h_+(\phi)=\sum_vh_ve^{iv\phi}$ with integer $v$. A component $h_v\neq0$ requires $v\alpha(\eta_2-b\eta_1)\equiv\alpha\pmod{2\pi}$, and for $h_-=h_+^*$ the right-hand side has the sign $-1$. With $\alpha=2\pi/n$,
		\begin{equation}\label{eq:gcd}
			(b-\eta_1\eta_2)\,v\equiv\pm1\pmod n\,.
		\end{equation}
		This has an integer solution if and only if
		\begin{itemize}
			\item $\gcd(n,\,b+1)=1$ for counter-rotating beams ($\eta_1\eta_2=-1$),
			\item $\gcd(n,\,b-1)=1$ for co-rotating beams ($\eta_1\eta_2=+1$).
		\end{itemize}
		This is a symmetry constraint on the full Floquet response: it holds to all orders in the drive amplitudes and in $\lambda$, and does not rely on a high-frequency expansion. It is a necessary and sufficient condition for a transverse component to be symmetry-allowed; it does not guarantee that its coefficient is nonzero, since additional cancellations can occur. The smallest allowed order is a symmetry-permitted minimum, assuming an analytic expansion; the actual leading order is higher if its coefficient vanishes. The rule relies on the $C_n$ symmetry of the full Hamiltonian: anisotropic dispersion, anisotropic spin--orbit coupling, or additional exchange harmonics break it and can lift the zeros, so the vanishing of the 34 excluded channels is a property of the present symmetric model. Each Fourier component of the transverse response must satisfy the congruence in Eq.~\eqref{eq:gcd}. The smallest allowed nonzero $|v|$ gives the lowest-order mixed contribution, which scales as $\mathcal{A}_1^{\,b|v|}\mathcal{A}_2^{\,|v|}$. More than one allowed harmonic can be present, however, and the observed phase winding is then determined by their relative weights. Thus, the gcd condition constrains the allowed phase harmonics but does not by itself fix the winding of the complete response. For the $p$-wave ($n=1$) the rule has no content, and $v=0$ is also allowed: it is the $\phi$-independent torque of each beam alone.
		
		For $n\ge2$, Eq.~\eqref{eq:gcd} excludes $b=n-1$ for counter-rotating beams and $b=n+1$ for co-rotating beams, because $\gcd(n,n)=n$. It excludes the even multipoles at $b=1$ with counter-rotating beams, and all $b=1$ channels with co-rotating beams. It allows the resonant channel $b=n$ for every $n$, since $\gcd(n,n\pm1)=1$. For in-plane $\boldsymbol{N}$ the rotation turns $\boldsymbol{N}$ into another in-plane direction and gives no rule.
		
		The rule is consistent with $n=0$. The rotation symmetry is then continuous, and Eq.~\eqref{eq:gcd} becomes $(b-\eta_1\eta_2)v=\pm1$, which is solvable only for co-rotating beams at $b=2$. For $\boldsymbol N\parallel\hat z$ the numerics give zero for the other 13 channels and $1.8\times10^{-7}\,u_0$ for co-rotating $b=2$ (Appendix~\ref{app:fm}). For in-plane $\boldsymbol N$ the torque of a Rashba ferromagnet is at most $4.6\times10^{-4}\,u_0$, about 26 times below the largest multipole torque, and scales as $1/\hbar\omega$.
		
		\subsection{Out-of-plane N\'eel vector}
		\label{sec:resultsz}
		
		\subsubsection{Channel map}
		\label{sec:mapz}
		
		We evaluate $\boldsymbol{h}_{\rm eff}$ from Eq.~\eqref{eq:heff} for $\boldsymbol{N}\parallel\hat z$ in all 84 combinations of multipole order ($n=1,\ldots,6$), harmonic number ($b=1,\ldots,7$), and helicity pair. We sample six relative phases, $\phi=2\pi m/6$. Figure~\ref{f2} shows the largest value of $|\tau_{\perp}|$ among these sampled phases. The 34 channels forbidden by Eq.~\eqref{eq:gcd} vanish, as required by symmetry. Of the 50 symmetry-allowed channels, 39 exceed the numerical threshold; the remaining 11 are also shown as zero. Thus, the gcd condition identifies the channels in which a response is allowed, while the numerical calculation determines its magnitude and reveals additional cancellations.\begin{figure*}[t]
			\centering
			\includegraphics[width=0.85\linewidth]{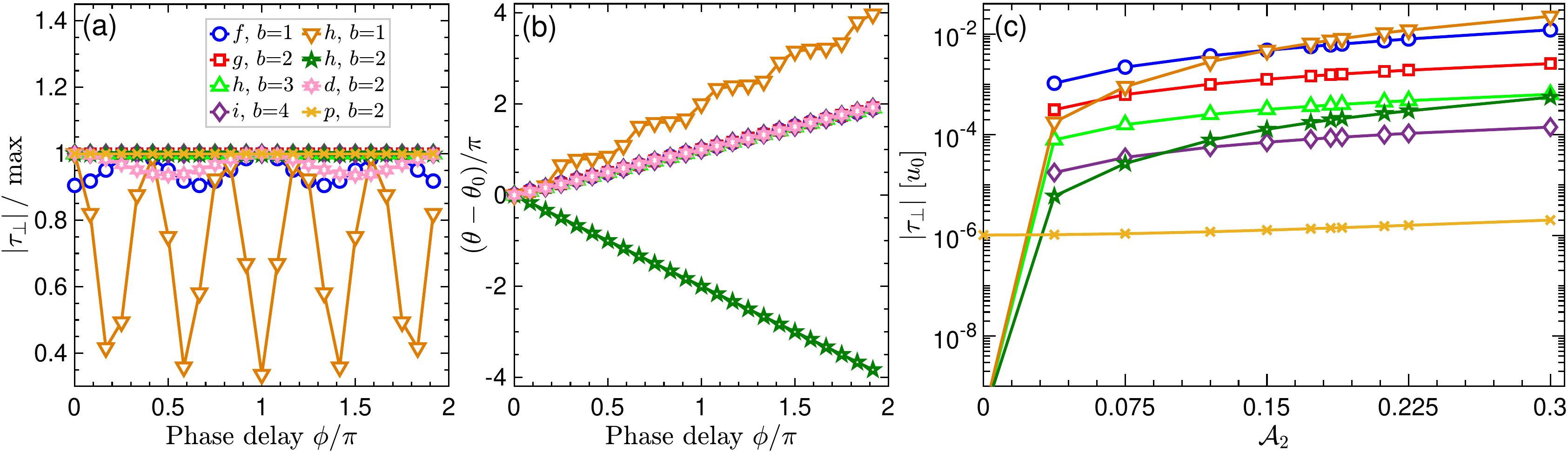}
			\caption{Phase steering for $\boldsymbol{N}\parallel\hat z$ and counter-rotating beams. (a)~$|\tau_{\perp}|$ divided by its maximum over the sampled phases (maximum in $u_0$ in the legend). (b)~Direction $\theta$ of $\tau_{\perp}$ relative to $\phi=0$~($\theta_0$), in units of $\pi$; the slope is the winding number. The $p$-wave torque does not depend on $\phi$ and appears only in (a). We fix $\mathcal{A}_1=\mathcal{A}_2=0.15$ and $\lambda=1$~meV. (c) Torque against $\mathcal{A}_2$ for $\boldsymbol{N}\parallel\hat z$ and counter-rotating beams at fixed phase: $\phi=\pi$ ($f$, $b=1$), $5\pi/3$ ($g$, $b=2$), $2\pi/3$ ($i$, $b=4$), 0 (others).}
			\label{f3}
		\end{figure*}
		
		\subsubsection{Two kinds of channels}
		\label{sec:classes}
		
		The nonzero responses fall into two groups with distinct frequency and parameter dependences. Table~\ref{tab:kres} summarizes the leading channels.\begin{table}[b!]
			\vspace*{-0.6em}\centering
			\caption{Leading channels for $\boldsymbol{N}\parallel\hat z$ (counter-rotating beams unless marked ``co''), $\mathcal{A}_1=\mathcal{A}_2=0.15$, $\lambda=1$~meV. Order: lowest order in the amplitudes; $|v|$: smallest allowed nonzero phase-harmonic index from Eq.~\eqref{eq:gcd}. Last columns: dependence on $\omega$ and exponent of $\lambda$ between 1 and 2~meV.}
			\label{tab:kres}
			\footnotesize
			\renewcommand{\arraystretch}{1.}
			\resizebox{\columnwidth}{!}{%
				\begin{tabular}{c|c|c|c|c|c}
					\hline\hline
					Class & Channel  & Order & $|v|$ & $\omega$ dependence & $\lambda$ exp. \\
					\hline
					I & $f$, $b=1$ &  2 & 1 & none ($\le1\%$) & 1.9 \\
					I & $h$, $b=1$ &  4 & 2 & none ($\le1\%$) & 1.1 \\
					I & $g$, $b=2$ & 3 & 1 & none ($\le2\%$) & 2.0 \\
					I & $h$, $b=3$ &  4 & 1 & none ($\le3\%$) & 2.0 \\
					I & $h$, $b=2$ &  6 & 2 & none ($\le5\%$) & 1.9 \\
					I & $i$, $b=4$ &  5 & 1 & none ($\le7\%$) & 2.0 \\
					\hline
					II & $i$, $b=2$, co  & 3 & 1 & $1/\hbar\omega$ & -- \\
					II & $h$, $b=2$, co &  3 & 1 & $1/\hbar\omega$ & -- \\
					II & $p$, $b=1$, co & 2 & 0 & $1/(\hbar\omega)^2$ & -- \\
					II & $p$, $b=2$ &  2 & 0 & $1/(\hbar\omega)^2$ & 3.1 \\
					II & $d$, $b=2$ &  3 & 1 & $1/\hbar\omega$ & 2.9 \\
					II & $f$, $b=3$ &  4 & 1 & $1/(\hbar\omega)^2$ & 2.7 \\
					II & $g$, $b=4$ &  5 & 1 & $1/(\hbar\omega)^2$ & 2.8 \\
					II & $h$, $b=5$ & 6 & 1 & $1/(\hbar\omega)^2$ & 2.8 \\
					II & $i$, $b=6$ & 7 & 1 & $1/(\hbar\omega)^2$ & -- \\
					\hline\hline
				\end{tabular}
			}
		\end{table}
		
		\textit{Class I} (counter-rotating beams, $b+1\le n$, $\gcd(n,b+1)=1$). The largest responses occur for the $f$- and $h$-wave cases at $b=1$. Since the two components then have the same frequency, equal amplitudes with counter-rotating helicities combine into linearly polarized light. At fixed dimensionless vector-potential amplitudes, these torques are nearly independent of $\hbar\omega$ (see below). They are therefore not photon-resonant responses. Instead, they arise from the cycle-averaged change in the exchange texture acting on spin states tilted by the Rashba coupling.
		
		\textit{Class II} is small. The largest are the co-rotating channels at $b=2$, the $p$-wave and the $d$-wave at $b=2$. The resonant channels $b=n\ge3$ lie between $f$-wave and $i$-wave. At fixed vector-potential amplitudes, class~II falls with $\hbar\omega$: as $1/\hbar\omega$ for the $d$-wave at $b=2$ and the co-rotating channels at $b=2$, and as $1/(\hbar\omega)^2$ for the $p$-wave and the resonant channels with $n=3$ to 6. For the same multipole, class~I is larger than the resonant torque by a factor between $2.2\times10^{3}$ ($h$-wave, $b=2$) and $8.5\times10^{4}$ ($h$-wave, $b=1$). The resonant channel, which has the most direct photon-counting meaning, is therefore not the one to look for in this model.
		
		The $p$-wave belongs to class II for all $b$ in this out-of-plane configuration. For $b\ge2$ its torque does not depend on $\phi$, because $v=0$ is allowed and the torques of the two beams add. The fundamental beam alone gives $1.0\times10^{-6}\,u_0$ and the harmonic beam alone gives $2.5\times10^{-7}$ ($b=2$) and $1.1\times10^{-7}\,u_0$ ($b=3$), so a single circularly polarized beam gives a torque. At $b=1$ the beams interfere: for counter-rotating beams the torque changes with $\phi$ between $1.9\times10^{-6}$ and $2.5\times10^{-6}\,u_0$, and for co-rotating beams the light is circular with amplitude $2\mathcal{A}|\cos(\phi/2)|$, so the torque vanishes at $\phi=\pi$. For $\boldsymbol{N}\parallel\hat z$ the $p$-wave is the only multipole with a torque for every harmonic and both helicity pairs.
		
		\subsubsection{Phase steering}
		\label{sec:phase}
		
		Equation~\eqref{eq:gcd} says that for counter-rotating beams $h_+=h_{{\rm eff},x}+ih_{{\rm eff},y}$ contains only the harmonics $e^{iv\phi}$ with $-(b+1)v\equiv1\pmod n$. We computed 24 phases for nine channels (eight in Fig.~\ref{f3}) and Fourier analyzed $h_+(\phi)$. For the $d$-, $f$-, $g$- and $i$-wave, the harmonics outside the allowed class are nearly zero. For the $h$-wave at $b=1,2,3$ we repeated the analysis with 60 phases, which resolves harmonics up to $\pm30$: the harmonics outside the allowed class are below $2\times10^{-17}$ in the units of $u_0$, i.e.\ $10^{-13}$ or less of the largest harmonic.
		
		The direction of $\tau_{\perp}$ turns once when $\phi$ increases by $2\pi$ for the $d$-wave at $b=2$, the $f$-wave at $b=1$, the $g$-wave at $b=2$, the $h$-wave at $b=3$ and the $i$-wave at $b=4$, as $|v|=1$ predicts. For the $f$-wave at $b=3$ (resonance) it turns once in the opposite sense. For the $h$-wave it turns twice, in the positive sense at $b=1$ and the negative sense at $b=2$. At $b=1$ the $h$-wave torque has two harmonics, $v=2$ and $v=-3$, so its magnitude changes with $\phi$ by a factor of three [Fig.~\ref{f3}(a)] while the larger harmonic sets the winding number [Fig.~\ref{f3}(b)]. In the other channels of Fig.~\ref{f3} the magnitude varies by less than 10\%. The phase delay can therefore rotate the torque in the plane at nearly fixed size.
		
		\subsubsection{Amplitude ratio}
		\label{sec:amp}
		
		Figure~\ref{f3}(c) shows $|\tau_{\perp}|$ against $\mathcal{A}_2$ between 0 and 0.3 at $\mathcal{A}_1=0.15$ and fixed phase, for seven representative channels. These include the ratios $\sqrt{4/3}$, $\sqrt{3/2}$ and $\sqrt{8/5}$, where the constant field has blindspots. Within this amplitude range, the torque grows monotonically with $\mathcal{A}_2$ and has no additional zero in any of the seven channels. Six channels are zero at $\mathcal{A}_2=0$ because they need both beams. The $p$-wave keeps $1.0\times10^{-6}\,u_0$, the torque of the fundamental beam alone. The channels with $|v|=1$ are proportional to $\mathcal{A}_2$ for $\mathcal{A}_2/\mathcal{A}_1\le1$, and the $h$-wave at $b=1$ and 2 ($|v|=2$) grow faster. For $\boldsymbol{N}\parallel\hat z$ the amplitude ratio therefore cannot switch the torque off. It tunes its size.
		
		\subsubsection{Dependence on $\lambda$, $\hbar\omega$, and $M$}
		\label{sec:scaling}
		
		\textit{Dependence on $\lambda$.} For $0.5\le\lambda\le4$~meV the class~I torque grows approximately as $\lambda^2$ [Fig.~\ref{f5}(a)], with an effective exponent between 1.7 and 2.0, and 1.1 for the $h$-wave at $b=1$. Equation~\eqref{eq:h0} reproduces the exact values over this range. At larger $\lambda$ the growth slows: at $\lambda=16$~meV, larger than $M\epsilon$, the $f$-wave torque is $0.60\,u_0$, 130 times its value at 1~meV. Class~II depends more strongly on $\lambda$, with an exponent of 2.7 to 3.1. We did not use $\lambda<0.5$~meV, because the angular windows around the nodes of $\cos n\theta$ become narrower than our grid for $n\ge5$ (Appendix~\ref{app:num}). Interfaces with heavy metals have Rashba coefficients of tens of meV\,nm, i.e.\ $\lambda$ of tens of meV for $k_{\rm F}\sim10^9~{\rm m^{-1}}$.\begin{figure*}[t]
			\centering
			\includegraphics[width=0.85\linewidth]{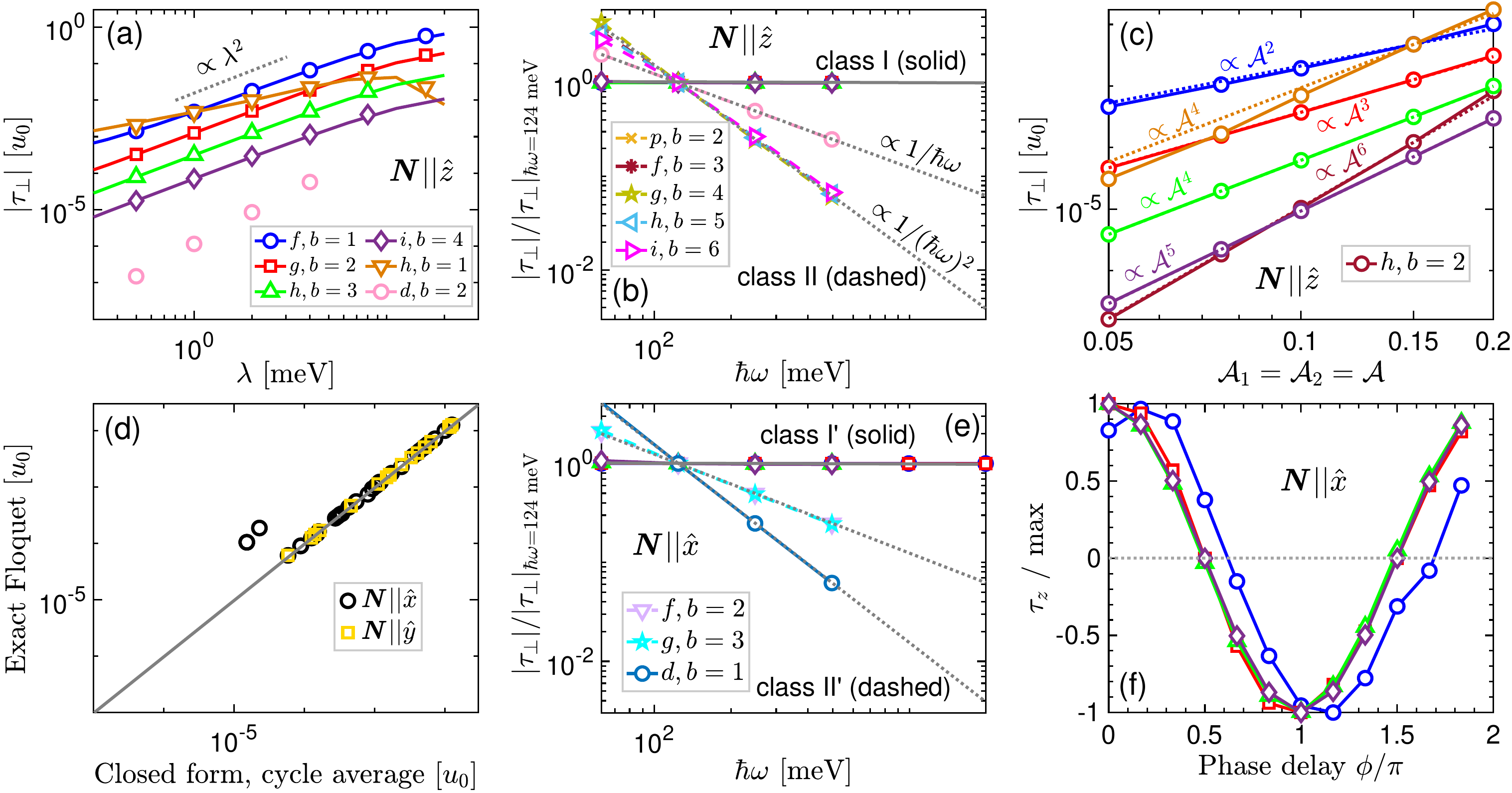}
			\caption{Scaling of the torque for counter-rotating beams. (a)--(c) $\bs N\parallel\hat z$; (d)--(f) $\bs N\parallel\hat x$. Unless varied, $\mathcal A_1=\mathcal A_2=0.15$, $\lambda=1$~meV and $\hbar\omega=124$~meV.
				(a) $\lambda$: exact Floquet states (points), Eq.~\eqref{eq:h0} (lines), $\lambda^2$ (dotted). Class~II (the $d$-wave) has no line because its cycle average is zero.
				(b) $\hbar\omega$ at fixed dimensionless vector-potential amplitudes, divided by the value at 124~meV: class I (solid), class II (dashed); dotted lines $1/\hbar\omega$ and $1/(\hbar\omega)^2$.
				(c) $\mathcal A_1=\mathcal A_2=\mathcal A$; dotted: $\mathcal A^{(b+1)|v|}$ through the value at $\mathcal A=0.15$.
				(d) Exact torque against Eq.~\eqref{eq:h0} for all channels with a finite cycle average, six phases each; $\bs N\parallel\hat x$ (circles) and $\bs N\parallel\hat y$ (squares); the line is equality.
				(e) $\hbar\omega$-dependent $\tau_{\perp} = (\tau_y^2 + \tau_z^2)^{1/2}$ at fixed dimensionless vector-potential amplitudes for $\bs N\parallel\hat x$, divided by the value at 124~meV: class I$'$ (solid), $b=n-1$ (dashed); dotted lines $1/\hbar\omega$ and $1/(\hbar\omega)^2$.
				(f) $h_{{\rm eff},y}$ (the field, so $\tau_z$) against $\phi$ for $\bs N\parallel\hat x$, divided by its largest value. For $\boldsymbol N\parallel\hat x$, $\tau_z$ (class~I$'$) rotates $\boldsymbol N$ within the plane and $\tau_y$ (class~II$'$) tilts it out of the plane.}
			\label{f5}
		\end{figure*}
		
		\textit{Dependence on $\hbar\omega$ and on the amplitudes.} At fixed dimensionless vector-potential amplitudes, class~I is nearly independent of $\hbar\omega$, whereas class~II falls as $1/\hbar\omega$ or $1/(\hbar\omega)^2$ [Fig.~\ref{f5}(b)]. The class~I torque is a power of the amplitudes with the exponent $(b+1)|v|$ [Fig.~\ref{f5}(c)]: between $\mathcal{A}=0.05$ and 0.1 the local exponent is 3.00 ($g$, $b=2$), 4.00 ($h$, $b=3$), 5.00 ($i$, $b=4$) and 6.00 ($h$, $b=2$), and 2.04 for the $f$-wave at $b=1$ between 0.05 and 0.075. The $h$-wave at $b=1$ (lowest order 4) has exponents of 4.2 to 5.0, because the next order is large. The $f$-wave exponent rises to 2.65 between 0.15 and 0.2, so $\mathcal{A}=0.15$ is at the upper end of the range where the lowest order describes the torque.
		
		\textit{Dependence on $M$.} Doubling $M$ at fixed $\epsilon$ changes the class~I torque of the $f$-, $g$- and $h$-wave (at $b=3$) by only 1.04 to 1.21, so these follow $|\tau_{\perp}|\propto\lambda^2$. Class~II depends on $M$ (factors 1.03 to 4.2), and so do the class~I channels $h$ at $b=1$ and 2 (factors 2.0 and 1.8).
		
		\subsection{In-plane N\'eel vector}
		\label{sec:inplane}
		
		\subsubsection{Channel map and the two kinds of channels}
		
		We repeated the calculation for $\boldsymbol{N}\parallel\hat x$ and $\hat y$, using the same 84 channels and six sampled phases. The torque is $\boldsymbol{N}\times\boldsymbol{h}_{\rm eff}$: $h_y\hat z-h_z\hat y$ for $\boldsymbol{N}\parallel\hat x$ and $h_z\hat x-h_x\hat z$ for $\boldsymbol{N}\parallel\hat y$. The rotation by $2\pi/n$ behind Eq.~\eqref{eq:gcd} turns an in-plane $\boldsymbol{N}$ into another in-plane direction, so there is no such rule. For $\boldsymbol{N}\parallel\hat x$, 83 channels exceed the numerical threshold; for $\boldsymbol{N}\parallel\hat y$, 62 do. The remaining channels are shown as zero. The largest torque is $1.2\times10^{-2}\,u_0$, for the $i$-wave at $b=1$ with counter-rotating beams and both orientations, 2.5 times the largest for $\boldsymbol{N}\parallel\hat z$. For in-plane $\boldsymbol N$ the maxima over $\phi$ are taken from 24 sampled phases with a trigonometric interpolation (Appendix~\ref{app:num}); the six-phase maximum used in the channel map of Fig.~\ref{f2} can be up to 49\% lower. For $n=4$ the two in-plane orientations are related by a $90^\circ$ rotation and agree to $6\%$ or better, the accuracy of the angular grid.
		
		The allowed channels again fall into two classes, which differ from those for $\boldsymbol{N}\parallel\hat z$. \textit{Class~I$'$} contains the ten counter-rotating channels with $b+1<n$ ($f$-wave at $b=1$; $g$-wave at $b=1,2$; $h$-wave at $b=1,2,3$; $i$-wave at $b=1$ to 4). Their torque ranges from $4.7\times10^{-5}$ to $1.2\times10^{-2}\,u_0$. At fixed vector-potential amplitudes, it is nearly independent of $\hbar\omega$ and follows approximately the same $\lambda^2$ dependence as class~I (Tab.~\ref{tab:ip}). The field is in the plane, perpendicular to $\boldsymbol N$ (along $\hat y$ for $\boldsymbol N\parallel\hat x$), so the torque is along $\hat z$ and rotates $\boldsymbol N$ within the plane. The cause is the same cycle-averaged exchange, and the six class~I channels are among the ten. For in-plane $\boldsymbol{N}$ it acts for every $b$ with $b+1<n$, not only when $\gcd(n,b+1)=1$. For example, the $g$-wave at $b=1$ gives $4.0\times10^{-3}\,u_0$ for $\boldsymbol{N}\parallel\hat x$ (0.9 times the $f$-wave) and zero for $\boldsymbol{N}\parallel\hat z$.\begin{table}[t!]
			\vspace*{-0.6em}\centering
			\caption{Counter-rotating channels with the largest in-plane torque $|\tau_{\perp}|$ at $\mathcal{A}_1=\mathcal{A}_2=0.15$, $\lambda=1$~meV, $\hbar\omega=124$~meV, and maximum over $\phi$, interpolated from 24 sampled phases (Appendix~\ref{app:num}); entries marked $^\dagger$ are maxima over six sampled phases. Class~I$'$: $b+1<n$. Class~II$'$: $b=n-1$. Last columns: dependence on $\omega$ and exponent of $\lambda$ between 1 and 2~meV for $\boldsymbol{N}\parallel\hat x$ (dashes: not computed, since $\lambda<1$~meV is not converged for $n\ge5$).}
			\label{tab:ip}
			\renewcommand{\arraystretch}{1.3}
			\resizebox{\columnwidth}{!}{%
				\begin{tabular}{c|c|c|c|c|c}
					\hline\hline
					Class & Channel & $\boldsymbol{N}\parallel\hat x$ & $\boldsymbol{N}\parallel\hat y$ & $\omega$ dependence & $\lambda$ exp. \\
					\hline
					I$'$ & $f$, $b=1$ & $4.5\!\times\!10^{-3}$& $2.3\!\times\!10^{-3}$ & none & 1.9 \\
					I$'$ & $g$, $b=1$ &$4.0\!\times\!10^{-3}$ & $4.0\!\times\!10^{-3}$& none & 1.7 \\
					I$'$ & $h$, $b=1$ & $1.1\!\times\!10^{-2}$ & $1.1\!\times\!10^{-2}$ & none & 1.6 \\
					I$'$ & $i$, $b=1$ & $1.2\!\times\!10^{-2}$ & $1.2\!\times\!10^{-2}$ & none & 1.7 \\
					I$'$ & $g$, $b=2$ & $1.1\!\times\!10^{-3}$ & $1.1\!\times\!10^{-3}$ & none & 2.0 \\
					I$'$ & $h$, $b=2$ & $1.5\!\times\!10^{-4}$ & $7.5\!\times\!10^{-5}$& none & -- \\
					I$'$ & $i$, $b=2$ & $1.0\!\times\!10^{-3}$ & $1.0\!\times\!10^{-3}$ & none & -- \\
					I$'$ & $h$, $b=3$ & $2.7\!\times\!10^{-4}$ & $2.6\!\times\!10^{-4}$ & none & 2.0 \\
					I$'$ & $i$, $b=3$ & $4.7\!\times\!10^{-5}$$^\dagger$ & $4.7\!\times\!10^{-5}$$^\dagger$ & none & -- \\
					I$'$ & $i$, $b=4$ & $6.4\!\times\!10^{-5}$ & $5.5\!\times\!10^{-5}$$^\dagger$ & none & 2.0 \\\hline
					II$'$$^\dagger$ & $d$, $b=1$ & $4.5\!\times\!10^{-5}$ & $4.5\!\times\!10^{-5}$ & $1/(\hbar\omega)^2$ & 1.9 \\
					II$'$$^\dagger$ & $f$, $b=2$ & $7.4\!\times\!10^{-5}$ & $7.4\!\times\!10^{-5}$ & $1/\hbar\omega$ & 2.0 \\
					II$'$$^\dagger$ & $g$, $b=3$ & $2.7\!\times\!10^{-5}$ & $2.7\!\times\!10^{-5}$ & $1/\hbar\omega$ & 2.0 \\
					II$'$$^\dagger$ & $h$, $b=4$ & $7.2\!\times\!10^{-6}$ & $7.2\!\times\!10^{-6}$ & $1/\hbar\omega$ & 2.0 \\
					II$'$$^\dagger$ & $i$, $b=5$ & $1.8\!\times\!10^{-6}$ & $1.8\!\times\!10^{-6}$ & $1/\hbar\omega$ & 2.0 \\
					\hline\hline
				\end{tabular}
			}
		\end{table}
		
		\subsubsection{Scaling and comparison with the cycle average}
		
		\textit{Class~I$'$.} The torque scales as $\mathcal{A}_1^{\,b}\mathcal{A}_2$ for the channels we tested ($g$, $b=2$; $h$, $b=3$; $i$, $b=4$). The $b=1$ channels are not pure power laws at $\mathcal{A}=0.15$. Figure~\ref{f5}(d) compares the exact torque with Eq.~\eqref{eq:h0}. For the $f$-wave at $b=1$, the $g$-wave at $b=1,2$, the $h$-wave at $b=3$ and the $i$-wave at $b=3,4$ they agree to $2.3\%$ at all six phases. For the $h$- and $i$-wave at $b=1$ the differences at individual phases are larger where the cycle-averaged torque is small, although the largest torques agree to about $1\%$. These channels grow more slowly than $\lambda^2$ and have a grid error of about $5\%$ and we cannot separate grid error from higher orders in $\lambda$. For the $h$- and $i$-wave at $b=2$ they agree to about $3\%$ once the amplitude ramp is resolved finely (Appendix~\ref{app:num}); with a coarse ramp the exact result is wrong at the few momenta where the ramp path nearly crosses a band touching. The $\omega$ dependence is flat: from 62 to 496~meV the channels of Tab.~\ref{tab:ip} change by at most $7\%$ [Fig.~\ref{f5}(e)]. For $\boldsymbol N\parallel\hat x$ ($\hat y$), the momentum-independent field has $h_{{\rm eff},y}=0$ ($h_{{\rm eff},x}=0$) for every multipole (Tab.~\ref{tab:effective_fields}), so it never gives $\tau_z$. The class~I$'$ torque along $\hat z$ therefore exists only in the Fermi-sea calculation and is a finite-momentum effect.
		
		\textit{Class~II$'$.} For $b=n-1$ the cycle-averaged contribution is zero, or at most $8\%$ of the full response in the channels examined. The remaining torque is therefore generated mainly by the time-dependent part of the drive. It comes mainly from $h_z$, so the torque lies in the plane, perpendicular to $\boldsymbol{N}$ (along $\hat y$ for $\boldsymbol N\parallel\hat x$), and tilts $\boldsymbol N$ out of the plane. For $n\ge3$ it scales as $\mathcal{A}_1^{\,n-1}\mathcal{A}_2\,\lambda^2/\hbar\omega$ (Tab.~\ref{tab:ip}), is between $1.7\times10^{-6}$ and $7.4\times10^{-5}\,u_0$ at 124~meV, and falls with $n$ by a factor of 3 to 4 per step. The $f$-wave at $b=2$ also has an $h_y$ component of $6.2\times10^{-6}\,u_0$ that does not depend on $\hbar\omega$ (the $\boldsymbol k$-independent part of the drive), so at large $\hbar\omega$ its torque tends to this value, not to zero. The $d$-wave at $b=1$ ($b=n-1$ for $n=2$) is different: its torque is in $h_y$, falls as $1/(\hbar\omega)^2$ and is $4.5\times10^{-5}\,u_0$. All other channels, including the co-rotating ones, are below $7.0\times10^{-6}\,u_0$.
		
		\textit{Phase dependence.} For $\boldsymbol{N}\parallel\hat x$ the first harmonic ($e^{\pm i\phi}$) is the largest component for all channels of Tab.~\ref{tab:ip} except the $h$-wave at $b=2$ and the $i$-wave at $b=3$ [Fig.~\ref{f5}(f)]. The torque of the $b=n-1$ channels and of the $i$-wave at $b=4$ changes sign when $\phi$ changes by $\pi$, as for $\cos(\phi-\phi_0)$. For the others the sign change is only approximate, because even harmonics are present. The phase therefore controls the sign and size of the torque, not its axis, which stays along $\hat z$ for class~I$'$ (a rotation of $\boldsymbol N$ within the plane). For $b=1$ this is a function of the polarization axis (Sec.~\ref{sec:discussion}).
		
		\textit{Single frequency.} A single circularly polarized beam also gives a $\phi$-independent torque on an in-plane $\boldsymbol{N}$, of order $\mathcal{A}^2$. For odd $n$ and $\boldsymbol{N}\parallel\hat x$ it falls as $1/(\hbar\omega)^2$ and at $\mathcal{A}=0.15$, $\hbar\omega=124$~meV it is $1.8\times10^{-6}\,u_0$ ($p$-wave), $5.7\times10^{-7}$ ($f$) and $3.2\times10^{-7}$ ($h$). For even $n$ and $\boldsymbol{N}\parallel\hat x$ it is below $7.4\times10^{-10}\,u_0$, and for odd $n$ and $\boldsymbol{N}\parallel\hat y$ it is zero. It is far below the class~I$'$ torque of the same $n$.

		\subsection{Comparison of three cases}
		\label{sec:decomp}
		
		To identify the roles of the different terms in the exchange texture, we perform two additional calculations. In the first, we retain the static exchange term ($j=0$) together with the $\boldsymbol{k}$-independent light-induced term ($j=n$); in the second, we retain the static term together with the $\boldsymbol{k}$-dependent light-induced terms ($1\le j<n$), following the decomposition introduced in Sec.~\ref{sec:model}. We compare both with the full calculation. These are separate driven Hamiltonians whose Floquet states are recalculated independently. In particular, the Fermi-sea torque in the $\boldsymbol{k}$-independent model is not the same quantity as the constant effective field $\boldsymbol N\times\boldsymbol h_{\rm eff}$ evaluated in Sec.~\ref{sec:kindep}.
		
		We express the constant field of Sec.~\ref{sec:kindep} and the three Fermi-sea results in the common unit $u_0$. We also compare calculations at $\hbar\omega=62$, 124, 248, and 496~meV. The main trends are as follows.
		
		In the resonant channels ($b=n$) the constant field is $10^{-3}$ to $10^{-2}\,u_0$ for $\boldsymbol N\parallel\hat z$ (largest for the $d$-wave, $1.0\times10^{-2}\,u_0$), but the Fermi-sea torque is $10^{3}$ to $10^{4}$ times smaller, because occupied-band cancellation restricts the contributing states to the shell between the two spin-split Fermi surfaces. The full suppression also includes spin-matrix-element and momentum-integration effects whose separate contributions we have not established. In class~I and I$'$ the order is reversed: the Fermi-sea torque is 0.7 to 8 times the constant-field torque for the $g$-, $h$- and $i$-wave at $\boldsymbol N\parallel\hat z$, and 99 times for the in-plane $i$-wave at $b=1$ ($1.2\times10^{-2}$ against $1.2\times10^{-4}\,u_0$). For the $f$- and $h$-wave at $b=1$ the constant field is zero by the selection rule, whereas the Fermi-sea torque is $4.7\times10^{-3}\,u_0$. The largest Fermi-sea torque ($1.2\times10^{-2}\,u_0$) is as large as the largest constant-field torque ($1.2\times10^{-2}\,u_0$).
		
		In class~I, class~I$'$ and class~II$'$ for $b=n-1\ge2$, the $\boldsymbol k$-dependent model gives the full torque within 2\%. The $\boldsymbol k$-independent model gives at most $10^{-4}$ of it for $\boldsymbol N\parallel\hat z$ and 0.4\% for the in-plane channels at $b=1$ with $n\ge3$.
		
		For $b=n$ neither model gives the full torque. For the $d$-wave the $\boldsymbol k$-dependent model gives zero, the $\boldsymbol k$-independent case $1.7\times10^{-6}\,u_0$ and the full model $1.2\times10^{-6}\,u_0$. For the $f$-wave the two parts are comparable, and for the $g$-wave the $\boldsymbol k$-dependent part is 80\%. These channels are $10^{3}$ to $10^{5}$ times smaller than class~I. The two reduced-model responses do not add to the full response. The difference between the full torque and their sum can be comparable to the full torque itself, because the Floquet states and occupations change when the Hamiltonian is changed. The reduced-model results should therefore be interpreted diagnostically, not as additive components.
		
		In the resonant channels the $\boldsymbol k$-independent torque falls by a factor of 14 ($d$-wave) and 64 ($f$-wave) from $\hbar\omega=62$ to 496~meV, i.e.\ as $1/\hbar\omega$ to the power 1.3 and 2.0. For the in-plane $f$-wave at $b=2$ the $\boldsymbol k$-independent part does not depend on $\hbar\omega$ (less than 1\%), whereas the $\boldsymbol k$-dependent part falls as $1/\hbar\omega$.
		
		The constant field of Sec.~\ref{sec:kindep} is not a general estimate of the full Fermi-sea torque. It captures the resonant channels but misses the dominant class~I and class~I$'$ responses, which arise mainly from the $\boldsymbol{k}$-dependent light-induced exchange terms and can be as large as the largest constant-field torque. The three-model comparison was performed at $\mathcal A_1=\mathcal A_2=0.15$ and $\lambda=1$~meV.
		
		For the parameters and channels examined here, the full Fermi-sea response differs from the momentum-independent effective field in four main respects. (i) For $\boldsymbol N\parallel\hat z$ the Fermi-sea torque obeys $\gcd(n,b\pm1)=1$, Eq.~\eqref{eq:gcd}, and not the photon-counting rule. For $b>n\ge2$ the momentum-independent field contains only $h_{\mathrm{IFE}}$. The full Fermi-sea response can nevertheless have symmetry-allowed transverse channels when the gcd condition is satisfied; in the parameter range studied, these channels are small. (ii) In the resonant channels the Fermi-sea torque is $5\times10^{-5}$ to $1.4\times10^{-3}$ of the field of Eq.~\eqref{eq:resonant_general}. It falls as $1/\hbar\omega$ ($d$-wave) or $1/(\hbar\omega)^2$ (the others), whereas the constant field is $\propto1/\hbar\omega$. (iii) The zero of the constant field at $\mathcal A_2/\mathcal A_1=\sqrt{c_1/c_2}$ is not a zero of the Fermi-sea torque [Fig.~\ref{f3}(c)], and neither is the $p$-wave zero at $\mathcal A_2=\sqrt b\,\mathcal A_1$ in our scan. (iv) The largest Fermi-sea torques (class~I for $\boldsymbol N\parallel\hat z$, class~I$'$ for in-plane $\boldsymbol N$) come from the terms with $j<n$ that the constant field does not contain. Co-rotating channels at $b=2$ are allowed for $\boldsymbol N\parallel\hat z$ but small (at most $5.5\times10^{-6}\,u_0$).
		
		These comparisons are specific to the model and parameter ranges studied. In particular, frequency scalings quoted above refer to calculations at fixed dimensionless vector-potential amplitudes, not at fixed electric-field amplitudes. The full Fermi-sea torque, as defined here, is an energy-valued electronic generalized torque per occupied electron; translating it into the dynamics of a local N\'eel order requires a microscopic coupling to the magnetic sublattices and a corresponding magnetic-dynamics model.
		
		\section{Experimental feasibility}
		\label{sec:discussion}
		
		The largest out-of-plane Fermi-sea torque in our calculations is
		$4.7\times10^{-3}\,u_0$ per occupied electron, corresponding to
		$0.19~\mu$eV for $u_0=40~\mu$eV. Expressed as an equivalent Zeeman
		energy for an electron spin with $g=2$, this is approximately $3.3$~mT, taking $h_{\rm eff}$ as the coefficient of $\boldsymbol\sigma$ in $(g\mu_B/2)\boldsymbol B\cdot\boldsymbol\sigma$ (the spin splitting is twice this). When the N\'eel vector lies in the plane, the largest value
		in the parameter range considered is $1.2\times10^{-2}\,u_0$. These values characterize the electronic
		generalized torque per occupied electron. They are not directly the
		torque on a local magnetic moment. A conversion to the local-moment
		torque requires the number of participating carriers per moment, the
		local spin $S$, and, more importantly, the microscopic coupling
		between the itinerant electrons and the magnetic sublattices. The
		calculation therefore identifies a possible steering response, but
		does not establish the size of a resulting N\'eel-vector tilt or
		switching threshold. The occupations are prescribed and no relaxation is included, so we do not claim that the response is purely reactive or dissipation-free; dissipative contributions are not calculated. For odd $n$ the two-component Hamiltonian at fixed $\boldsymbol N$ has ordinary spinful time-reversal symmetry, $\Theta=i\sigma_yK$, so $\boldsymbol N$ is not a conventional time-reversal-odd N\'eel vector. For the $p$-wave case it is the normal $\hat{{\bs J}}\times\hat{{\bs J}}'$ ~(moment directions) of the plane of coplanar moments, which is even under time reversal, as we show for a four-band model in Appendix~\ref{app:pwave}. For $n=3$ and 5 it remains a phenomenological spin-texture parameter, and our results for these cases are generalized forces on this parameter. We do not calculate energy
		transfer to the local moments or their subsequent dynamics.
		
		The dimensionless amplitudes are
		$\mathcal{A}_1=eE_1/(\hbar\omega k_{\rm F})$ and
		$\mathcal{A}_2=eE_2/(\hbar b\omega k_{\rm F})$. For
		$k_{\rm F}\sim10^{9}~{\rm m}^{-1}$ and
		$\omega/2\pi\simeq30$~THz ($\hbar\omega=124$~meV), the amplitude
		$\mathcal{A}_1=0.15$ corresponds to
		$E_1\simeq0.19$~MV/cm. If the two dimensionless amplitudes are equal,
		then $E_2=bE_1$: the second component requires approximately
		$0.37$~MV/cm at $b=2$ (60~THz) and $0.56$~MV/cm at $b=3$
		(90~THz). Fields of a few MV/cm have been demonstrated in the
		terahertz and mid-infrared ranges~\cite{fulop2020strongfield,kim2024ionizing}.
		Thus, the field strengths used in our model are within the range of
		reported strong-field experiments. This comparison alone, however,
		does not establish experimental feasibility in a particular
		heterostructure. Pulse duration, absorption, substrate effects, and
		thermal management must also be considered. In metallic films,
		intraband absorption can cause heating, while in semiconductors
		interband absorption becomes important when $b\hbar\omega$ exceeds
		the gap~\cite{oka2019floquet,rudner2020band}. Short pulses, of the
		order of a picosecond or less, may help limit heating
		\cite{kirilyuk2010ultrafast}. These processes, as well as relaxation
		and heating during the pulse, are not included in our calculation.
		
		The large responses obtained for the $f$- and $h$-wave cases at
		$b=1$ do not require two distinct optical frequencies. For
		$b=1$, $\eta_1=-\eta_2$, and
		$\mathcal{A}_1=\mathcal{A}_2=\mathcal{A}$, the two components combine
		to give linearly polarized light with amplitude $2\mathcal{A}$ and
		polarization axis at angle $-\phi/2$ relative to $\hat x$. Rotating
		the polarization axis by an angle $\theta$ is equivalent to changing
		$\phi$ by $-2\theta$. Accordingly, the calculated torque direction
		for the $f$-wave case, where one harmonic ($|v|=1$) dominates, rotates by $2\theta$. For the $h$-wave at $b=1$ two harmonics ($v=2$ and $-3$) are present, so the net winding is two but the direction does not rotate uniformly. This provides a direct way to
		test the predicted phase dependence using a single-frequency source
		and a rotatable polarization axis, for example with a half-wave
		plate. For the parameters above, the required field amplitude is
		$E=2E_1\simeq0.37$~MV/cm at 30~THz. The predicted response should
		be interpreted as an electronic torque; whether it produces a
		measurable N\'eel-vector rotation depends on the magnetic dynamics
		of the sample.
		
		In the present model, every transverse torque requires RSOC
		\cite{manchon2015new,amin2020interfacial}. Altermagnetic monolayers
		on heavy-metal substrates such as Pt, W, and Ta, as well as 2D Janus
		structures, are possible settings in which interfacial or intrinsic
		structural inversion breaking can produce appreciable RSOC
		\cite{song2021spintorques,chen2021soc2d,zollner2025soc}. A gate
		voltage may also provide a means of tuning the coupling. For
		$k_{\rm F}\sim10^{9}~{\rm m}^{-1}$, $\lambda=1$~meV corresponds to
		$\alpha_R\sim1$~meV\,nm, which is at the weaker end of reported
		Rashba energy scales; heavy-metal interfaces can have values of
		several tens of meV\,nm. In our calculations, the class-I torque
		grows approximately as $\lambda^2$ up to about $\lambda=4$~meV.
		For the $f$-wave case, increasing $\lambda$ to 16~meV gives a
		reported torque of $0.60\,u_0$, about 130 times its value at $\lambda=1$~meV, corresponding to an equivalent electron-spin Zeeman field of approximately 0.4~T, with the convention used above. This extrapolation should be treated with care:
		large RSOC can modify the band structure and the electronic states,
		whereas our estimate assumes that changing the interface affects
		only $\lambda$ and leaves the other model parameters unchanged.
		
		Our model is intended as a minimal description of 2D magnets and
		thin-film magnetic systems~\cite{smejkal2022emerging,song2025altermagnets}.
		For $p$-wave systems ($n=1$), metallic magnets with an electrically switchable $p$-wave magnet have been
		reported~\cite{yamada2025pwave,song2025electrical}. For $d$-wave
		systems ($n=2$), RuO$_2$ is a widely discussed candidate, although
		its magnetic order remains debated~\cite{lin2024ruo2,smejkal2022emerging}.
		MnTe~\cite{Krempasky2024} and CrSb~\cite{Long2026,Reimers2024}
		have been discussed in connection with $g$-wave order ($n=4$),
		but applying the present 2D model requires establishing that the
		relevant film or surface has the in-plane exchange texture represented
		by Eq.~\eqref{eq_2d}. First-principles calculations have predicted
		$h$-wave magnetism in FeSe~\cite{zk69-k6b2,8hq8-38qs}. We are not
		aware of established experimental realizations of the $f$- and
		$i$-wave cases. In addition, the higher odd-parity textures in our
		model are phenomenological extensions of the reduced $p$-wave
		description; their microscopic realization in a specific material
		remains to be established. The material examples should therefore
		be viewed as possible directions for testing the underlying ideas,
		rather than as direct material-specific predictions.
		
		The relevant observable is the light-induced change in the N\'eel
		vector. Time-resolved magneto-optical Kerr measurements are one
		possible probe~\cite{razdolski2017trmoke,huang2021trmoke}, provided
		the measured Kerr signal contains a component that is odd under
		reversal of $\boldsymbol N$. The model suggests several trends that
		could be tested.
		
		For
		$\boldsymbol N\parallel\hat z$, the direction of the calculated
		transverse response winds by an integer $v$ as $\phi$ is varied through $2\pi$. Eq.~\eqref{eq:gcd} selects the allowed harmonics and the dominant one sets $v$ (Table~\ref{tab:kres}); where two harmonics compete, as for the $h$-wave at $b=1$, the rotation is not uniform.
		Measuring this phase dependence would test the predicted
		connection between the harmonic content of the drive and the
		direction of the electronic response.
		
		For an out-of-plane
		N\'eel vector, the condition $\gcd(n,b\pm1)=1$ predicts a
		vanishing transverse response in 34 of the 84 channels
		considered. Examples for counter-rotating beams include the
		$d$-wave at $b=1$, the $f$-wave at $b=2$, and the $g$-wave at
		$b=3$. For co-rotating beams, the condition excludes every
		$n\geq2$ case at $b=1$. Comparing a symmetry-forbidden channel
		with an allowed harmonic in the same sample would provide a
		useful null test.
		
		The large class-I response identified
		for an out-of-plane N\'eel vector occurs for counter-rotating
		beams. Reversing the helicity of one component changes the
		driving channel and is predicted to suppress this response to
		the much smaller class-II values in the cases listed in
		Table~\ref{tab:kres}.
		
		At fixed dimensionless vector-potential
		amplitudes, the class-I torque is approximately independent of
		$\hbar\omega$, whereas the class-II contribution decreases as
		the frequency is increased. At fixed electric-field amplitude,
		however, $\mathcal{A}_{1,2}$ also depend on frequency. The
		resulting experimental frequency dependence therefore combines
		the intrinsic frequency scaling with the frequency dependence
		of the optical amplitudes and with absorption and relaxation.
		
		In the dominant
		$\boldsymbol{k}$-dependent channels, the calculated torque has
		no amplitude-ratio blindspot over the parameter range studied.
		Changing $\mathcal{A}_2/\mathcal{A}_1$ therefore tunes the
		response magnitude without producing the zeros found in the
		$\boldsymbol{k}$-independent effective field.
		
		Rotating the N\'eel
		vector from out of plane into the plane changes the set of
		active channels and can generate an out-of-plane component of
		the electronic torque. For example, the $g$-wave at $b=1$
		gives a vanishing response for
		$\boldsymbol N\parallel\hat z$ but a torque of
		$4\times10^{-3}\,u_0$ for
		$\boldsymbol N\parallel\hat x$.
		
		These tests concern the symmetry and parameter dependence of the
		calculated electronic response. Predicting the magnitude and time
		dependence of an experimentally measured N\'eel-vector tilt will
		require a material-specific model that includes the coupling to the
		local moments, magnetic anisotropy, damping, and the effects of
		heating and relaxation.		
		
		\section{Conclusion}
		\label{sec:conclusion}
		
		We have studied the optical spin-orbit torque generated by bichromatic $(\omega,b\omega)$ light in two-dimensional Rashba-coupled magnetic multipoles, covering orders $n=1$--$6$, from $p$-wave magnets to $i$-wave altermagnets. By combining analytical results for the momentum-independent light-induced field with numerical calculations based on the exact Floquet states of the occupied electronic states, we have identified how the multipole order, harmonic number, relative phase, beam amplitudes, and helicities determine the transverse electronic response.
		
		The two contributions to the light-induced response have distinct properties. The momentum-independent effective field follows photon-counting selection rules. Its transverse component has a leading contribution at resonance ($b=n$), depends on the relative phase and helicities, and can vanish at specific amplitude ratios (optical blindspots). This single-particle picture alone, however, does not describe the full Fermi-sea torque. When integrated over occupied states, occupied-band cancellation restricts the contributing states and, together with spin-matrix-element and momentum-integration effects that we do not separate, suppresses the resonant contribution by two to four orders of magnitude. Instead, the net response is dominated by momentum-dependent exchange terms generated by the drive at finite momentum. These terms govern several nonresonant channels and remove the amplitude-ratio blindspots found for the momentum-independent field. Furthermore, the largest out-of-plane and in-plane responses arise from different sets of channels, demonstrating the essential role of the N\'eel-vector orientation.
		
		For an out-of-plane N\'eel vector, the transverse Fermi-sea torque is constrained by the rotational symmetry of the model. A nonzero response is symmetry-allowed only when $\gcd(n,b+1)=1$ for counter-rotating beams or $\gcd(n,b-1)=1$ for co-rotating beams. This condition excludes 34 of the 84 channels considered. It is a symmetry-based condition for a response to be allowed; it does not by itself guarantee a finite or large torque in every allowed channel. In the allowed channels, the relative phase controls the direction of the torque---rotating it with a discrete winding number $v$ per $2\pi$ cycle---while the beam amplitudes determine its magnitude.
		
		The allowed out-of-plane channels have two main origins. Class~I comprises six channels associated with the cycle-averaged part of the driven exchange texture. These channels require counter-rotating beams, are approximately independent of $\hbar\omega$ at fixed dimensionless vector-potential amplitudes, and grow approximately as $\lambda^2$ in the parameter range studied. The largest responses in this class occur for the $f$- and $h$-wave cases at $b=1$. In this limit, the two frequency components have the same frequency and, for equal amplitudes and opposite helicities, combine to form a single linearly polarized field. Class~II channels depend on the oscillating part of the drive and decrease with increasing frequency, with scalings of either $1/\hbar\omega$ or $1/(\hbar\omega)^2$ in the regimes identified in the calculations.
		
		For an in-plane N\'eel vector, the out-of-plane gcd selection rule no longer applies, and additional channels become available. In particular, ten counter-rotating channels with $b+1<n$ (class~I$'$) exhibit an approximately frequency-independent response at fixed dimensionless vector-potential amplitudes with a field perpendicular to $\boldsymbol N$ in the plane; the torque is along $\hat z$ and rotates $\boldsymbol N$ within the plane. The channels with $b=n-1$ produce a smaller response that decreases approximately as $1/\hbar\omega$ and whose torque tilts $\boldsymbol N$ out of the plane. Thus, changing the N\'eel-vector orientation changes not only the direction of the response but also the set of harmonics that can generate it.
		
		These results suggest several experimentally testable trends: phase-dependent steering of the transverse response, helicity-dependent selection of the allowed channels, the absence of amplitude-ratio blindspots in the dominant $\boldsymbol{k}$-dependent contribution, and distinct frequency dependences for the different channel classes. They also indicate that a single linearly polarized beam may generate a substantial transverse electronic response in the $f$- and $h$-wave cases. The predicted responses could be investigated through time-resolved measurements of the N\'eel-vector orientation, provided that the material has a suitable magneto-optical signal.
		
		Several limitations should be kept in mind. The calculation uses a minimal one-band model with a single pair of spin-split bands, occupations assigned from the undriven bands, and no relaxation or heating. The odd-parity multipole description is based on a reduced model; its extension beyond the $p$-wave case is a modeling assumption rather than a microscopic derivation for a specific material. In addition, the calculated quantity is an energy-valued electronic generalized torque per occupied electron. Its conversion into a torque acting on local moments depends on the microscopic coupling to the magnetic sublattices, the number of carriers per moment, and the magnetic dynamics. The present results therefore identify possible electronic steering responses rather than demonstrate N\'eel-vector switching. A material-specific treatment that includes these effects, together with relaxation and heating, will be needed to assess the resulting magnetic dynamics quantitatively. Within these limits, our results establish a connection between magnetic multipole order and the symmetry and frequency content of optically induced electronic torques. We hope they provide useful guidance for future theoretical and experimental studies of light-driven control in two-dimensional magnetic systems.
		
		A Rashba ferromagnet ($n=0$) has one symmetry-allowed out-of-plane channel and only a weak in-plane response without a selection rule, so the large torques require the momentum-dependent exchange of the multipoles.
		
		The calculated electronic torque provides a route to steering
		the Néel vector on ultrafast timescales. Whether this torque
		produces a stable reorientation or switching of the local
		moments depends on the magnetic dynamics and material-specific
		parameters, which are not included in the present model.
		
		\section*{acknowledgments}
		M.,Y. was supported by the Department of Energy, Office of Basic Energy Sciences, Division of Materials Sciences and Engineering, under Contract No. DE-FG02-08ER46542 for the idea, formal development, analytical and numerical work, and writing of the manuscript.
		
		\section*{Data availability}
		The data that support the findings of this article are not publicly available. The data are available from the authors upon reasonable request.
		
		\appendix
		
		\section{Minimal model for the odd-parity $p$-wave magnet}
		\label{app:pwave}
		
		In Sec.~\ref{sec:model}, the odd-parity multipoles are described by an exchange term of the form $\epsilon\mathcal{M}_n(\boldsymbol{k})\,\boldsymbol{N}\cdot\boldsymbol{\sigma}$. A real odd-parity magnet generally has several magnetic sublattices. Here we illustrate how the $p$-wave form can arise from a coplanar magnet. In this appendix, momentum is expressed in physical units.
		
		We follow the coplanar $p$-wave magnet of Ref.~\cite{p21c-rtv7}, with two sublattices ($\zeta$) and spin ($\sigma$), hopping $t$, exchange hopping $t_J$, and moment directions $\hat{{\bs J}}=(\cos\varphi,\sin\varphi,0)$ and $\hat{{\bs J}}'=(\cos\varphi',\sin\varphi',0)$ on the two sublattices (lattice constant set to 1). We add the Rashba term, taken to be the same on both sublattices,
		\begin{align}
			\label{eq:fb_H}
			H_{\boldsymbol k}={}&2t\Big(\cos\tfrac{k_x}{2}\,\zeta_1+\cos k_y\Big)\notag \\ {} &+2t_J\Big(\sin\tfrac{k_x}{2}\,\zeta_2\,\hat{{\bs J}}\!\cdot\!\boldsymbol\sigma+\cos k_y\,\zeta_3\,\hat{{\bs J}}'\!\cdot\!\boldsymbol\sigma\Big)\notag \\ {} &+\lambda\,\zeta_0\,(\sin k_x\,\sigma_y-\sin k_y\,\sigma_x).
		\end{align}
		
		Without the exchange term the sublattice states with $\zeta_1=\pm1$ are eigenstates, with energies $2t\cos k_y\pm2t\cos(k_x/2)$ and splitting $\Delta E_{\boldsymbol k}=4t\cos(k_x/2)$. In the basis $|\zeta_1=\pm1\rangle$ we have $\langle+|\zeta_2|-\rangle=i$ and $\langle+|\zeta_3|-\rangle=1$, while $\zeta_0$ is unchanged. Then
		\begin{subequations}
			\begin{align}
				H_{\boldsymbol k}&=\begin{pmatrix}h_++\lambda_{\boldsymbol k}&V_{\boldsymbol k}\\V_{\boldsymbol k}^\dagger&h_-+\lambda_{\boldsymbol k}\end{pmatrix},\\
				h_\pm&=2t\cos k_y\pm\tfrac12\Delta E_{\boldsymbol k},\quad\lambda_{\boldsymbol k}=\lambda(\sin k_x\,\sigma_y-\sin k_y\,\sigma_x),\\
				V_{\boldsymbol k}&=2t_J\Big(i\sin\tfrac{k_x}{2}\,\hat{{\bs J}}\!\cdot\!\boldsymbol\sigma+\cos k_y\,\hat{{\bs J}}'\!\cdot\!\boldsymbol\sigma\Big).
			\end{align}
		\end{subequations}
		The Rashba term is diagonal in $\zeta_1$ and identical in both blocks, so it does not couple them at first order. The blocks couple only through $V_{\boldsymbol k}$.
		
		For $|V_{\boldsymbol k}|,\lambda\ll|\Delta E_{\boldsymbol k}|$~($t_J, \lambda \ll t$), the effective Hamiltonian of the $\zeta_1=+1$ block is
		\begin{equation}
			H^{\rm eff}_{\boldsymbol k}=h_++\lambda_{\boldsymbol k}+\frac{V_{\boldsymbol k}V_{\boldsymbol k}^\dagger}{\Delta E_{\boldsymbol k}}+\mathcal O\!\Big(\frac{t_J^4}{t^3},\frac{\lambda t_J^2}{t^2}\Big).
		\end{equation}
		With $A=\hat{{\bs J}}\!\cdot\!\boldsymbol\sigma$, $B=\hat{{\bs J}}'\!\cdot\!\boldsymbol\sigma$ and $[A,B]=2i(\hat{{\bs J}}\times\hat{{\bs J}}')\!\cdot\!\boldsymbol\sigma$,
		\begin{align}
			&V_{\boldsymbol k}V_{\boldsymbol k}^\dagger=4t_J^2\Big[\sin^2\tfrac{k_x}{2}+\cos^2k_y+i\sin\tfrac{k_x}{2}\cos k_y\,[A,B]\Big]\notag\\
			&=4t_J^2\Big[\sin^2\tfrac{k_x}{2}+\cos^2k_y-2\sin\tfrac{k_x}{2}\cos k_y\,(\hat{{\bs J}}\times\hat{{\bs J}}')\!\cdot\!\boldsymbol\sigma\Big].
		\end{align}
		Hence
		\begin{subequations}
			\begin{align}
				H^{\rm eff}_{\boldsymbol k}&=a_{\boldsymbol k}+b_{\boldsymbol k}\,(\hat{{\bs J}}\times\hat{{\bs J}}')\!\cdot\!\boldsymbol\sigma+\lambda(\sin k_x\,\sigma_y-\sin k_y\,\sigma_x),\\
				a_{\boldsymbol k}&=2t\cos k_y+2t\cos\tfrac{k_x}{2}+\frac{t_J^2(\sin^2\frac{k_x}{2}+\cos^2k_y)}{t\cos(k_x/2)},\\
				b_{\boldsymbol k}&=-\frac{2t_J^2}{t}\,\frac{\sin(k_x/2)\cos k_y}{\cos(k_x/2)}.
			\end{align}
		\end{subequations}

		For $|k|\ll1$ we use $\sin(k_x/2)/\cos(k_x/2)\simeq k_x/2$, $\cos k_y\simeq1$ and $\sin k\simeq k$, so that $b_{\boldsymbol k}\simeq-(t_J^2/t)\,k_x$ and
		\begin{equation}
			H^{\rm eff}_{\boldsymbol k}\simeq a_{\boldsymbol k}-\frac{t_J^2}{t}\,k_x\,\boldsymbol N\!\cdot\!\boldsymbol\sigma+\lambda(k_x\sigma_y-k_y\sigma_x),\,\boldsymbol N=\hat{\bs J}\times\hat{\bs J}'.
		\end{equation}
		The $p$-wave exchange term of Eq.~\eqref{eq:hamiltonian} is $\epsilon\mathcal M_1(\boldsymbol k)\,\boldsymbol N\!\cdot\!\boldsymbol\sigma=M\epsilon\,(k_x/k_{\rm F})\,\boldsymbol N\!\cdot\!\boldsymbol\sigma$ in physical units. Matching the coefficients of $k_x$ gives $M\epsilon/k_{\rm F}=-t_J^2/t$, where the sign can be absorbed into $\boldsymbol N\to-\boldsymbol N$. We take $\varphi'-\varphi=\pi/2$, so $|\hat{\bs J}\times\hat{\bs J}'|=1$ and $\boldsymbol N\parallel\hat j$; the in-plane orientations follow by rotating $\hat{\bs J}$ and $\hat{\bs J}'$ in spin space at fixed Rashba term. Expanding $a_{\boldsymbol k}$ gives $4t+t_J^2/t+(-t/4+3t_J^2/8t)k_x^2+(-t-t_J^2/t)k_y^2$, which is anisotropic, whereas Eq.~\eqref{eq:hamiltonian} assumes an isotropic dispersion. The projection therefore motivates the $p$-wave exchange structure and a leading uniform Rashba term; the isotropic continuum model is an additional simplification.
		
		Under time reversal $\hat{{\bs J}}\to-\hat{{\bs J}}$ and $\hat{{\bs J}}'\to-\hat{{\bs J}}'$, so $\boldsymbol N=\hat{{\bs J}}\times\hat{{\bs J}}'$ is even. This agrees with the ordinary spinful time-reversal symmetry $\Theta=i\sigma_yK$ of $H^{\rm eff}_{\boldsymbol k}$, since $b_{\boldsymbol k}$ and the Rashba term are both odd in $\boldsymbol k$: $\Theta H^{\rm eff}_{\boldsymbol k}\Theta^{-1}=H^{\rm eff}_{-\boldsymbol k}$. The time-reversal-odd quantities are $\hat{{\bs J}}$ and $\hat{{\bs J}}'$ separately, and the antiunitary symmetry of the magnet is time reversal combined with a spin rotation and a translation~\cite{p21c-rtv7}. Thus $\boldsymbol N$ is the normal of the plane of the coplanar moments. A common rotation of $\hat{{\bs J}}$ and $\hat{{\bs J}}'$ rotates $\boldsymbol N$. Without Rashba coupling this rotation is a symmetry of $H$; with Rashba coupling the orientation of $\boldsymbol N$ matters, which is the dependence studied in the main text.
		
		With the drive, $\boldsymbol k\to\boldsymbol k+\boldsymbol{\mathcal A}(t)$ in Eq.~\eqref{eq:fb_H}. The elimination above is static and holds if $\hbar\omega\ll|\Delta E_{\boldsymbol k}|=4|t\cos(k_x/2)|$ over the momenta visited during the cycle, and if the driven momenta stay away from $k_x=\pm\pi$, where $\Delta E_{\boldsymbol k}\to0$. We have not tested the driven four-band model; this is an assumption of the reduced calculation. We also assume that the Rashba term is the same on both sublattices; a staggered Rashba term would couple the blocks at first order. The derivation covers $n=1$ only. The $f$- and $h$-wave textures remain phenomenological.
		
		\section{Fermi-sea calculation}
		\label{app:num}
		
		We calculate the Floquet states of the driven $2\times2$ Hamiltonian independently at each momentum, without using the high-frequency expansion. The momentum sum in Eq.~\eqref{eq:heff} is restricted to the shell in which at least one undriven band lies within $12T$ of the chemical potential $\mu$, where $T$ denotes temperature in energy units. Outside this shell, both static bands are either occupied or empty, and their contributions cancel for the occupation prescription used here. The shell is determined at $\lambda=0$; increasing its width from $12T$ to $16T$ does not change the reported results for $\lambda\le4$~meV. For $\lambda=16$~meV the Rashba shift of the Fermi surfaces exceeds $12T$, and the shell must be widened to $24T$ or more; this raises the $f$-wave torque.
		
		We use a polar momentum grid with $N_\theta$ angular points and radial spacing $T/3$. For $\boldsymbol{N}\parallel\hat{z}$, we take $N_\theta=144$ for $n\leq3$, $N_\theta=288$ for $n=4,6$, and $N_\theta=300$ for the $h$-wave ($n=5$). The choice $N_\theta=300$ is a multiple of five and resolves the angular structure of the $h$-wave texture. For in-plane $\boldsymbol{N}$, we use $N_\theta=288$. The angular grid must resolve the narrow regions around the nodes of $\cos n\theta$, whose characteristic width scales as $\lambda/(M\epsilon n)$. The continuum model is evaluated for $k\leq1.5k_{\rm F}$. For $n=6$, the lower band turns over near $2.4k_{\rm F}$, so the additional Fermi surface lies outside the momentum range retained in the model and is excluded by this cutoff. For $n\leq5$, this cutoff does not change the results reported here.
		
		The one-period evolution operator is calculated using a fourth-order Magnus propagator with two Gauss points per time step. We use $\max(120,40b)$ steps per period. Floquet states are tracked by maximum overlap as the two drive amplitudes are increased in eight equal steps, each lasting one period, for $\boldsymbol N\parallel\hat z$. For in-plane $\boldsymbol N$ the effective vector stays in a plane and nearly crosses zero at some momenta, so eight steps assign the wrong branch there. We therefore use 128 amplitude steps for in-plane $\boldsymbol N$, with a coarse time step ($\max(24,8b)$ per period) to track the branch and one full-accuracy period at full amplitude. With eight steps the $i$-wave torque at $b=2$ was too large by a factor of 1.9 and the $h$-wave torque at $b=2$ too small by 21\%. This procedure defines the state-continuation and occupation prescription used in the calculation; it should not be interpreted as a demonstration of adiabatic following for a physical pulse. The occupations $n_s(\boldsymbol{k})$ and the normalization $N_{\rm el}$ are assigned from the undriven bands for the chosen orientation of $\boldsymbol{N}$, as described in Sec.~\ref{sec:model}. No relaxation, heating, or redistribution among Floquet states is included.

		We performed the following numerical convergence checks. Doubling $N_\theta$ changes the out-of-plane torques of the $d$-, $f$-, $g$-, and $h$-wave channels in Table~\ref{tab:kres} by at most $0.06\%$, and the $i$-wave result at $b=4$ by $1.7\%$. Doubling $N_s$, increasing the number of amplitude-ramp steps from 8 to 16, and widening the momentum shell change the nine tested channels by at most $0.01\%$. For in-plane $\boldsymbol{N}$, doubling $N_\theta$ changes the $f$-wave result at $b=1$ by $1.0\%$ and the $g$-wave result at $b=2$ by $3.1\%$. For the in-plane $h$- and $i$-wave cases at $b=1$, the values at $N_\theta=288$, 576, and 1152 are, respectively, $(1.02,1.02,0.99)\times10^{-2}u_0$ for the $h$-wave and $1.09\times10^{-2}u_0$ at $N_\theta=288$ for the $i$-wave. We therefore quote the large in-plane torques (above $2.5\times10^{-3}u_0$) to approximately $5\%$ and the other converged values to $\approx 3\%$. 
		
		For $\boldsymbol N\parallel\hat z$ the maximum over six phases equals the maximum over 24 phases to five digits in the six channels tested ($f$-wave $b=1$; $h$-wave $b=1,2,3$; $g$-wave $b=2$; $i$-wave $b=4$). For in-plane $\boldsymbol N$ it does not: the six-phase maximum is 1\% to 14\% below the interpolated maximum over $\phi$ in 15 of 16 channels, and 49\% below for the $i$-wave at $b=2$ with $\boldsymbol N\parallel\hat x$. Table~\ref{tab:ip} therefore lists the maximum of a trigonometric interpolation of $|\tau_\perp|^2$ built from 24 phases, which exceeds the largest of the 24 sampled values by at most 7\%. Evaluating the torque directly at the interpolated maximum gives $1.171\times10^{-2}\,u_0$ for the $i$-wave at $b=1$ (interpolant $1.184\times10^{-2}$, largest sampled value $1.185\times10^{-2}$) and $1.028\times10^{-3}\,u_0$ for the $i$-wave at $b=2$ (interpolant $1.030\times10^{-3}$), both for $\boldsymbol N\parallel\hat x$. The interpolation is therefore accurate to about $1\%$, and for the $i$-wave at $b=1$ the quoted value is the largest sampled one. The in-plane channels $i$-wave $b=3,4$ ($\boldsymbol N\parallel\hat y$), $i$-wave $b=3$ ($\boldsymbol N\parallel\hat x$) and class~II$'$ were not resampled; the $i$-wave $b=4$ channel for $\boldsymbol N\parallel\hat x$ and the $h$-wave $b=3$ channel for $\boldsymbol N\parallel\hat x$ did not change.
		
		For nine allowed $\boldsymbol N\parallel\hat z$ channels from $10^{-10}$ to $3\times10^{-4}\,u_0$, doubling $N_\theta$ changes the torque by less than $2\times10^{-6}\,u_0$ (largest for the $i$-wave $b=4$ and $h$-wave $b=3$ channels, at $1.7\%$ and $0.5\%$) and by less than $5\times10^{-11}\,u_0$ for the channels below $10^{-7}\,u_0$ (0.003\% to 2\%). The channels forbidden by the selection rule are at most $10^{-16}\,u_0$; allowed channels below $10^{-15}\,u_0$ are at this noise floor and are shown as zero in the maps. At $\lambda=0.25$~meV, the in-plane $h$-wave torque at $b=1$ still changes by $40\%$ between $N_\theta=1200$ and 2400. At $\lambda=1$~meV, the corresponding change is $0.5\%$ between 300 and 600 angular points. We therefore do not use the under-resolved values at $\lambda=0.25$~meV in the quantitative comparisons. The reported calculations use $\lambda\geq0.5$~meV for $n\leq4$ and $\lambda\geq1$~meV for $n\geq5$.
		
		We compared Eq.~\eqref{eq:heff} with a numerical derivative of the sum of occupied quasienergies with respect to the Néel-vector orientation, at fixed occupations. For $\boldsymbol{N}\parallel\hat{z}$ in the $d$- and $g$-wave cases at $b=2$, and for $\boldsymbol{N}\parallel\hat{y}$ in the $g$-wave case at $b=2$, the two methods agree to better than $0.03\%$. For $\boldsymbol{N}\parallel\hat{x}$ in the $f$-wave case at $b=1$, the numerical derivatives for tilt angles $0.01$, $0.004$, and $0.0015$ are $4.44$, $4.54$, and $4.55$, respectively, in units of $10^{-3}u_0$, compared with $4.62\times10^{-3}u_0$ from Eq.~\eqref{eq:heff}. The difference at the smallest tilt is $1.3\%$. A tilt of $0.02$ is too large for this finite-difference check because it is comparable to the angular width of the nodal regions.
		
		At $\lambda=0$, the torque vanishes for all channels tested. We also checked the selection rule $\gcd(n,b\mp1)=1$ over 504 combinations. The Fourier components forbidden by the rule, extracted from data sampled at 24 phases, are at most $3\times10^{-10}$ of the allowed components, except for the aliased $h$-wave data, where the ratio is $3\times10^{-3}$.
		
		At $\hbar\omega=992$ and 1984~meV, the exact torque for $\boldsymbol{N}\parallel\hat{z}$ approaches the leading high-frequency result of Eq.~\eqref{eq:h0}: the relative differences are $0.06\%$ for the $f$-wave at $b=1$, $0.006\%$ for the $g$-wave at $b=2$, and $0.7\%$ for the $h$-wave at $b=1$.
		
		At fixed $M\epsilon$, halving $\epsilon$ doubles the torque in each of the 11 channels tested. This scaling follows from the normalization and parameter dependence of the model and is not asserted beyond the tested channels.

		We did not test amplitude-ramp convergence for drive amplitudes above $0.2$ (or above $0.3$ for a single beam), where Floquet-level crossings may compromise overlap-based state continuation. We also did not systematically vary $M\epsilon/\hbar\omega$ independently of the frequency scans in Fig.~\ref{f5}, or study the dependence on temperature and chemical potential. Only the principal orientations $\boldsymbol{N}\parallel\hat{x},\hat{y},\hat{z}$ were considered. The calculations neglect relaxation and heating and use occupations inherited from the undriven bands. These restrictions should be kept in mind when interpreting the numerical torque magnitudes.
		
		\section{Exact Floquet calculation}
		\label{app:floquet}
		
		To assess the leading high-frequency approximation for the momentum-independent field, we calculate the one-period propagator of the driven $2\times2$ Hamiltonian at $\boldsymbol{k}=0$ without expanding in $1/\omega$. We extract the transverse component of the stroboscopic Floquet Hamiltonian and average it over the starting time $t_0$. This tests the effective-field expansion at the zone center, but does not establish its accuracy at all occupied momenta.
		
		The propagator is constructed using midpoint time steps and the exact exponential of the Hamiltonian at each step. We use 6000--12000 steps per period; increasing this number to 48000 changes the reported values by less than three significant digits. For each starting time $t_0$, we write the traceless part of the one-period propagator as
		\begin{equation}
			U(t_0+\mathcal{T},t_0)
			=\exp\left[-i\theta\,\hat{\boldsymbol n}\cdot\boldsymbol\sigma\right],
			\qquad 0\leq\theta\leq\pi,
		\end{equation}
		where $\mathcal{T}=2\pi/\omega$. The spin-dependent part of the corresponding stroboscopic Floquet Hamiltonian is $\frac{\hbar\theta}{\mathcal{T}}
		\hat{\boldsymbol n}\cdot\boldsymbol\sigma.$
		We average this field over eight equally spaced starting times. The factor of $\hbar$ converts the frequency $\theta/\mathcal{T}$ into an energy. Since the Floquet Hamiltonian is defined modulo $\hbar\omega$, when $\max_t|\boldsymbol{h}(t)|/\hbar\omega\gtrsim1$, we use it only to assess the breakdown of the first-order high-frequency result.
		
		For small $M\epsilon=10$~meV, $\lambda=0.5$~meV, and $\hbar\omega=124$~meV, we consider counter-rotating light with $\boldsymbol{N}\parallel\hat{z}$ and $\mathcal{A}_1=\mathcal{A}_2=0.5$. Here, $\max_t|\boldsymbol{h}(t)|/\hbar\omega\leq0.08$ for all multipoles and harmonic numbers considered. The exact and first-order transverse fields agree within $1.5\%$ for all channels with a nonzero first-order field, except for the $g$-wave at $b=2$, where the difference is $2.7\%$. The maximum deviation, taken over the channels and five relative phases, increases with drive amplitude. For equal amplitudes, it is approximately $0.5\%$, $2.7\%$, $11\%$, and $81\%$ at $\mathcal{A}_1=\mathcal{A}_2=0.3$, 0.5, 0.7, and 1, respectively. The corresponding maximum values of $\max_t|\boldsymbol h(t)|/\hbar\omega$ are 0.05, 0.08, 0.6, and 5.2. Thus, the first-order expansion is reliable for the weak-drive parameters used in the main comparison, but not at large amplitudes.
		
		The exact calculation also reproduces the amplitude-ratio blindspots. For $\mathcal{A}_1=0.4$ and $u_0 = M \epsilon \lambda/\hbar \omega$, the transverse field vanishes at $\mathcal{A}_2/\mathcal{A}_1=1.414$ for the $p$-wave at $b=2$, 1.155 for the $g$-wave at $b=2$, 1.224 for the $h$-wave at $b=3$, and 1.264 for the $i$-wave at $b=4$. These values agree with the first-order predictions $\sqrt{2}$, $\sqrt{4/3}$, $\sqrt{3/2}$, and $\sqrt{8/5}$, respectively. Channels with a vanishing first-order transverse field remain below $10^{-3}u_0$ in the exact calculation at $\mathcal{A}_1=\mathcal{A}_2=0.5$ and $\phi=0.7$, for both helicity configurations. By contrast, the allowed counter-rotating channels considered here exceed $0.018u_0$. Thus, the selection rules hold as leading-order rules at the zone center, with only small higher-order leakage for these parameters.
		
		For large $M\epsilon=0.5$~eV, with the same $\lambda$ and $\hbar\omega$ and $\mathcal{A}_1=\mathcal{A}_2=1$, $\max_t|\boldsymbol{h}(t)|/\hbar\omega$ reaches 8 for the $p$-wave and 129 for the $h$-wave. In this regime, the first-order field averaged over $b=1,\ldots,6$ exceeds the exact result by factors of approximately 50 and 200 for the $p$- and $h$-wave cases, respectively. The blindspots of the virtual-pair channels ($g$-wave at $b=2$, $h$-wave at $b=3$, and $i$-wave at $b=4$) remain within approximately $2\%$ of the first-order values $\sqrt{c_1/c_2}$ for $\mathcal{A}_1=0.1$--0.3. In contrast, the $p$-wave zero $\mathcal{A}_2=\sqrt{b}\,\mathcal{A}_1$ shifts by 2--4\% at $\mathcal{A}_1=0.1$ and 10--18\% at 0.2; at 0.3, the minimum of $|\boldsymbol h_\perp|$ is no longer close to zero. These results show that the leading-order selection rules and virtual-pair blindspots remain robust at $\boldsymbol{k}=0$, whereas quantitative field amplitudes require the exact calculation when the drive is strong compared with the relevant dressed energy scales.
		
		\section{Derivation of the momentum-independent effective fields}
		\label{app:derivations}
		
		Setting $\boldsymbol{k}=0$ in the Peierls-substituted Hamiltonian isolates the momentum-independent part of the driven exchange texture. For $\boldsymbol{N}\parallel\hat{{\bs J}}$, the Hamiltonian can be written as
		\begin{subequations}
			\label{eq:A1}
			\begin{align}
				H(t)&=\epsilon\boldsymbol{\mathcal A}^{2}\sigma_0+S(t)+f(t)\sigma_j,\\
				S(t)&=\lambda\left[\mathcal A_x(t)\sigma_y-\mathcal A_y(t)\sigma_x\right],
				\qquad
				f(t)=\epsilon\mathcal M_n\big(\boldsymbol{\mathcal A}(t)\big).
			\end{align}
		\end{subequations}
		The first term is proportional to the identity and commutes with the remaining terms, so it does not contribute to the effective spin field. All multipole textures in Eq.~\eqref{eq:multipoles} have the form $M\operatorname{Re}(k_x+ik_y)^n$. Thus,
		\begin{align}
			f(t)=\epsilon M\operatorname{Re}[z(t)^n],
		\end{align}where $z(t)=\mathcal A_x(t)+i\mathcal A_y(t)
		=\mathcal A_1e^{i\eta_1\omega t}
		+\mathcal A_2e^{i\eta_2(b\omega t+\phi)}.$
		
		We use the Fourier convention
		\begin{equation}
			H(t)=\sum_{m=-\infty}^{\infty}H_m e^{-im\omega t},
			\,
			\mathcal A_\alpha(t)=\sum_{m=-\infty}^{\infty}
			\mathcal A_{\alpha,m}e^{-im\omega t}.
		\end{equation}
		The nonzero Fourier coefficients of the vector potential in Eq.~\eqref{eq:vector_potential} are
		\begin{subequations}
			\label{eq:A2}
			\begin{align}
				\mathcal A_{x,\pm1}&=\frac{\mathcal A_1}{2},
				&\mathcal A_{y,\pm1}&=\mp\frac{\eta_1\mathcal A_1}{2i},\\
				\mathcal A_{x,\pm b}&=\frac{\mathcal A_2e^{\mp i\phi}}{2},
				&\mathcal A_{y,\pm b}&=\mp\frac{\eta_2\mathcal A_2e^{\mp i\phi}}{2i}.
			\end{align}
		\end{subequations}
		For $b=1$, the coefficients at $m=\pm1$ are the sums of the contributions from the two optical components. Defining
		$S_m=\lambda(\mathcal A_{x,m}\sigma_y-\mathcal A_{y,m}\sigma_x)$
		and denoting the Fourier coefficients of $f(t)$ by $f_m$, the first-order van Vleck high-frequency Hamiltonian is
		\begin{equation}
			\label{eq:A3}
			H_{\rm eff}=H_0+\sum_{m\geq1}
			\frac{[H_{-m},H_{m}]}{m\hbar\omega},
		\end{equation}
		where $H_m=S_m+f_m\sigma_j$ and $H_0=f_0\sigma_j$, after dropping the scalar terms. Since $[\sigma_j,\sigma_j]=0$,
		\begin{equation}
			[H_{-m},H_m]
			=[S_{-m},S_{m}]
			-f_{-m}[S_m,\sigma_j]
			+f_m[S_{-m},\sigma_j].
		\end{equation}
		The resulting effective field contains a cycle-averaged exchange term, an IFE term, and terms transverse to the N\'eel vector.
		
		The zeroth Fourier component $H_0=f_0\sigma_j$ is the time-averaged exchange term. It is parallel to $\boldsymbol N$, contains no factor of $\lambda$, and is not suppressed by $1/\omega$.
		
		Using $[\sigma_y,\sigma_x]=-2i\sigma_z$, we obtain
		\begin{equation}
			[S_{-m},S_{m}]
			=-2i\lambda^2\left(
			\mathcal A_{x,m}\mathcal A_{y,-m}
			-\mathcal A_{y,m}\mathcal A_{x,-m}
			\right)\sigma_z.
		\end{equation}
		For $b\neq1$, the contributions from $m=1$ and $m=b$ are, respectively, $-\lambda^2\eta_1\mathcal A_1^2\sigma_z$ and $-\lambda^2\eta_2\mathcal A_2^2\sigma_z$. After division by $m\hbar\omega$ and summation, they give $h_{\rm IFE}\sigma_z$, where $h_{\rm IFE}=-\lambda^2\Omega_b$ and $\Omega_b$ is defined in Eq.~\eqref{eq:Omega_b_def}. For $b=1$, the two optical components contribute to the same Fourier harmonic, and their interference produces the cross term proportional to $\mathcal A_1\mathcal A_2(\eta_1+\eta_2)\cos\phi$ in $\Omega_b$. This contribution is independent of the multipole order.
		
		For $\boldsymbol N\parallel\hat z$, using $[\sigma_y,\sigma_z]=2i\sigma_x$ and $[\sigma_x,\sigma_z]=-2i\sigma_y$, the in-plane components ($\alpha=x,y$) are
		\begin{equation}
			\label{eq:A4}
			h_{{\rm eff},\alpha}^{(N_z)}
			=\sum_{m\in\{1,b\}}
			\frac{2i\lambda}{m\hbar\omega}
			\left(f_{-m}\mathcal A_{\alpha,m}
			-f_m\mathcal A_{\alpha,-m}\right).
		\end{equation}
		When $b=1$, the set $\{1,b\}$ contains one harmonic, whose Fourier coefficient includes both optical components. For $\boldsymbol N\parallel\hat x$ or $\hat y$, the commutators
		$[S_m,\sigma_x]=-2i\lambda\mathcal A_{x,m}\sigma_z$
		and
		$[S_m,\sigma_y]=-2i\lambda\mathcal A_{y,m}\sigma_z$
		produce only a $z$ component. Consequently,
		\begin{equation}
			\label{eq:A5}
			h_{{\rm eff},z}^{(N_x)}-h_{\rm IFE}
			=-h_{{\rm eff},x}^{(N_z)},
			\qquad
			h_{{\rm eff},z}^{(N_y)}-h_{\rm IFE}
			=-h_{{\rm eff},y}^{(N_z)}.
		\end{equation}
		The remaining in-plane component vanishes for these principal orientations: $h_{{\rm eff},y}^{(N_x)}=0$ and $h_{{\rm eff},x}^{(N_y)}=0$. These relations generate the in-plane-N\'eel-vector entries of Table~\ref{tab:effective_fields} from the $N_z$ entries.
		
		For a unit vector $\boldsymbol N$, the exchange term is $f(t)\boldsymbol N\cdot\boldsymbol\sigma$. The commutators involving the exchange term are linear in $\boldsymbol N$, whereas the IFE contribution is independent of $\boldsymbol N$. The resulting field can therefore be written as
		\begin{equation}
			\boldsymbol h_{\rm eff}
			=h_{\rm IFE}\hat{\boldsymbol z}
			+h_\parallel\boldsymbol N
			+\boldsymbol w\times\boldsymbol N,
			\,
			\boldsymbol w=
			\left(-h_{{\rm eff},y}^{(N_z)},
			h_{{\rm eff},x}^{(N_z)},0\right),
		\end{equation}
		where $h_\parallel$ is the cycle-averaged longitudinal contribution. This expression reproduces the three principal orientations considered in the main text.
		
		Expanding $z(t)^n$ gives the Fourier component $f_m$:
		{\small\begin{align}
				\label{eq:A6}
				f_m={}&\frac{\epsilon M}{2}\Bigg[
				\sum_{q:\,(n-q)\eta_1+qb\eta_2=m}
				\binom nq\mathcal A_1^{n-q}\mathcal A_2^q
				e^{iq\eta_2\phi}\notag\\
				&\hspace{1.2cm}+
				\sum_{q:\,(n-q)\eta_1+qb\eta_2=-m}
				\binom nq\mathcal A_1^{n-q}\mathcal A_2^q
				e^{-iq\eta_2\phi}
				\Bigg].
		\end{align}}Only $m=1$ and $m=b$ enter Eq.~\eqref{eq:A4}, while $m=0$ gives the cycle-averaged longitudinal term. The conditions on the sums therefore give the photon-counting selection rules discussed in Sec.~\ref{sec:photon_counting}.
		
		For $n=1$, $f_m=\epsilon M\mathcal A_{x,m}$. Substitution into Eq.~\eqref{eq:A4} gives $h_{{\rm eff},x}^{(N_z)}=0$ and
		\begin{equation}
			h_{{\rm eff},y}^{(N_z)}
			=-\frac{\epsilon M}{\lambda}h_{\rm IFE}
			=M\epsilon\lambda\Omega_b,
		\end{equation}
		for every $b$. Equation~\eqref{eq:A5} then gives
		$h_{{\rm eff},z}^{(N_y)}=h_{\rm IFE}-M\epsilon\lambda\Omega_b$
		and
		$h_{{\rm eff},z}^{(N_x)}=h_{\rm IFE}$.
		
		For counter-rotating beams, the cycle average can be nonzero when $q=n/(b+1)$ is an integer. In that case, $b+1$ divides $n$ and
		\begin{equation}
			f_0=\epsilon M\binom nq
			\mathcal A_1^{n-q}\mathcal A_2^q\cos(q\phi).
		\end{equation}
		For co-rotating beams, no such zero-frequency combination exists for positive $n$ and $b$, so $f_0=0$.
		
		For counter-rotating beams, the component at $m=1$ receives a contribution from $q=1$, with amplitude proportional to $n\mathcal A_1^{n-1}\mathcal A_2$. In Eq.~\eqref{eq:A4}, this produces the terms proportional to $n$. The component at $m=b=n$ receives a contribution from $q=0$ for either helicity configuration; its amplitude is proportional to $\mathcal A_1^n$, and the factor $1/m$ in Eq.~\eqref{eq:A4} produces the terms proportional to $1/n$. Combining these contributions gives Eq.~\eqref{eq:resonant_general}. No other value of $q$ contributes to these harmonics for $n\geq2$.
		
		For counter-rotating beams, the harmonic $m=1$ receives the $q=1$ contribution, proportional to $n\mathcal A_1^{n-1}\mathcal A_2$, while the harmonic $m=b$ receives the $q=2$ contribution, proportional to $\binom n2\mathcal A_1^{n-2}\mathcal A_2^2$. Including the $1/b$ factor in Eq.~\eqref{eq:A4}, the resulting transverse field for $\boldsymbol N\parallel\hat z$ is{\small\begin{subequations}
				\begin{align}
					h_{{\rm eff},x}^{(N_z)}=&-
					\frac{\lambda\epsilon M}{\hbar\omega}
					\frac{1-\eta_1\eta_2}{2}
					\mathcal A_1^{n-2}\mathcal A_2
					\left[n\mathcal A_1^2
					-\frac{n(n-1)}{2(n-2)}\mathcal A_2^2\right]\sin\phi,
					\\
					h_{{\rm eff},y}^{(N_z)}=&
					\frac{\lambda\epsilon M}{\hbar\omega}
					\frac{\eta_1-\eta_2}{2}
					\mathcal A_1^{n-2}\mathcal A_2
					\left[n\mathcal A_1^2
					-\frac{n(n-1)}{2(n-2)}\mathcal A_2^2\right]\cos\phi.
				\end{align}
		\end{subequations}}
		The bracketed amplitude factor is proportional to
		$c_1\mathcal A_1^2-c_2\mathcal A_2^2$, with
		$c_1:c_2=2(n-2):(n-1)$. It therefore vanishes at the amplitude ratio given in the main text.
		
		For $m\in\{1,b\}$, none of the integers $q=0,\ldots,n$ satisfies the frequency-matching conditions in Eq.~\eqref{eq:A6}. Thus $f_{\pm1}=f_{\pm b}=0$, and the only remaining contribution to the momentum-independent field is the IFE term. This statement applies to the $\boldsymbol{k}$-independent field; it is not a selection rule for the full Fermi-sea response, which also contains the $\boldsymbol{k}$-dependent exchange terms.
		
		We evaluated Eq.~\eqref{eq:A3} symbolically for each multipole order, principal N\'eel-vector orientation, and harmonic number to obtain Table~\ref{tab:effective_fields}. We also compared the analytical fields with a direct evaluation of Eq.~\eqref{eq:A3} using the Fourier components of the time-dependent Hamiltonian. For $n=1,\ldots,6$, $b=1,\ldots,7$, all three principal orientations, all four helicity combinations, and randomly selected amplitudes and phases, the largest discrepancy was approximately $10^{-14}$, consistent with machine precision.
		
		\section{Rashba ferromagnet ($n=0$)}
		\label{app:fm}
		To test what the multipole structure contributes, we repeat the occupied-state calculation of Sec.~\ref{sec:both} for $n=0$, i.e.\ a uniform exchange $M\epsilon\,\boldsymbol N\!\cdot\!\boldsymbol\sigma$ with Rashba coupling (a Rashba ferromagnet with magnetization along $\boldsymbol N$). We use the same parameters, grids, and amplitude ramp as in Appendix~\ref{app:num}, $b=1$ to 7, both helicities, and six phases.
		
		In the case of out-of-plane $\boldsymbol N\parallel\hat z$, for $n=0$ the rule gives $\gcd(0,b+1)=b+1>1$ for counter-rotating beams, so all seven counter-rotating channels are forbidden, and $\gcd(0,b-1)=1$ only for $b=2$ with co-rotating beams. Of the 14 channels (seven harmonics, two helicity configurations), the numerics give zero (below $10^{-15}\,u_0$) for 13. The co-rotating $b=2$ channel is allowed and is finite: $1.8\times10^{-7}\,u_0$ at $\mathcal A_1=\mathcal A_2=0.15$, $\lambda=1$~meV, independent of $N_\theta$ (144 to 576) and of the momentum shell. Its magnitude does not depend on the phase, but its direction rotates with it: $h_+=h_x+ih_y\propto-i\,e^{-i\phi}$, as required by continuous rotational covariance for co-rotating beams, so the torque points along $+\hat x$ at $\phi=0$ and rotates clockwise as $\phi$ increases (checked at six phases). It scales as $\lambda^3$, as $\mathcal A^3$, and approximately as $(\hbar\omega)^{-2}$ (it falls by a factor of 4.6 from 124 to 248~meV). This is four to five orders of magnitude below the largest multipole torques ($4.8\times10^{-3}\,u_0$ out of plane, $1.2\times10^{-2}\,u_0$ in plane).\begin{table}[t!]
			\vspace*{-0.6em}\centering
			\caption{Maximum over phases of $|\tau_\perp|$ ($u_0$) for $n=0$, $\boldsymbol N\parallel\hat x$ (same for $\hat y$). For $\boldsymbol N\parallel\hat z$ only the co-rotating $b=2$ channel is finite, $1.8\times10^{-7}$.}
			\label{tab:fm}
			\renewcommand{\arraystretch}{1.35}
			\resizebox{\columnwidth}{!}{%
				\begin{tabular}{c|c|c|c|c|c|c|c}
					\hline\hline
					$b$ & 1 & 2 & 3 & 4 & 5 & 6 & 7 \\\hline
					counter-rot. ($10^{-5}$) & 3.2 & 5.9 & 7.8 & 8.7 & 9.3 & 9.7 & 9.9 \\\hline
					co-rot. ($10^{-5}$) & 46 & 17 & 15 & 14 & 14 & 13 & 13 \\
					\hline\hline
				\end{tabular}
			}
		\end{table}
		
		For in-plane $\boldsymbol N\parallel\hat x$ or $\hat y$, finite torques appear, but they are small and unstructured: the maximum over phases is $3\times10^{-5}$ to $1\times10^{-4}\,u_0$ for counter-rotating beams and $1.3\times10^{-4}$ to $4.6\times10^{-4}\,u_0$ for co-rotating beams (Tab.~\ref{tab:fm}). They vary slowly with $b$ and have no selection rule. Checks at $b=1$ (co-rotating): the torque scales as $\lambda^2$ and falls as $1/\hbar\omega$ (halved from 124 to 248~meV). It is therefore a class-II type response from the time-dependent part of the drive, with no cycle-averaged ($\omega$-independent) channel. The largest value, $4.6\times10^{-4}\,u_0$, is 26 times smaller than the largest in-plane multipole torque ($1.2\times10^{-2}\,u_0$) and similar to the weakest class~I$'$ channels. The results for $\hat x$ and $\hat y$ are identical, as expected for a uniform exchange.
		
		The $\bs k$-dependent exchange of the multipoles is therefore what produces the large, selection-rule-controlled torques of Secs.~\ref{sec:kdep} and~\ref{sec:both}; for a Rashba ferromagnet the response is small (one allowed out-of-plane channel at $1.8\times10^{-7}\,u_0$, a weak in-plane response) and has no selection rule. We treat $n=0$ as a ferromagnet; an antiferromagnetic reading would need an explicit reduced model and order-parameter coupling, which we do not construct.
	}
	\bibliography{bib}

\begin{thebibliography}{80}%
\makeatletter
\providecommand \@ifxundefined [1]{%
 \@ifx{#1\undefined}
}%
\providecommand \@ifnum [1]{%
 \ifnum #1\expandafter \@firstoftwo
 \else \expandafter \@secondoftwo
 \fi
}%
\providecommand \@ifx [1]{%
 \ifx #1\expandafter \@firstoftwo
 \else \expandafter \@secondoftwo
 \fi
}%
\providecommand \natexlab [1]{#1}%
\providecommand \enquote  [1]{``#1''}%
\providecommand \bibnamefont  [1]{#1}%
\providecommand \bibfnamefont [1]{#1}%
\providecommand \citenamefont [1]{#1}%
\providecommand \href@noop [0]{\@secondoftwo}%
\providecommand \href [0]{\begingroup \@sanitize@url \@href}%
\providecommand \@href[1]{\@@startlink{#1}\@@href}%
\providecommand \@@href[1]{\endgroup#1\@@endlink}%
\providecommand \@sanitize@url [0]{\catcode `\\12\catcode `\$12\catcode
  `\&12\catcode `\#12\catcode `\^12\catcode `\_12\catcode `\%12\relax}%
\providecommand \@@startlink[1]{}%
\providecommand \@@endlink[0]{}%
\providecommand \url  [0]{\begingroup\@sanitize@url \@url }%
\providecommand \@url [1]{\endgroup\@href {#1}{\urlprefix }}%
\providecommand \urlprefix  [0]{URL }%
\providecommand \Eprint [0]{\href }%
\providecommand \doibase [0]{https://doi.org/}%
\providecommand \selectlanguage [0]{\@gobble}%
\providecommand \bibinfo  [0]{\@secondoftwo}%
\providecommand \bibfield  [0]{\@secondoftwo}%
\providecommand \translation [1]{[#1]}%
\providecommand \BibitemOpen [0]{}%
\providecommand \bibitemStop [0]{}%
\providecommand \bibitemNoStop [0]{.\EOS\space}%
\providecommand \EOS [0]{\spacefactor3000\relax}%
\providecommand \BibitemShut  [1]{\csname bibitem#1\endcsname}%
\let\auto@bib@innerbib\@empty
\bibitem [{\citenamefont {{\v{S}}mejkal}\ \emph
  {et~al.}(2022{\natexlab{a}})\citenamefont {{\v{S}}mejkal}, \citenamefont
  {Sinova},\ and\ \citenamefont {Jungwirth}}]{smejkal2022beyond}%
  \BibitemOpen
  \bibfield  {author} {\bibinfo {author} {\bibfnamefont {L.}~\bibnamefont
  {{\v{S}}mejkal}}, \bibinfo {author} {\bibfnamefont {J.}~\bibnamefont
  {Sinova}},\ and\ \bibinfo {author} {\bibfnamefont {T.}~\bibnamefont
  {Jungwirth}},\ }\bibfield  {title} {\bibinfo {title} {{Beyond Conventional
  Ferromagnetism and Antiferromagnetism: A Phase with Nonrelativistic Spin and
  Crystal Rotation Symmetry}},\ }\href
  {https://doi.org/10.1103/PhysRevX.12.031042} {\bibfield  {journal} {\bibinfo
  {journal} {Physical Review X}\ }\textbf {\bibinfo {volume} {12}},\ \bibinfo
  {pages} {031042} (\bibinfo {year} {2022}{\natexlab{a}})}\BibitemShut
  {NoStop}%
\bibitem [{\citenamefont {{\v{S}}mejkal}\ \emph
  {et~al.}(2022{\natexlab{b}})\citenamefont {{\v{S}}mejkal}, \citenamefont
  {Sinova},\ and\ \citenamefont {Jungwirth}}]{smejkal2022emerging}%
  \BibitemOpen
  \bibfield  {author} {\bibinfo {author} {\bibfnamefont {L.}~\bibnamefont
  {{\v{S}}mejkal}}, \bibinfo {author} {\bibfnamefont {J.}~\bibnamefont
  {Sinova}},\ and\ \bibinfo {author} {\bibfnamefont {T.}~\bibnamefont
  {Jungwirth}},\ }\bibfield  {title} {\bibinfo {title} {{Emerging Research
  Landscape of Altermagnetism}},\ }\href
  {https://doi.org/10.1103/PhysRevX.12.040501} {\bibfield  {journal} {\bibinfo
  {journal} {Physical Review X}\ }\textbf {\bibinfo {volume} {12}},\ \bibinfo
  {pages} {040501} (\bibinfo {year} {2022}{\natexlab{b}})}\BibitemShut
  {NoStop}%
\bibitem [{\citenamefont {Šmejkal}\ \emph {et~al.}(2020)\citenamefont
  {Šmejkal}, \citenamefont {González-Hernández}, \citenamefont {Jungwirth},\
  and\ \citenamefont {Sinova}}]{doi:10.1126/sciadv.aaz8809}%
  \BibitemOpen
  \bibfield  {author} {\bibinfo {author} {\bibfnamefont {L.}~\bibnamefont
  {Šmejkal}}, \bibinfo {author} {\bibfnamefont {R.}~\bibnamefont
  {González-Hernández}}, \bibinfo {author} {\bibfnamefont {T.}~\bibnamefont
  {Jungwirth}},\ and\ \bibinfo {author} {\bibfnamefont {J.}~\bibnamefont
  {Sinova}},\ }\bibfield  {title} {\bibinfo {title} {Crystal time-reversal
  symmetry breaking and spontaneous hall effect in collinear
  antiferromagnets},\ }\href {https://doi.org/10.1126/sciadv.aaz8809}
  {\bibfield  {journal} {\bibinfo  {journal} {Science Advances}\ }\textbf
  {\bibinfo {volume} {6}},\ \bibinfo {pages} {eaaz8809} (\bibinfo {year}
  {2020})}\BibitemShut {NoStop}%
\bibitem [{\citenamefont {Shen}\ \emph {et~al.}(2026)\citenamefont {Shen},
  \citenamefont {Hembeck}, \citenamefont {Campos}, \citenamefont {Dong},
  \citenamefont {Li}, \citenamefont {Li}, \citenamefont {Wang}, \citenamefont
  {Priessnitz}, \citenamefont {Gao}, \citenamefont {Wei}, \citenamefont
  {Jiang}, \citenamefont {Zhang}, \citenamefont {Wang}, \citenamefont {Zhang},
  \citenamefont {Shi}, \citenamefont {Liu}, \citenamefont {Šmejkal},
  \citenamefont {Chen},\ and\ \citenamefont
  {Liu}}]{shen2026roomtemperatureferroelectricpwavemagnet}%
  \BibitemOpen
  \bibfield  {author} {\bibinfo {author} {\bibfnamefont {K.}~\bibnamefont
  {Shen}}, \bibinfo {author} {\bibfnamefont {N.}~\bibnamefont {Hembeck}},
  \bibinfo {author} {\bibfnamefont {W.~H.}\ \bibnamefont {Campos}}, \bibinfo
  {author} {\bibfnamefont {J.}~\bibnamefont {Dong}}, \bibinfo {author}
  {\bibfnamefont {B.}~\bibnamefont {Li}}, \bibinfo {author} {\bibfnamefont
  {Y.}~\bibnamefont {Li}}, \bibinfo {author} {\bibfnamefont {G.}~\bibnamefont
  {Wang}}, \bibinfo {author} {\bibfnamefont {J.}~\bibnamefont {Priessnitz}},
  \bibinfo {author} {\bibfnamefont {Y.}~\bibnamefont {Gao}}, \bibinfo {author}
  {\bibfnamefont {Y.}~\bibnamefont {Wei}}, \bibinfo {author} {\bibfnamefont
  {H.}~\bibnamefont {Jiang}}, \bibinfo {author} {\bibfnamefont
  {X.}~\bibnamefont {Zhang}}, \bibinfo {author} {\bibfnamefont
  {M.}~\bibnamefont {Wang}}, \bibinfo {author} {\bibfnamefont {J.}~\bibnamefont
  {Zhang}}, \bibinfo {author} {\bibfnamefont {W.}~\bibnamefont {Shi}}, \bibinfo
  {author} {\bibfnamefont {Q.}~\bibnamefont {Liu}}, \bibinfo {author}
  {\bibfnamefont {L.}~\bibnamefont {Šmejkal}}, \bibinfo {author}
  {\bibfnamefont {X.}~\bibnamefont {Chen}},\ and\ \bibinfo {author}
  {\bibfnamefont {M.}~\bibnamefont {Liu}},\ }\bibfield  {title} {\bibinfo
  {title} {Room-temperature ferroelectric $p$-wave magnet},\ }\href
  {https://arxiv.org/abs/2610.12892} {\bibfield  {journal} {\bibinfo  {journal}
  {arXiv preprint arXiv:2610.12892}\ } (\bibinfo {year} {2026})}\BibitemShut
  {NoStop}%
\bibitem [{\citenamefont {Chakraborty}\ \emph {et~al.}(2025)\citenamefont
  {Chakraborty}, \citenamefont {Birk~Hellenes}, \citenamefont
  {Jaeschke-Ubiergo}, \citenamefont {Jungwirth}, \citenamefont
  {{\v{S}}mejkal},\ and\ \citenamefont {Sinova}}]{Chakraborty2025}%
  \BibitemOpen
  \bibfield  {author} {\bibinfo {author} {\bibfnamefont {A.}~\bibnamefont
  {Chakraborty}}, \bibinfo {author} {\bibfnamefont {A.}~\bibnamefont
  {Birk~Hellenes}}, \bibinfo {author} {\bibfnamefont {R.}~\bibnamefont
  {Jaeschke-Ubiergo}}, \bibinfo {author} {\bibfnamefont {T.}~\bibnamefont
  {Jungwirth}}, \bibinfo {author} {\bibfnamefont {L.}~\bibnamefont
  {{\v{S}}mejkal}},\ and\ \bibinfo {author} {\bibfnamefont {J.}~\bibnamefont
  {Sinova}},\ }\bibfield  {title} {\bibinfo {title} {Highly efficient
  non-relativistic edelstein effect in nodal $p$-wave magnets},\ }\href
  {https://doi.org/10.1038/s41467-025-62516-0} {\bibfield  {journal} {\bibinfo
  {journal} {Nature Communications}\ }\textbf {\bibinfo {volume} {16}},\
  \bibinfo {pages} {7270} (\bibinfo {year} {2025})}\BibitemShut {NoStop}%
\bibitem [{\citenamefont {Hellenes}\ \emph {et~al.}(2024)\citenamefont
  {Hellenes}, \citenamefont {Jungwirth}, \citenamefont {Jaeschke-Ubiergo},
  \citenamefont {Chakraborty}, \citenamefont {Sinova},\ and\ \citenamefont
  {Šmejkal}}]{hellenes2024pwavemagnets}%
  \BibitemOpen
  \bibfield  {author} {\bibinfo {author} {\bibfnamefont {A.~B.}\ \bibnamefont
  {Hellenes}}, \bibinfo {author} {\bibfnamefont {T.}~\bibnamefont {Jungwirth}},
  \bibinfo {author} {\bibfnamefont {R.}~\bibnamefont {Jaeschke-Ubiergo}},
  \bibinfo {author} {\bibfnamefont {A.}~\bibnamefont {Chakraborty}}, \bibinfo
  {author} {\bibfnamefont {J.}~\bibnamefont {Sinova}},\ and\ \bibinfo {author}
  {\bibfnamefont {L.}~\bibnamefont {Šmejkal}},\ }\bibfield  {title} {\bibinfo
  {title} {$p$-wave magnets},\ }\href {https://arxiv.org/abs/2309.01607}
  {\bibfield  {journal} {\bibinfo  {journal} {arXiv preprint arXiv:2309.01607}\
  } (\bibinfo {year} {2024})}\BibitemShut {NoStop}%
\bibitem [{\citenamefont {Liu}\ \emph {et~al.}(2026{\natexlab{a}})\citenamefont
  {Liu}, \citenamefont {Yu}, \citenamefont {Zhang},\ and\ \citenamefont
  {Liu}}]{liu2026oddparity}%
  \BibitemOpen
  \bibfield  {author} {\bibinfo {author} {\bibfnamefont {Y.}~\bibnamefont
  {Liu}}, \bibinfo {author} {\bibfnamefont {J.}~\bibnamefont {Yu}}, \bibinfo
  {author} {\bibfnamefont {P.}~\bibnamefont {Zhang}},\ and\ \bibinfo {author}
  {\bibfnamefont {C.-C.}\ \bibnamefont {Liu}},\ }\bibfield  {title} {\bibinfo
  {title} {{Complete Hierarchy of Nonrelativistic Odd-Parity Spin Splitting in
  Collinear Magnets}},\ }\href {https://arxiv.org/abs/2607.19303} {\bibfield
  {journal} {\bibinfo  {journal} {arXiv preprint arXiv:2607.19303}\ } (\bibinfo
  {year} {2026}{\natexlab{a}})}\BibitemShut {NoStop}%
\bibitem [{\citenamefont {Yamada}\ \emph {et~al.}(2025)\citenamefont {Yamada},
  \citenamefont {Birch}, \citenamefont {Baral}, \citenamefont {Okumura},
  \citenamefont {Nakano}, \citenamefont {Gao}, \citenamefont {Ezawa},
  \citenamefont {Nomoto}, \citenamefont {Masell}, \citenamefont {Ishihara},
  \citenamefont {Kolincio}, \citenamefont {Belopolski}, \citenamefont
  {Sagayama}, \citenamefont {Nakao}, \citenamefont {Ohishi}, \citenamefont
  {Ohhara} \emph {et~al.}}]{yamada2025pwave}%
  \BibitemOpen
  \bibfield  {author} {\bibinfo {author} {\bibfnamefont {R.}~\bibnamefont
  {Yamada}}, \bibinfo {author} {\bibfnamefont {M.~T.}\ \bibnamefont {Birch}},
  \bibinfo {author} {\bibfnamefont {P.~R.}\ \bibnamefont {Baral}}, \bibinfo
  {author} {\bibfnamefont {S.}~\bibnamefont {Okumura}}, \bibinfo {author}
  {\bibfnamefont {R.}~\bibnamefont {Nakano}}, \bibinfo {author} {\bibfnamefont
  {S.}~\bibnamefont {Gao}}, \bibinfo {author} {\bibfnamefont {M.}~\bibnamefont
  {Ezawa}}, \bibinfo {author} {\bibfnamefont {T.}~\bibnamefont {Nomoto}},
  \bibinfo {author} {\bibfnamefont {J.}~\bibnamefont {Masell}}, \bibinfo
  {author} {\bibfnamefont {Y.}~\bibnamefont {Ishihara}}, \bibinfo {author}
  {\bibfnamefont {K.~K.}\ \bibnamefont {Kolincio}}, \bibinfo {author}
  {\bibfnamefont {I.}~\bibnamefont {Belopolski}}, \bibinfo {author}
  {\bibfnamefont {H.}~\bibnamefont {Sagayama}}, \bibinfo {author}
  {\bibfnamefont {H.}~\bibnamefont {Nakao}}, \bibinfo {author} {\bibfnamefont
  {K.}~\bibnamefont {Ohishi}}, \bibinfo {author} {\bibfnamefont
  {T.}~\bibnamefont {Ohhara}}, \emph {et~al.},\ }\bibfield  {title} {\bibinfo
  {title} {{A Metallic $p$-Wave Magnet with Commensurate Spin Helix}},\ }\href
  {https://doi.org/10.1038/s41586-025-09633-4} {\bibfield  {journal} {\bibinfo
  {journal} {Nature}\ }\textbf {\bibinfo {volume} {646}},\ \bibinfo {pages}
  {837} (\bibinfo {year} {2025})}\BibitemShut {NoStop}%
\bibitem [{\citenamefont {Song}\ \emph
  {et~al.}(2025{\natexlab{a}})\citenamefont {Song}, \citenamefont
  {Stavri{\'c}}, \citenamefont {Barone}, \citenamefont {Droghetti},
  \citenamefont {Antonenko}, \citenamefont {Venderbos}, \citenamefont
  {Occhialini}, \citenamefont {Ilyas}, \citenamefont {Erge{\c{c}}en},
  \citenamefont {Gedik}, \citenamefont {Cheong}, \citenamefont {Fernandes},
  \citenamefont {Picozzi},\ and\ \citenamefont {Comin}}]{song2025electrical}%
  \BibitemOpen
  \bibfield  {author} {\bibinfo {author} {\bibfnamefont {Q.}~\bibnamefont
  {Song}}, \bibinfo {author} {\bibfnamefont {S.}~\bibnamefont {Stavri{\'c}}},
  \bibinfo {author} {\bibfnamefont {P.}~\bibnamefont {Barone}}, \bibinfo
  {author} {\bibfnamefont {A.}~\bibnamefont {Droghetti}}, \bibinfo {author}
  {\bibfnamefont {D.~S.}\ \bibnamefont {Antonenko}}, \bibinfo {author}
  {\bibfnamefont {J.~W.~F.}\ \bibnamefont {Venderbos}}, \bibinfo {author}
  {\bibfnamefont {C.~A.}\ \bibnamefont {Occhialini}}, \bibinfo {author}
  {\bibfnamefont {B.}~\bibnamefont {Ilyas}}, \bibinfo {author} {\bibfnamefont
  {E.}~\bibnamefont {Erge{\c{c}}en}}, \bibinfo {author} {\bibfnamefont
  {N.}~\bibnamefont {Gedik}}, \bibinfo {author} {\bibfnamefont {S.-W.}\
  \bibnamefont {Cheong}}, \bibinfo {author} {\bibfnamefont {R.~M.}\
  \bibnamefont {Fernandes}}, \bibinfo {author} {\bibfnamefont {S.}~\bibnamefont
  {Picozzi}},\ and\ \bibinfo {author} {\bibfnamefont {R.}~\bibnamefont
  {Comin}},\ }\bibfield  {title} {\bibinfo {title} {{Electrical Switching of a
  $p$-Wave Magnet}},\ }\href {https://doi.org/10.1038/s41586-025-09034-7}
  {\bibfield  {journal} {\bibinfo  {journal} {Nature}\ }\textbf {\bibinfo
  {volume} {642}},\ \bibinfo {pages} {64} (\bibinfo {year}
  {2025}{\natexlab{a}})}\BibitemShut {NoStop}%
\bibitem [{\citenamefont {Fedchenko}\ \emph {et~al.}(2024)\citenamefont
  {Fedchenko}, \citenamefont {Minár}, \citenamefont {Akashdeep}, \citenamefont
  {D’Souza}, \citenamefont {Vasilyev}, \citenamefont {Tkach}, \citenamefont
  {Odenbreit}, \citenamefont {Nguyen}, \citenamefont {Kutnyakhov},
  \citenamefont {Wind}, \citenamefont {Wenthaus}, \citenamefont {Scholz},
  \citenamefont {Rossnagel}, \citenamefont {Hoesch}, \citenamefont
  {Aeschlimann}, \citenamefont {Stadtmüller}, \citenamefont {Kläui},
  \citenamefont {Schönhense}, \citenamefont {Jungwirth}, \citenamefont
  {Hellenes}, \citenamefont {Jakob}, \citenamefont {Šmejkal}, \citenamefont
  {Sinova},\ and\ \citenamefont {Elmers}}]{doi:10.1126/sciadv.adj4883}%
  \BibitemOpen
  \bibfield  {author} {\bibinfo {author} {\bibfnamefont {O.}~\bibnamefont
  {Fedchenko}}, \bibinfo {author} {\bibfnamefont {J.}~\bibnamefont {Minár}},
  \bibinfo {author} {\bibfnamefont {A.}~\bibnamefont {Akashdeep}}, \bibinfo
  {author} {\bibfnamefont {S.~W.}\ \bibnamefont {D’Souza}}, \bibinfo {author}
  {\bibfnamefont {D.}~\bibnamefont {Vasilyev}}, \bibinfo {author}
  {\bibfnamefont {O.}~\bibnamefont {Tkach}}, \bibinfo {author} {\bibfnamefont
  {L.}~\bibnamefont {Odenbreit}}, \bibinfo {author} {\bibfnamefont
  {Q.}~\bibnamefont {Nguyen}}, \bibinfo {author} {\bibfnamefont
  {D.}~\bibnamefont {Kutnyakhov}}, \bibinfo {author} {\bibfnamefont
  {N.}~\bibnamefont {Wind}}, \bibinfo {author} {\bibfnamefont {L.}~\bibnamefont
  {Wenthaus}}, \bibinfo {author} {\bibfnamefont {M.}~\bibnamefont {Scholz}},
  \bibinfo {author} {\bibfnamefont {K.}~\bibnamefont {Rossnagel}}, \bibinfo
  {author} {\bibfnamefont {M.}~\bibnamefont {Hoesch}}, \bibinfo {author}
  {\bibfnamefont {M.}~\bibnamefont {Aeschlimann}}, \bibinfo {author}
  {\bibfnamefont {B.}~\bibnamefont {Stadtmüller}}, \bibinfo {author}
  {\bibfnamefont {M.}~\bibnamefont {Kläui}}, \bibinfo {author} {\bibfnamefont
  {G.}~\bibnamefont {Schönhense}}, \bibinfo {author} {\bibfnamefont
  {T.}~\bibnamefont {Jungwirth}}, \bibinfo {author} {\bibfnamefont {A.~B.}\
  \bibnamefont {Hellenes}}, \bibinfo {author} {\bibfnamefont {G.}~\bibnamefont
  {Jakob}}, \bibinfo {author} {\bibfnamefont {L.}~\bibnamefont {Šmejkal}},
  \bibinfo {author} {\bibfnamefont {J.}~\bibnamefont {Sinova}},\ and\ \bibinfo
  {author} {\bibfnamefont {H.-J.}\ \bibnamefont {Elmers}},\ }\bibfield  {title}
  {\bibinfo {title} {Observation of time-reversal symmetry breaking in the band
  structure of altermagnetic {R}u{O}$_2$},\ }\href
  {https://doi.org/10.1126/sciadv.adj4883} {\bibfield  {journal} {\bibinfo
  {journal} {Science Advances}\ }\textbf {\bibinfo {volume} {10}},\ \bibinfo
  {pages} {eadj4883} (\bibinfo {year} {2024})}\BibitemShut {NoStop}%
\bibitem [{\citenamefont {Feng}\ \emph {et~al.}(2022)\citenamefont {Feng},
  \citenamefont {Zhou}, \citenamefont {{\v{S}}mejkal}, \citenamefont {Wu},
  \citenamefont {Zhu}, \citenamefont {Guo}, \citenamefont
  {Gonz{\'a}lez-Hern{\'a}ndez}, \citenamefont {Wang}, \citenamefont {Yan},
  \citenamefont {Qin}, \citenamefont {Zhang}, \citenamefont {Wu}, \citenamefont
  {Chen}, \citenamefont {Meng}, \citenamefont {Liu}, \citenamefont {Xia},
  \citenamefont {Sinova}, \citenamefont {Jungwirth},\ and\ \citenamefont
  {Liu}}]{Feng2022}%
  \BibitemOpen
  \bibfield  {author} {\bibinfo {author} {\bibfnamefont {Z.}~\bibnamefont
  {Feng}}, \bibinfo {author} {\bibfnamefont {X.}~\bibnamefont {Zhou}}, \bibinfo
  {author} {\bibfnamefont {L.}~\bibnamefont {{\v{S}}mejkal}}, \bibinfo {author}
  {\bibfnamefont {L.}~\bibnamefont {Wu}}, \bibinfo {author} {\bibfnamefont
  {Z.}~\bibnamefont {Zhu}}, \bibinfo {author} {\bibfnamefont {H.}~\bibnamefont
  {Guo}}, \bibinfo {author} {\bibfnamefont {R.}~\bibnamefont
  {Gonz{\'a}lez-Hern{\'a}ndez}}, \bibinfo {author} {\bibfnamefont
  {X.}~\bibnamefont {Wang}}, \bibinfo {author} {\bibfnamefont {H.}~\bibnamefont
  {Yan}}, \bibinfo {author} {\bibfnamefont {P.}~\bibnamefont {Qin}}, \bibinfo
  {author} {\bibfnamefont {X.}~\bibnamefont {Zhang}}, \bibinfo {author}
  {\bibfnamefont {H.}~\bibnamefont {Wu}}, \bibinfo {author} {\bibfnamefont
  {H.}~\bibnamefont {Chen}}, \bibinfo {author} {\bibfnamefont {Z.}~\bibnamefont
  {Meng}}, \bibinfo {author} {\bibfnamefont {L.}~\bibnamefont {Liu}}, \bibinfo
  {author} {\bibfnamefont {Z.}~\bibnamefont {Xia}}, \bibinfo {author}
  {\bibfnamefont {J.}~\bibnamefont {Sinova}}, \bibinfo {author} {\bibfnamefont
  {T.}~\bibnamefont {Jungwirth}},\ and\ \bibinfo {author} {\bibfnamefont
  {Z.}~\bibnamefont {Liu}},\ }\bibfield  {title} {\bibinfo {title} {An
  anomalous hall effect in altermagnetic ruthenium dioxide},\ }\href
  {https://doi.org/10.1038/s41928-022-00866-z} {\bibfield  {journal} {\bibinfo
  {journal} {Nature Electronics}\ }\textbf {\bibinfo {volume} {5}},\ \bibinfo
  {pages} {735} (\bibinfo {year} {2022})}\BibitemShut {NoStop}%
\bibitem [{\citenamefont {{\ifmmode \check{S}\else \v{S}\fi{}}mejkal}\ \emph
  {et~al.}(2023)\citenamefont {{\ifmmode \check{S}\else \v{S}\fi{}}mejkal},
  \citenamefont {Marmodoro}, \citenamefont {Ahn}, \citenamefont
  {Gonz{\'a}lez-Hern{\'a}ndez}, \citenamefont {Turek}, \citenamefont
  {Mankovsky}, \citenamefont {Ebert}, \citenamefont {D'Souza}, \citenamefont
  {{\ifmmode \check{S}\else \v{S}\fi{}}ipr}, \citenamefont {Sinova},\ and\
  \citenamefont {Jungwirth}}]{PhysRevLett.131.256703}%
  \BibitemOpen
  \bibfield  {author} {\bibinfo {author} {\bibfnamefont {L.}~\bibnamefont
  {{\ifmmode \check{S}\else \v{S}\fi{}}mejkal}}, \bibinfo {author}
  {\bibfnamefont {A.}~\bibnamefont {Marmodoro}}, \bibinfo {author}
  {\bibfnamefont {K.-H.}\ \bibnamefont {Ahn}}, \bibinfo {author} {\bibfnamefont
  {R.}~\bibnamefont {Gonz{\'a}lez-Hern{\'a}ndez}}, \bibinfo {author}
  {\bibfnamefont {I.}~\bibnamefont {Turek}}, \bibinfo {author} {\bibfnamefont
  {S.}~\bibnamefont {Mankovsky}}, \bibinfo {author} {\bibfnamefont
  {H.}~\bibnamefont {Ebert}}, \bibinfo {author} {\bibfnamefont {S.~W.}\
  \bibnamefont {D'Souza}}, \bibinfo {author} {\bibfnamefont {O.}~\bibnamefont
  {{\ifmmode \check{S}\else \v{S}\fi{}}ipr}}, \bibinfo {author} {\bibfnamefont
  {J.}~\bibnamefont {Sinova}},\ and\ \bibinfo {author} {\bibfnamefont
  {T.}~\bibnamefont {Jungwirth}},\ }\bibfield  {title} {\bibinfo {title}
  {Chiral magnons in altermagnetic ${\mathrm{ruo}}_{2}$},\ }\href
  {https://doi.org/10.1103/PhysRevLett.131.256703} {\bibfield  {journal}
  {\bibinfo  {journal} {Phys. Rev. Lett.}\ }\textbf {\bibinfo {volume} {131}},\
  \bibinfo {pages} {256703} (\bibinfo {year} {2023})}\BibitemShut {NoStop}%
\bibitem [{\citenamefont {Zhou}\ \emph {et~al.}(2024)\citenamefont {Zhou},
  \citenamefont {Feng}, \citenamefont {Zhang}, \citenamefont {{\ifmmode
  \check{S}\else \v{S}\fi{}}mejkal}, \citenamefont {Sinova}, \citenamefont
  {Mokrousov},\ and\ \citenamefont {Yao}}]{PhysRevLett.132.056701}%
  \BibitemOpen
  \bibfield  {author} {\bibinfo {author} {\bibfnamefont {X.}~\bibnamefont
  {Zhou}}, \bibinfo {author} {\bibfnamefont {W.}~\bibnamefont {Feng}}, \bibinfo
  {author} {\bibfnamefont {R.-W.}\ \bibnamefont {Zhang}}, \bibinfo {author}
  {\bibfnamefont {L.}~\bibnamefont {{\ifmmode \check{S}\else
  \v{S}\fi{}}mejkal}}, \bibinfo {author} {\bibfnamefont {J.}~\bibnamefont
  {Sinova}}, \bibinfo {author} {\bibfnamefont {Y.}~\bibnamefont {Mokrousov}},\
  and\ \bibinfo {author} {\bibfnamefont {Y.}~\bibnamefont {Yao}},\ }\bibfield
  {title} {\bibinfo {title} {Crystal thermal transport in altermagnetic
  ${\mathrm{ruo}}_{2}$},\ }\href
  {https://doi.org/10.1103/PhysRevLett.132.056701} {\bibfield  {journal}
  {\bibinfo  {journal} {Phys. Rev. Lett.}\ }\textbf {\bibinfo {volume} {132}},\
  \bibinfo {pages} {056701} (\bibinfo {year} {2024})}\BibitemShut {NoStop}%
\bibitem [{\citenamefont {Reichlova}\ \emph {et~al.}(2024)\citenamefont
  {Reichlova}, \citenamefont {Lopes~Seeger}, \citenamefont
  {Gonz{\'a}lez-Hern{\'a}ndez}, \citenamefont {Kounta}, \citenamefont
  {Schlitz}, \citenamefont {Kriegner}, \citenamefont {Ritzinger}, \citenamefont
  {Lammel}, \citenamefont {Leivisk{\"a}}, \citenamefont {Birk~Hellenes},
  \citenamefont {Olejn{\'i}k}, \citenamefont {Pet{\v{r}}i{\v{c}}ek},
  \citenamefont {Dole{\v{z}}al}, \citenamefont {Horak}, \citenamefont
  {Schmoranzerova}, \citenamefont {Badura}, \citenamefont {Bertaina},
  \citenamefont {Thomas}, \citenamefont {Baltz}, \citenamefont {Michez},
  \citenamefont {Sinova}, \citenamefont {Goennenwein}, \citenamefont
  {Jungwirth},\ and\ \citenamefont {{\v{S}}mejkal}}]{Reichlova2024}%
  \BibitemOpen
  \bibfield  {author} {\bibinfo {author} {\bibfnamefont {H.}~\bibnamefont
  {Reichlova}}, \bibinfo {author} {\bibfnamefont {R.}~\bibnamefont
  {Lopes~Seeger}}, \bibinfo {author} {\bibfnamefont {R.}~\bibnamefont
  {Gonz{\'a}lez-Hern{\'a}ndez}}, \bibinfo {author} {\bibfnamefont
  {I.}~\bibnamefont {Kounta}}, \bibinfo {author} {\bibfnamefont
  {R.}~\bibnamefont {Schlitz}}, \bibinfo {author} {\bibfnamefont
  {D.}~\bibnamefont {Kriegner}}, \bibinfo {author} {\bibfnamefont
  {P.}~\bibnamefont {Ritzinger}}, \bibinfo {author} {\bibfnamefont
  {M.}~\bibnamefont {Lammel}}, \bibinfo {author} {\bibfnamefont
  {M.}~\bibnamefont {Leivisk{\"a}}}, \bibinfo {author} {\bibfnamefont
  {A.}~\bibnamefont {Birk~Hellenes}}, \bibinfo {author} {\bibfnamefont
  {K.}~\bibnamefont {Olejn{\'i}k}}, \bibinfo {author} {\bibfnamefont
  {V.}~\bibnamefont {Pet{\v{r}}i{\v{c}}ek}}, \bibinfo {author} {\bibfnamefont
  {P.}~\bibnamefont {Dole{\v{z}}al}}, \bibinfo {author} {\bibfnamefont
  {L.}~\bibnamefont {Horak}}, \bibinfo {author} {\bibfnamefont
  {E.}~\bibnamefont {Schmoranzerova}}, \bibinfo {author} {\bibfnamefont
  {A.}~\bibnamefont {Badura}}, \bibinfo {author} {\bibfnamefont
  {S.}~\bibnamefont {Bertaina}}, \bibinfo {author} {\bibfnamefont
  {A.}~\bibnamefont {Thomas}}, \bibinfo {author} {\bibfnamefont
  {V.}~\bibnamefont {Baltz}}, \bibinfo {author} {\bibfnamefont
  {L.}~\bibnamefont {Michez}}, \bibinfo {author} {\bibfnamefont
  {J.}~\bibnamefont {Sinova}}, \bibinfo {author} {\bibfnamefont {S.~T.~B.}\
  \bibnamefont {Goennenwein}}, \bibinfo {author} {\bibfnamefont
  {T.}~\bibnamefont {Jungwirth}},\ and\ \bibinfo {author} {\bibfnamefont
  {L.}~\bibnamefont {{\v{S}}mejkal}},\ }\bibfield  {title} {\bibinfo {title}
  {Observation of a spontaneous anomalous hall response in the {M}n$_5${S}i$_3$
  $d$-wave altermagnet candidate},\ }\href
  {https://doi.org/10.1038/s41467-024-48493-w} {\bibfield  {journal} {\bibinfo
  {journal} {Nature Communications}\ }\textbf {\bibinfo {volume} {15}},\
  \bibinfo {pages} {4961} (\bibinfo {year} {2024})}\BibitemShut {NoStop}%
\bibitem [{\citenamefont {Adamantopoulos}\ \emph {et~al.}(2024)\citenamefont
  {Adamantopoulos}, \citenamefont {Merte}, \citenamefont {Freimuth},
  \citenamefont {Go}, \citenamefont {Zhang}, \citenamefont
  {Le{\v{z}}ai{\'{c}}}, \citenamefont {Feng}, \citenamefont {Yao},
  \citenamefont {Sinova}, \citenamefont {{\v{S}}mejkal}, \citenamefont
  {Bl{\"u}gel},\ and\ \citenamefont {Mokrousov}}]{Adamantopoulos2024}%
  \BibitemOpen
  \bibfield  {author} {\bibinfo {author} {\bibfnamefont {T.}~\bibnamefont
  {Adamantopoulos}}, \bibinfo {author} {\bibfnamefont {M.}~\bibnamefont
  {Merte}}, \bibinfo {author} {\bibfnamefont {F.}~\bibnamefont {Freimuth}},
  \bibinfo {author} {\bibfnamefont {D.}~\bibnamefont {Go}}, \bibinfo {author}
  {\bibfnamefont {L.}~\bibnamefont {Zhang}}, \bibinfo {author} {\bibfnamefont
  {M.}~\bibnamefont {Le{\v{z}}ai{\'{c}}}}, \bibinfo {author} {\bibfnamefont
  {W.}~\bibnamefont {Feng}}, \bibinfo {author} {\bibfnamefont {Y.}~\bibnamefont
  {Yao}}, \bibinfo {author} {\bibfnamefont {J.}~\bibnamefont {Sinova}},
  \bibinfo {author} {\bibfnamefont {L.}~\bibnamefont {{\v{S}}mejkal}}, \bibinfo
  {author} {\bibfnamefont {S.}~\bibnamefont {Bl{\"u}gel}},\ and\ \bibinfo
  {author} {\bibfnamefont {Y.}~\bibnamefont {Mokrousov}},\ }\bibfield  {title}
  {\bibinfo {title} {Spin and orbital magnetism by light in rutile
  altermagnets},\ }\href {https://doi.org/10.1038/s44306-024-00053-0}
  {\bibfield  {journal} {\bibinfo  {journal} {npj Spintronics}\ }\textbf
  {\bibinfo {volume} {2}},\ \bibinfo {pages} {46} (\bibinfo {year}
  {2024})}\BibitemShut {NoStop}%
\bibitem [{\citenamefont {Cheong}\ and\ \citenamefont
  {Huang}(2025)}]{cheong2025classification}%
  \BibitemOpen
  \bibfield  {author} {\bibinfo {author} {\bibfnamefont {S.-W.}\ \bibnamefont
  {Cheong}}\ and\ \bibinfo {author} {\bibfnamefont {F.-T.}\ \bibnamefont
  {Huang}},\ }\bibfield  {title} {\bibinfo {title} {{Altermagnetism
  Classification}},\ }\href {https://doi.org/10.1038/s41535-025-00756-5}
  {\bibfield  {journal} {\bibinfo  {journal} {npj Quantum Materials}\ }\textbf
  {\bibinfo {volume} {10}},\ \bibinfo {pages} {38} (\bibinfo {year}
  {2025})}\BibitemShut {NoStop}%
\bibitem [{\citenamefont {Xiao}\ \emph {et~al.}(2024)\citenamefont {Xiao},
  \citenamefont {Zhao}, \citenamefont {Li}, \citenamefont {Shindou},\ and\
  \citenamefont {Song}}]{xiao2024spin}%
  \BibitemOpen
  \bibfield  {author} {\bibinfo {author} {\bibfnamefont {Z.}~\bibnamefont
  {Xiao}}, \bibinfo {author} {\bibfnamefont {J.}~\bibnamefont {Zhao}}, \bibinfo
  {author} {\bibfnamefont {Y.}~\bibnamefont {Li}}, \bibinfo {author}
  {\bibfnamefont {R.}~\bibnamefont {Shindou}},\ and\ \bibinfo {author}
  {\bibfnamefont {Z.-D.}\ \bibnamefont {Song}},\ }\bibfield  {title} {\bibinfo
  {title} {{Spin Space Groups: Full Classification and Applications}},\ }\href
  {https://doi.org/10.1103/PhysRevX.14.031037} {\bibfield  {journal} {\bibinfo
  {journal} {Physical Review X}\ }\textbf {\bibinfo {volume} {14}},\ \bibinfo
  {pages} {031037} (\bibinfo {year} {2024})}\BibitemShut {NoStop}%
\bibitem [{\citenamefont {Jungwirth}\ \emph {et~al.}(2016)\citenamefont
  {Jungwirth}, \citenamefont {Marti}, \citenamefont {Wadley},\ and\
  \citenamefont {Wunderlich}}]{jungwirth2016antiferromagnetic}%
  \BibitemOpen
  \bibfield  {author} {\bibinfo {author} {\bibfnamefont {T.}~\bibnamefont
  {Jungwirth}}, \bibinfo {author} {\bibfnamefont {X.}~\bibnamefont {Marti}},
  \bibinfo {author} {\bibfnamefont {P.}~\bibnamefont {Wadley}},\ and\ \bibinfo
  {author} {\bibfnamefont {J.}~\bibnamefont {Wunderlich}},\ }\bibfield  {title}
  {\bibinfo {title} {{Antiferromagnetic spintronics}},\ }\href
  {https://doi.org/10.1038/nnano.2016.18} {\bibfield  {journal} {\bibinfo
  {journal} {Nature Nanotechnology}\ }\textbf {\bibinfo {volume} {11}},\
  \bibinfo {pages} {231} (\bibinfo {year} {2016})}\BibitemShut {NoStop}%
\bibitem [{\citenamefont {Baltz}\ \emph {et~al.}(2018)\citenamefont {Baltz},
  \citenamefont {Manchon}, \citenamefont {Tsoi}, \citenamefont {Moriyama},
  \citenamefont {Ono},\ and\ \citenamefont
  {Tserkovnyak}}]{jungwirth2018review}%
  \BibitemOpen
  \bibfield  {author} {\bibinfo {author} {\bibfnamefont {V.}~\bibnamefont
  {Baltz}}, \bibinfo {author} {\bibfnamefont {A.}~\bibnamefont {Manchon}},
  \bibinfo {author} {\bibfnamefont {M.}~\bibnamefont {Tsoi}}, \bibinfo {author}
  {\bibfnamefont {T.}~\bibnamefont {Moriyama}}, \bibinfo {author}
  {\bibfnamefont {T.}~\bibnamefont {Ono}},\ and\ \bibinfo {author}
  {\bibfnamefont {Y.}~\bibnamefont {Tserkovnyak}},\ }\bibfield  {title}
  {\bibinfo {title} {{Antiferromagnetic spintronics}},\ }\href
  {https://doi.org/10.1103/RevModPhys.90.015005} {\bibfield  {journal}
  {\bibinfo  {journal} {Reviews of Modern Physics}\ }\textbf {\bibinfo {volume}
  {90}},\ \bibinfo {pages} {015005} (\bibinfo {year} {2018})}\BibitemShut
  {NoStop}%
\bibitem [{\citenamefont {Železný}\ \emph {et~al.}(2018)\citenamefont
  {Železný}, \citenamefont {Wadley}, \citenamefont {Olejník}, \citenamefont
  {Hoffmann},\ and\ \citenamefont {Ohno}}]{zelezny2018spintransport}%
  \BibitemOpen
  \bibfield  {author} {\bibinfo {author} {\bibfnamefont {J.}~\bibnamefont
  {Železný}}, \bibinfo {author} {\bibfnamefont {P.}~\bibnamefont {Wadley}},
  \bibinfo {author} {\bibfnamefont {K.}~\bibnamefont {Olejník}}, \bibinfo
  {author} {\bibfnamefont {A.}~\bibnamefont {Hoffmann}},\ and\ \bibinfo
  {author} {\bibfnamefont {H.}~\bibnamefont {Ohno}},\ }\bibfield  {title}
  {\bibinfo {title} {{Spin transport and spin torque in antiferromagnetic
  devices}},\ }\href {https://doi.org/10.1038/s41567-018-0062-7} {\bibfield
  {journal} {\bibinfo  {journal} {Nature Physics}\ }\textbf {\bibinfo {volume}
  {14}},\ \bibinfo {pages} {220} (\bibinfo {year} {2018})}\BibitemShut
  {NoStop}%
\bibitem [{\citenamefont {Song}\ \emph
  {et~al.}(2025{\natexlab{b}})\citenamefont {Song}, \citenamefont {Bai},
  \citenamefont {Zhou}, \citenamefont {Han}, \citenamefont {Reichlova},
  \citenamefont {Dil}, \citenamefont {Liu}, \citenamefont {Chen},\ and\
  \citenamefont {Pan}}]{song2025altermagnets}%
  \BibitemOpen
  \bibfield  {author} {\bibinfo {author} {\bibfnamefont {C.}~\bibnamefont
  {Song}}, \bibinfo {author} {\bibfnamefont {H.}~\bibnamefont {Bai}}, \bibinfo
  {author} {\bibfnamefont {Z.}~\bibnamefont {Zhou}}, \bibinfo {author}
  {\bibfnamefont {L.}~\bibnamefont {Han}}, \bibinfo {author} {\bibfnamefont
  {H.}~\bibnamefont {Reichlova}}, \bibinfo {author} {\bibfnamefont {J.~H.}\
  \bibnamefont {Dil}}, \bibinfo {author} {\bibfnamefont {J.}~\bibnamefont
  {Liu}}, \bibinfo {author} {\bibfnamefont {X.}~\bibnamefont {Chen}},\ and\
  \bibinfo {author} {\bibfnamefont {F.}~\bibnamefont {Pan}},\ }\bibfield
  {title} {\bibinfo {title} {{Altermagnets as a new class of functional
  materials}},\ }\href {https://doi.org/10.1038/s41578-025-00779-1} {\bibfield
  {journal} {\bibinfo  {journal} {Nature Reviews Materials}\ }\textbf {\bibinfo
  {volume} {10}},\ \bibinfo {pages} {473} (\bibinfo {year}
  {2025}{\natexlab{b}})}\BibitemShut {NoStop}%
\bibitem [{\citenamefont {Guo}\ \emph {et~al.}(2025)\citenamefont {Guo},
  \citenamefont {Wang}, \citenamefont {Wang}, \citenamefont {Zhang},
  \citenamefont {Zhou},\ and\ \citenamefont {Cheng}}]{guo2025spinpolarized}%
  \BibitemOpen
  \bibfield  {author} {\bibinfo {author} {\bibfnamefont {Z.}~\bibnamefont
  {Guo}}, \bibinfo {author} {\bibfnamefont {X.}~\bibnamefont {Wang}}, \bibinfo
  {author} {\bibfnamefont {W.}~\bibnamefont {Wang}}, \bibinfo {author}
  {\bibfnamefont {G.}~\bibnamefont {Zhang}}, \bibinfo {author} {\bibfnamefont
  {X.}~\bibnamefont {Zhou}},\ and\ \bibinfo {author} {\bibfnamefont
  {Z.}~\bibnamefont {Cheng}},\ }\bibfield  {title} {\bibinfo {title}
  {{Spin-Polarized Antiferromagnets for Spintronics}},\ }\href
  {https://doi.org/10.1002/adma.202505779} {\bibfield  {journal} {\bibinfo
  {journal} {Advanced Materials}\ }\textbf {\bibinfo {volume} {37}},\ \bibinfo
  {pages} {2505779} (\bibinfo {year} {2025})}\BibitemShut {NoStop}%
\bibitem [{\citenamefont {Kimel}\ and\ \citenamefont
  {Li}(2019)}]{kimel2019writing}%
  \BibitemOpen
  \bibfield  {author} {\bibinfo {author} {\bibfnamefont {A.~V.}\ \bibnamefont
  {Kimel}}\ and\ \bibinfo {author} {\bibfnamefont {M.}~\bibnamefont {Li}},\
  }\bibfield  {title} {\bibinfo {title} {{Writing magnetic memory with
  ultrashort light pulses}},\ }\href
  {https://doi.org/10.1038/s41578-019-0086-3} {\bibfield  {journal} {\bibinfo
  {journal} {Nature Reviews Materials}\ }\textbf {\bibinfo {volume} {4}},\
  \bibinfo {pages} {189} (\bibinfo {year} {2019})}\BibitemShut {NoStop}%
\bibitem [{\citenamefont {N{\v{e}}mec}\ \emph {et~al.}(2018)\citenamefont
  {N{\v{e}}mec}, \citenamefont {Fiebig}, \citenamefont {Kampfrath},\ and\
  \citenamefont {Kimel}}]{nemec2018opto}%
  \BibitemOpen
  \bibfield  {author} {\bibinfo {author} {\bibfnamefont {P.}~\bibnamefont
  {N{\v{e}}mec}}, \bibinfo {author} {\bibfnamefont {M.}~\bibnamefont {Fiebig}},
  \bibinfo {author} {\bibfnamefont {T.}~\bibnamefont {Kampfrath}},\ and\
  \bibinfo {author} {\bibfnamefont {A.~V.}\ \bibnamefont {Kimel}},\ }\bibfield
  {title} {\bibinfo {title} {{Antiferromagnetic opto-spintronics}},\ }\href
  {https://doi.org/10.1038/s41567-018-0051-x} {\bibfield  {journal} {\bibinfo
  {journal} {Nature Physics}\ }\textbf {\bibinfo {volume} {14}},\ \bibinfo
  {pages} {229} (\bibinfo {year} {2018})}\BibitemShut {NoStop}%
\bibitem [{\citenamefont {Kirilyuk}\ \emph {et~al.}(2010)\citenamefont
  {Kirilyuk}, \citenamefont {Kimel},\ and\ \citenamefont
  {Rasing}}]{kirilyuk2010ultrafast}%
  \BibitemOpen
  \bibfield  {author} {\bibinfo {author} {\bibfnamefont {A.}~\bibnamefont
  {Kirilyuk}}, \bibinfo {author} {\bibfnamefont {A.~V.}\ \bibnamefont
  {Kimel}},\ and\ \bibinfo {author} {\bibfnamefont {T.}~\bibnamefont
  {Rasing}},\ }\bibfield  {title} {\bibinfo {title} {{Ultrafast optical
  manipulation of magnetic order}},\ }\href
  {https://doi.org/10.1103/RevModPhys.82.2731} {\bibfield  {journal} {\bibinfo
  {journal} {Reviews of Modern Physics}\ }\textbf {\bibinfo {volume} {82}},\
  \bibinfo {pages} {2731} (\bibinfo {year} {2010})}\BibitemShut {NoStop}%
\bibitem [{\citenamefont {Oka}\ and\ \citenamefont
  {Kitamura}(2019)}]{oka2019floquet}%
  \BibitemOpen
  \bibfield  {author} {\bibinfo {author} {\bibfnamefont {T.}~\bibnamefont
  {Oka}}\ and\ \bibinfo {author} {\bibfnamefont {S.}~\bibnamefont {Kitamura}},\
  }\bibfield  {title} {\bibinfo {title} {{Floquet Engineering of Quantum
  Materials}},\ }\href
  {https://doi.org/10.1146/annurev-conmatphys-031218-013423} {\bibfield
  {journal} {\bibinfo  {journal} {Annual Review of Condensed Matter Physics}\
  }\textbf {\bibinfo {volume} {10}},\ \bibinfo {pages} {387} (\bibinfo {year}
  {2019})}\BibitemShut {NoStop}%
\bibitem [{\citenamefont {Bukov}\ \emph {et~al.}(2015)\citenamefont {Bukov},
  \citenamefont {D'Alessio},\ and\ \citenamefont
  {Polkovnikov}}]{bukov2015universal}%
  \BibitemOpen
  \bibfield  {author} {\bibinfo {author} {\bibfnamefont {M.}~\bibnamefont
  {Bukov}}, \bibinfo {author} {\bibfnamefont {L.}~\bibnamefont {D'Alessio}},\
  and\ \bibinfo {author} {\bibfnamefont {A.}~\bibnamefont {Polkovnikov}},\
  }\bibfield  {title} {\bibinfo {title} {{Universal high-frequency behavior of
  periodically driven systems: from dynamical stabilization to {Floquet}
  engineering}},\ }\href {https://doi.org/10.1080/00018732.2015.1055918}
  {\bibfield  {journal} {\bibinfo  {journal} {Advances in Physics}\ }\textbf
  {\bibinfo {volume} {64}},\ \bibinfo {pages} {139} (\bibinfo {year}
  {2015})}\BibitemShut {NoStop}%
\bibitem [{\citenamefont {Pershan}\ \emph {et~al.}(1966)\citenamefont
  {Pershan}, \citenamefont {van~der Ziel},\ and\ \citenamefont
  {Malmstrom}}]{pershan1966inverse}%
  \BibitemOpen
  \bibfield  {author} {\bibinfo {author} {\bibfnamefont {P.~S.}\ \bibnamefont
  {Pershan}}, \bibinfo {author} {\bibfnamefont {J.~P.}\ \bibnamefont {van~der
  Ziel}},\ and\ \bibinfo {author} {\bibfnamefont {L.~D.}\ \bibnamefont
  {Malmstrom}},\ }\bibfield  {title} {\bibinfo {title} {{Theoretical Discussion
  of the Inverse Faraday Effect, Raman Scattering, and Related Phenomena}},\
  }\href {https://doi.org/10.1103/PhysRev.143.574} {\bibfield  {journal}
  {\bibinfo  {journal} {Physical Review}\ }\textbf {\bibinfo {volume} {143}},\
  \bibinfo {pages} {574} (\bibinfo {year} {1966})}\BibitemShut {NoStop}%
\bibitem [{\citenamefont {Battiato}\ \emph {et~al.}(2014)\citenamefont
  {Battiato}, \citenamefont {Barbalinardo},\ and\ \citenamefont
  {Oppeneer}}]{battiato2014quantum}%
  \BibitemOpen
  \bibfield  {author} {\bibinfo {author} {\bibfnamefont {M.}~\bibnamefont
  {Battiato}}, \bibinfo {author} {\bibfnamefont {G.}~\bibnamefont
  {Barbalinardo}},\ and\ \bibinfo {author} {\bibfnamefont {P.~M.}\ \bibnamefont
  {Oppeneer}},\ }\bibfield  {title} {\bibinfo {title} {{Quantum theory of the
  inverse Faraday effect}},\ }\href
  {https://doi.org/10.1103/PhysRevB.89.014413} {\bibfield  {journal} {\bibinfo
  {journal} {Physical Review B}\ }\textbf {\bibinfo {volume} {89}},\ \bibinfo
  {pages} {014413} (\bibinfo {year} {2014})}\BibitemShut {NoStop}%
\bibitem [{\citenamefont {Yarmohammadi}\ \emph
  {et~al.}(2026{\natexlab{a}})\citenamefont {Yarmohammadi}, \citenamefont
  {Berritta}, \citenamefont {Bukov}, \citenamefont {{\ifmmode \check{S}\else
  \v{S}\fi{}}mejkal}, \citenamefont {Linder},\ and\ \citenamefont
  {Oppeneer}}]{xt23-9pnv}%
  \BibitemOpen
  \bibfield  {author} {\bibinfo {author} {\bibfnamefont {M.}~\bibnamefont
  {Yarmohammadi}}, \bibinfo {author} {\bibfnamefont {M.}~\bibnamefont
  {Berritta}}, \bibinfo {author} {\bibfnamefont {M.}~\bibnamefont {Bukov}},
  \bibinfo {author} {\bibfnamefont {L.}~\bibnamefont {{\ifmmode \check{S}\else
  \v{S}\fi{}}mejkal}}, \bibinfo {author} {\bibfnamefont {J.}~\bibnamefont
  {Linder}},\ and\ \bibinfo {author} {\bibfnamefont {P.~M.}\ \bibnamefont
  {Oppeneer}},\ }\bibfield  {title} {\bibinfo {title} {Spin polarization
  engineering in $d$-wave altermagnets},\ }\href
  {https://doi.org/10.1103/xt23-9pnv} {\bibfield  {journal} {\bibinfo
  {journal} {Phys. Rev. B}\ }\textbf {\bibinfo {volume} {113}},\ \bibinfo
  {pages} {L060403} (\bibinfo {year} {2026}{\natexlab{a}})}\BibitemShut
  {NoStop}%
\bibitem [{\citenamefont {Yarmohammadi}\ \emph {et~al.}(2025)\citenamefont
  {Yarmohammadi}, \citenamefont {Z{\"u}licke}, \citenamefont {Berakdar},
  \citenamefont {Linder},\ and\ \citenamefont {Freericks}}]{k3xb-8pts}%
  \BibitemOpen
  \bibfield  {author} {\bibinfo {author} {\bibfnamefont {M.}~\bibnamefont
  {Yarmohammadi}}, \bibinfo {author} {\bibfnamefont {U.}~\bibnamefont
  {Z{\"u}licke}}, \bibinfo {author} {\bibfnamefont {J.}~\bibnamefont
  {Berakdar}}, \bibinfo {author} {\bibfnamefont {J.}~\bibnamefont {Linder}},\
  and\ \bibinfo {author} {\bibfnamefont {J.~K.}\ \bibnamefont {Freericks}},\
  }\bibfield  {title} {\bibinfo {title} {Anisotropic light-tailored {RKKY}
  interaction in two-dimensional $d$-wave altermagnets},\ }\href
  {https://doi.org/10.1103/k3xb-8pts} {\bibfield  {journal} {\bibinfo
  {journal} {Phys. Rev. B}\ }\textbf {\bibinfo {volume} {111}},\ \bibinfo
  {pages} {224412} (\bibinfo {year} {2025})}\BibitemShut {NoStop}%
\bibitem [{\citenamefont {Iurov}\ \emph {et~al.}(2026)\citenamefont {Iurov},
  \citenamefont {Zhemchuzhna},\ and\ \citenamefont
  {Danner-Jackson}}]{zp5c-vpd9}%
  \BibitemOpen
  \bibfield  {author} {\bibinfo {author} {\bibfnamefont {A.}~\bibnamefont
  {Iurov}}, \bibinfo {author} {\bibfnamefont {L.}~\bibnamefont {Zhemchuzhna}},\
  and\ \bibinfo {author} {\bibfnamefont {T.}~\bibnamefont {Danner-Jackson}},\
  }\bibfield  {title} {\bibinfo {title} {Exploring nontrivial band structure,
  berry curvature, spin polarizations, and spin currents in $d$-wave
  altermagnets tailored by anisotropic optical fields},\ }\href
  {https://doi.org/10.1103/zp5c-vpd9} {\bibfield  {journal} {\bibinfo
  {journal} {Phys. Rev. B}\ }\textbf {\bibinfo {volume} {114}},\ \bibinfo
  {pages} {125411} (\bibinfo {year} {2026})}\BibitemShut {NoStop}%
\bibitem [{\citenamefont {Cheraghchi}(2026{\natexlab{a}})}]{mslj-zz6t}%
  \BibitemOpen
  \bibfield  {author} {\bibinfo {author} {\bibfnamefont {H.}~\bibnamefont
  {Cheraghchi}},\ }\bibfield  {title} {\bibinfo {title} {Floquet-engineered
  chern insulator in two-dimensional ${d}_{{x}^{2}\ensuremath{-}{y}^{2}}$-wave
  altermagnets},\ }\href {https://doi.org/10.1103/mslj-zz6t} {\bibfield
  {journal} {\bibinfo  {journal} {Phys. Rev. B}\ }\textbf {\bibinfo {volume}
  {113}},\ \bibinfo {pages} {155439} (\bibinfo {year}
  {2026}{\natexlab{a}})}\BibitemShut {NoStop}%
\bibitem [{\citenamefont
  {Shah}(2026)}]{shah2026floquetengineeringtopologicalphases}%
  \BibitemOpen
  \bibfield  {author} {\bibinfo {author} {\bibfnamefont {M.}~\bibnamefont
  {Shah}},\ }\bibfield  {title} {\bibinfo {title} {Floquet engineering of
  topological phases and magneto-optical response in a driven $d$-wave
  altermagnet},\ }\href {https://arxiv.org/abs/2608.11192} {\bibfield
  {journal} {\bibinfo  {journal} {arXiv preprint arXiv:2608.11192}\ } (\bibinfo
  {year} {2026})}\BibitemShut {NoStop}%
\bibitem [{\citenamefont {Liu}\ \emph {et~al.}(2026{\natexlab{b}})\citenamefont
  {Liu}, \citenamefont {Zhu},\ and\ \citenamefont
  {Zhang}}]{liu2026linearlypolarizedlightinducedanomalous}%
  \BibitemOpen
  \bibfield  {author} {\bibinfo {author} {\bibfnamefont {Y.}~\bibnamefont
  {Liu}}, \bibinfo {author} {\bibfnamefont {T.}~\bibnamefont {Zhu}},\ and\
  \bibinfo {author} {\bibfnamefont {H.}~\bibnamefont {Zhang}},\ }\bibfield
  {title} {\bibinfo {title} {Linearly polarized light-induced anomalous hall
  effect and topological phase transitions in an altermagnetic topological
  insulator},\ }\href {https://arxiv.org/abs/2603.06486} {\bibfield  {journal}
  {\bibinfo  {journal} {arXiv preprint arXiv:2603.06486}\ } (\bibinfo {year}
  {2026}{\natexlab{b}})}\BibitemShut {NoStop}%
\bibitem [{\citenamefont {Li}\ \emph {et~al.}(2026)\citenamefont {Li},
  \citenamefont {Linder},\ and\ \citenamefont
  {Sukhachov}}]{li2026dynamicalunmaskingpwavemagnetism}%
  \BibitemOpen
  \bibfield  {author} {\bibinfo {author} {\bibfnamefont {Y.}~\bibnamefont
  {Li}}, \bibinfo {author} {\bibfnamefont {J.}~\bibnamefont {Linder}},\ and\
  \bibinfo {author} {\bibfnamefont {P.}~\bibnamefont {Sukhachov}},\ }\bibfield
  {title} {\bibinfo {title} {Dynamical unmasking of $p$-wave magnetism by ac
  spin and dc orbital polarizations},\ }\href
  {https://arxiv.org/abs/2607.15353} {\bibfield  {journal} {\bibinfo  {journal}
  {arXiv preprint arXiv:2607.15353}\ } (\bibinfo {year} {2026})}\BibitemShut
  {NoStop}%
\bibitem [{\citenamefont
  {Cheraghchi}(2026{\natexlab{b}})}]{cheraghchi2026floquettopologicalphasesanomalous}%
  \BibitemOpen
  \bibfield  {author} {\bibinfo {author} {\bibfnamefont {H.}~\bibnamefont
  {Cheraghchi}},\ }\bibfield  {title} {\bibinfo {title} {Floquet topological
  phases and anomalous hall signatures in irradiated two-dimensional
  $d_{xy}$-wave altermagnets},\ }\href {https://arxiv.org/abs/2606.26667}
  {\bibfield  {journal} {\bibinfo  {journal} {arXiv preprint arXiv:2606.26667}\
  } (\bibinfo {year} {2026}{\natexlab{b}})}\BibitemShut {NoStop}%
\bibitem [{\citenamefont {Mori}(2023)}]{mori2023floquetopen}%
  \BibitemOpen
  \bibfield  {author} {\bibinfo {author} {\bibfnamefont {T.}~\bibnamefont
  {Mori}},\ }\bibfield  {title} {\bibinfo {title} {{Floquet States in Open
  Quantum Systems}},\ }\href
  {https://doi.org/10.1146/annurev-conmatphys-040721-015537} {\bibfield
  {journal} {\bibinfo  {journal} {Annual Review of Condensed Matter Physics}\
  }\textbf {\bibinfo {volume} {14}},\ \bibinfo {pages} {35} (\bibinfo {year}
  {2023})}\BibitemShut {NoStop}%
\bibitem [{\citenamefont {Dannegger}\ \emph {et~al.}(2021)\citenamefont
  {Dannegger}, \citenamefont {Berritta}, \citenamefont {Carva}, \citenamefont
  {Selzer}, \citenamefont {Ritzmann}, \citenamefont {Oppeneer},\ and\
  \citenamefont {Nowak}}]{dannegger2021switching}%
  \BibitemOpen
  \bibfield  {author} {\bibinfo {author} {\bibfnamefont {T.}~\bibnamefont
  {Dannegger}}, \bibinfo {author} {\bibfnamefont {M.}~\bibnamefont {Berritta}},
  \bibinfo {author} {\bibfnamefont {K.}~\bibnamefont {Carva}}, \bibinfo
  {author} {\bibfnamefont {S.}~\bibnamefont {Selzer}}, \bibinfo {author}
  {\bibfnamefont {U.}~\bibnamefont {Ritzmann}}, \bibinfo {author}
  {\bibfnamefont {P.~M.}\ \bibnamefont {Oppeneer}},\ and\ \bibinfo {author}
  {\bibfnamefont {U.}~\bibnamefont {Nowak}},\ }\bibfield  {title} {\bibinfo
  {title} {{Ultrafast coherent all-optical switching of an antiferromagnet with
  the inverse Faraday effect}},\ }\href
  {https://doi.org/10.1103/PhysRevB.104.L060413} {\bibfield  {journal}
  {\bibinfo  {journal} {Physical Review B}\ }\textbf {\bibinfo {volume}
  {104}},\ \bibinfo {pages} {L060413} (\bibinfo {year} {2021})}\BibitemShut
  {NoStop}%
\bibitem [{\citenamefont {Wang}\ and\ \citenamefont
  {Liu}(2020)}]{wang2020ultrafast}%
  \BibitemOpen
  \bibfield  {author} {\bibinfo {author} {\bibfnamefont {C.}~\bibnamefont
  {Wang}}\ and\ \bibinfo {author} {\bibfnamefont {Y.}~\bibnamefont {Liu}},\
  }\bibfield  {title} {\bibinfo {title} {{Ultrafast optical manipulation of
  magnetic order in ferromagnetic materials}},\ }\href
  {https://doi.org/10.1186/s40580-020-00246-3} {\bibfield  {journal} {\bibinfo
  {journal} {Nano Convergence}\ }\textbf {\bibinfo {volume} {7}},\ \bibinfo
  {pages} {35} (\bibinfo {year} {2020})}\BibitemShut {NoStop}%
\bibitem [{\citenamefont {Trevisan}\ \emph {et~al.}(2022)\citenamefont
  {Trevisan}, \citenamefont {Villar~Arribi}, \citenamefont {Heinonen},
  \citenamefont {Slager},\ and\ \citenamefont {Orth}}]{trevisan2022bicircular}%
  \BibitemOpen
  \bibfield  {author} {\bibinfo {author} {\bibfnamefont {T.~V.}\ \bibnamefont
  {Trevisan}}, \bibinfo {author} {\bibfnamefont {P.}~\bibnamefont
  {Villar~Arribi}}, \bibinfo {author} {\bibfnamefont {O.}~\bibnamefont
  {Heinonen}}, \bibinfo {author} {\bibfnamefont {R.-J.}\ \bibnamefont
  {Slager}},\ and\ \bibinfo {author} {\bibfnamefont {P.~P.}\ \bibnamefont
  {Orth}},\ }\bibfield  {title} {\bibinfo {title} {{Bicircular Light Floquet
  Engineering of Magnetic Symmetry and Topology and Its Application to the
  Dirac Semimetal Cd$_3$As$_2$}},\ }\href
  {https://doi.org/10.1103/PhysRevLett.128.066602} {\bibfield  {journal}
  {\bibinfo  {journal} {Physical Review Letters}\ }\textbf {\bibinfo {volume}
  {128}},\ \bibinfo {pages} {066602} (\bibinfo {year} {2022})}\BibitemShut
  {NoStop}%
\bibitem [{\citenamefont {Yarmohammadi}\ \emph
  {et~al.}(2026{\natexlab{b}})\citenamefont {Yarmohammadi}, \citenamefont {Jo},
  \citenamefont {Berritta}, \citenamefont {{\v{S}}mejkal}, \citenamefont
  {Freericks},\ and\ \citenamefont {Oppeneer}}]{yarmohammadi2026giant}%
  \BibitemOpen
  \bibfield  {author} {\bibinfo {author} {\bibfnamefont {M.}~\bibnamefont
  {Yarmohammadi}}, \bibinfo {author} {\bibfnamefont {D.}~\bibnamefont {Jo}},
  \bibinfo {author} {\bibfnamefont {M.}~\bibnamefont {Berritta}}, \bibinfo
  {author} {\bibfnamefont {L.}~\bibnamefont {{\v{S}}mejkal}}, \bibinfo {author}
  {\bibfnamefont {J.~K.}\ \bibnamefont {Freericks}},\ and\ \bibinfo {author}
  {\bibfnamefont {P.~M.}\ \bibnamefont {Oppeneer}},\ }\bibfield  {title}
  {\bibinfo {title} {{Giant perpendicular Edelstein polarization in 2D
  compensated magnets via bichromatic Floquet driving}},\ }\href
  {https://arxiv.org/abs/2606.31867} {\bibfield  {journal} {\bibinfo  {journal}
  {arXiv preprint arXiv:2606.31867}\ } (\bibinfo {year}
  {2026}{\natexlab{b}})}\BibitemShut {NoStop}%
\bibitem [{\citenamefont {Yarmohammadi}\ \emph
  {et~al.}(2026{\natexlab{c}})\citenamefont {Yarmohammadi}, \citenamefont
  {Gunnink}, \citenamefont {Sinova},\ and\ \citenamefont
  {Freericks}}]{yarmohammadi2026amr}%
  \BibitemOpen
  \bibfield  {author} {\bibinfo {author} {\bibfnamefont {M.}~\bibnamefont
  {Yarmohammadi}}, \bibinfo {author} {\bibfnamefont {P.~M.}\ \bibnamefont
  {Gunnink}}, \bibinfo {author} {\bibfnamefont {J.}~\bibnamefont {Sinova}},\
  and\ \bibinfo {author} {\bibfnamefont {J.~K.}\ \bibnamefont {Freericks}},\
  }\bibfield  {title} {\bibinfo {title} {{Floquet-induced anisotropic
  magnetoresistance and anomalous Hall effect in 2D $d$-wave altermagnets with
  Rashba spin-orbit coupling}},\ }\href {https://arxiv.org/abs/2606.21281}
  {\bibfield  {journal} {\bibinfo  {journal} {arXiv preprint arXiv:2606.21281}\
  } (\bibinfo {year} {2026}{\natexlab{c}})}\BibitemShut {NoStop}%
\bibitem [{\citenamefont {Ganguli}\ \emph {et~al.}(2026)\citenamefont
  {Ganguli}, \citenamefont {Jana},\ and\ \citenamefont {Narayan}}]{6jv2-nmnn}%
  \BibitemOpen
  \bibfield  {author} {\bibinfo {author} {\bibfnamefont {M.}~\bibnamefont
  {Ganguli}}, \bibinfo {author} {\bibfnamefont {A.}~\bibnamefont {Jana}},\ and\
  \bibinfo {author} {\bibfnamefont {A.}~\bibnamefont {Narayan}},\ }\bibfield
  {title} {\bibinfo {title} {Tunable topology, hall response, and spin textures
  in bicircularly polarized light illuminated altermagnets},\ }\href
  {https://doi.org/10.1103/6jv2-nmnn} {\bibfield  {journal} {\bibinfo
  {journal} {Phys. Rev. B}\ }\textbf {\bibinfo {volume} {114}},\ \bibinfo
  {pages} {065411} (\bibinfo {year} {2026})}\BibitemShut {NoStop}%
\bibitem [{\citenamefont {Amin}\ \emph {et~al.}(2020)\citenamefont {Amin},
  \citenamefont {Haney},\ and\ \citenamefont {Stiles}}]{amin2020interfacial}%
  \BibitemOpen
  \bibfield  {author} {\bibinfo {author} {\bibfnamefont {V.~P.}\ \bibnamefont
  {Amin}}, \bibinfo {author} {\bibfnamefont {P.~M.}\ \bibnamefont {Haney}},\
  and\ \bibinfo {author} {\bibfnamefont {M.~D.}\ \bibnamefont {Stiles}},\
  }\bibfield  {title} {\bibinfo {title} {{Interfacial spin-orbit torques}},\
  }\href {https://doi.org/10.1063/5.0024019} {\bibfield  {journal} {\bibinfo
  {journal} {Journal of Applied Physics}\ }\textbf {\bibinfo {volume} {128}},\
  \bibinfo {pages} {151101} (\bibinfo {year} {2020})}\BibitemShut {NoStop}%
\bibitem [{\citenamefont {Hellman}\ \emph {et~al.}(2017)\citenamefont
  {Hellman}, \citenamefont {Hoffmann}, \citenamefont {Tserkovnyak},
  \citenamefont {Beach}, \citenamefont {Fullerton}, \citenamefont {Leighton},
  \citenamefont {MacDonald}, \citenamefont {Ralph}, \citenamefont {Arena},
  \citenamefont {Dürr}, \citenamefont {Fischer}, \citenamefont {Grollier},
  \citenamefont {Heremans}, \citenamefont {Jungwirth} \emph
  {et~al.}}]{hellman2017interface}%
  \BibitemOpen
  \bibfield  {author} {\bibinfo {author} {\bibfnamefont {F.}~\bibnamefont
  {Hellman}}, \bibinfo {author} {\bibfnamefont {A.}~\bibnamefont {Hoffmann}},
  \bibinfo {author} {\bibfnamefont {Y.}~\bibnamefont {Tserkovnyak}}, \bibinfo
  {author} {\bibfnamefont {G.~S.~D.}\ \bibnamefont {Beach}}, \bibinfo {author}
  {\bibfnamefont {E.~E.}\ \bibnamefont {Fullerton}}, \bibinfo {author}
  {\bibfnamefont {C.}~\bibnamefont {Leighton}}, \bibinfo {author}
  {\bibfnamefont {A.~H.}\ \bibnamefont {MacDonald}}, \bibinfo {author}
  {\bibfnamefont {D.~C.}\ \bibnamefont {Ralph}}, \bibinfo {author}
  {\bibfnamefont {D.~A.}\ \bibnamefont {Arena}}, \bibinfo {author}
  {\bibfnamefont {H.~A.}\ \bibnamefont {Dürr}}, \bibinfo {author}
  {\bibfnamefont {P.}~\bibnamefont {Fischer}}, \bibinfo {author} {\bibfnamefont
  {J.}~\bibnamefont {Grollier}}, \bibinfo {author} {\bibfnamefont {J.~P.}\
  \bibnamefont {Heremans}}, \bibinfo {author} {\bibfnamefont {T.}~\bibnamefont
  {Jungwirth}}, \emph {et~al.},\ }\bibfield  {title} {\bibinfo {title}
  {{Interface-induced phenomena in magnetism}},\ }\href
  {https://doi.org/10.1103/RevModPhys.89.025006} {\bibfield  {journal}
  {\bibinfo  {journal} {Reviews of Modern Physics}\ }\textbf {\bibinfo {volume}
  {89}},\ \bibinfo {pages} {025006} (\bibinfo {year} {2017})}\BibitemShut
  {NoStop}%
\bibitem [{\citenamefont {Wang}\ \emph {et~al.}(2025)\citenamefont {Wang},
  \citenamefont {Chen}, \citenamefont {Yang}, \citenamefont {Hu}, \citenamefont
  {Li}, \citenamefont {Wang}, \citenamefont {Zhang},\ and\ \citenamefont
  {Jiang}}]{wang2025orbitronics}%
  \BibitemOpen
  \bibfield  {author} {\bibinfo {author} {\bibfnamefont {P.}~\bibnamefont
  {Wang}}, \bibinfo {author} {\bibfnamefont {F.}~\bibnamefont {Chen}}, \bibinfo
  {author} {\bibfnamefont {Y.}~\bibnamefont {Yang}}, \bibinfo {author}
  {\bibfnamefont {S.}~\bibnamefont {Hu}}, \bibinfo {author} {\bibfnamefont
  {Y.}~\bibnamefont {Li}}, \bibinfo {author} {\bibfnamefont {W.}~\bibnamefont
  {Wang}}, \bibinfo {author} {\bibfnamefont {D.}~\bibnamefont {Zhang}},\ and\
  \bibinfo {author} {\bibfnamefont {Y.}~\bibnamefont {Jiang}},\ }\bibfield
  {title} {\bibinfo {title} {{Orbitronics: Mechanisms, Materials and
  Devices}},\ }\href {https://doi.org/10.1002/aelm.202400554} {\bibfield
  {journal} {\bibinfo  {journal} {Advanced Electronic Materials}\ }\textbf
  {\bibinfo {volume} {11}},\ \bibinfo {pages} {2400554} (\bibinfo {year}
  {2025})}\BibitemShut {NoStop}%
\bibitem [{\citenamefont {Song}\ \emph {et~al.}(2021)\citenamefont {Song},
  \citenamefont {Zhang}, \citenamefont {Liao}, \citenamefont {Zhou},
  \citenamefont {Zhou}, \citenamefont {Chen}, \citenamefont {You},
  \citenamefont {Chen},\ and\ \citenamefont {Pan}}]{song2021spintorques}%
  \BibitemOpen
  \bibfield  {author} {\bibinfo {author} {\bibfnamefont {C.}~\bibnamefont
  {Song}}, \bibinfo {author} {\bibfnamefont {R.}~\bibnamefont {Zhang}},
  \bibinfo {author} {\bibfnamefont {L.}~\bibnamefont {Liao}}, \bibinfo {author}
  {\bibfnamefont {Y.}~\bibnamefont {Zhou}}, \bibinfo {author} {\bibfnamefont
  {X.}~\bibnamefont {Zhou}}, \bibinfo {author} {\bibfnamefont {R.}~\bibnamefont
  {Chen}}, \bibinfo {author} {\bibfnamefont {Y.}~\bibnamefont {You}}, \bibinfo
  {author} {\bibfnamefont {X.}~\bibnamefont {Chen}},\ and\ \bibinfo {author}
  {\bibfnamefont {F.}~\bibnamefont {Pan}},\ }\bibfield  {title} {\bibinfo
  {title} {{Spin-orbit torques: Materials, mechanisms, performances, and
  potential applications}},\ }\href
  {https://doi.org/10.1016/j.pmatsci.2020.100761} {\bibfield  {journal}
  {\bibinfo  {journal} {Progress in Materials Science}\ }\textbf {\bibinfo
  {volume} {118}},\ \bibinfo {pages} {100761} (\bibinfo {year}
  {2021})}\BibitemShut {NoStop}%
\bibitem [{\citenamefont {Manchon}\ \emph {et~al.}(2019)\citenamefont
  {Manchon}, \citenamefont {Železn{\'y}}, \citenamefont {Miron}, \citenamefont
  {Jungwirth}, \citenamefont {Sinova}, \citenamefont {Thiaville}, \citenamefont
  {Garello},\ and\ \citenamefont {Gambardella}}]{manchon2019current}%
  \BibitemOpen
  \bibfield  {author} {\bibinfo {author} {\bibfnamefont {A.}~\bibnamefont
  {Manchon}}, \bibinfo {author} {\bibfnamefont {J.}~\bibnamefont
  {Železn{\'y}}}, \bibinfo {author} {\bibfnamefont {I.~M.}\ \bibnamefont
  {Miron}}, \bibinfo {author} {\bibfnamefont {T.}~\bibnamefont {Jungwirth}},
  \bibinfo {author} {\bibfnamefont {J.}~\bibnamefont {Sinova}}, \bibinfo
  {author} {\bibfnamefont {A.}~\bibnamefont {Thiaville}}, \bibinfo {author}
  {\bibfnamefont {K.}~\bibnamefont {Garello}},\ and\ \bibinfo {author}
  {\bibfnamefont {P.}~\bibnamefont {Gambardella}},\ }\bibfield  {title}
  {\bibinfo {title} {{Current-induced spin-orbit torques in ferromagnetic and
  antiferromagnetic systems}},\ }\href
  {https://doi.org/10.1103/RevModPhys.91.035004} {\bibfield  {journal}
  {\bibinfo  {journal} {Reviews of Modern Physics}\ }\textbf {\bibinfo {volume}
  {91}},\ \bibinfo {pages} {035004} (\bibinfo {year} {2019})}\BibitemShut
  {NoStop}%
\bibitem [{\citenamefont {Shao}\ \emph {et~al.}(2021)\citenamefont {Shao},
  \citenamefont {Li}, \citenamefont {Liu}, \citenamefont {Yang}, \citenamefont
  {Fukami}, \citenamefont {Razavi}, \citenamefont {Wu}, \citenamefont {Wang},
  \citenamefont {Freimuth}, \citenamefont {Mokrousov}, \citenamefont {Stiles},
  \citenamefont {Emori}, \citenamefont {Hoffmann}, \citenamefont {Åkerman}
  \emph {et~al.}}]{shao2021roadmap}%
  \BibitemOpen
  \bibfield  {author} {\bibinfo {author} {\bibfnamefont {Q.}~\bibnamefont
  {Shao}}, \bibinfo {author} {\bibfnamefont {P.}~\bibnamefont {Li}}, \bibinfo
  {author} {\bibfnamefont {L.}~\bibnamefont {Liu}}, \bibinfo {author}
  {\bibfnamefont {H.}~\bibnamefont {Yang}}, \bibinfo {author} {\bibfnamefont
  {S.}~\bibnamefont {Fukami}}, \bibinfo {author} {\bibfnamefont
  {A.}~\bibnamefont {Razavi}}, \bibinfo {author} {\bibfnamefont
  {H.}~\bibnamefont {Wu}}, \bibinfo {author} {\bibfnamefont {K.~L.}\
  \bibnamefont {Wang}}, \bibinfo {author} {\bibfnamefont {F.}~\bibnamefont
  {Freimuth}}, \bibinfo {author} {\bibfnamefont {Y.}~\bibnamefont {Mokrousov}},
  \bibinfo {author} {\bibfnamefont {M.~D.}\ \bibnamefont {Stiles}}, \bibinfo
  {author} {\bibfnamefont {S.}~\bibnamefont {Emori}}, \bibinfo {author}
  {\bibfnamefont {A.}~\bibnamefont {Hoffmann}}, \bibinfo {author}
  {\bibfnamefont {J.}~\bibnamefont {Åkerman}}, \emph {et~al.},\ }\bibfield
  {title} {\bibinfo {title} {Roadmap of {s}pin-{o}rbit {t}orques},\ }\href
  {https://doi.org/10.1109/TMAG.2021.3078583} {\bibfield  {journal} {\bibinfo
  {journal} {IEEE Transactions on Magnetics}\ }\textbf {\bibinfo {volume}
  {57}},\ \bibinfo {pages} {1} (\bibinfo {year} {2021})}\BibitemShut {NoStop}%
\bibitem [{\citenamefont {Gao}\ and\ \citenamefont
  {Ando}(2020)}]{gao2020spintorques}%
  \BibitemOpen
  \bibfield  {author} {\bibinfo {author} {\bibfnamefont {T.}~\bibnamefont
  {Gao}}\ and\ \bibinfo {author} {\bibfnamefont {K.}~\bibnamefont {Ando}},\
  }\bibfield  {title} {\bibinfo {title} {{Spin-orbit torques}},\ }\href
  {https://doi.org/10.1016/bs.hmm.2020.09.001} {\bibfield  {journal} {\bibinfo
  {journal} {Handbook of Magnetic Materials}\ }\textbf {\bibinfo {volume}
  {29}},\ \bibinfo {pages} {1} (\bibinfo {year} {2020})}\BibitemShut {NoStop}%
\bibitem [{\citenamefont {Ji}\ \emph {et~al.}(2024)\citenamefont {Ji},
  \citenamefont {Zhang}, \citenamefont {Chai},\ and\ \citenamefont
  {Nan}}]{ji2024controlling}%
  \BibitemOpen
  \bibfield  {author} {\bibinfo {author} {\bibfnamefont {G.}~\bibnamefont
  {Ji}}, \bibinfo {author} {\bibfnamefont {Y.}~\bibnamefont {Zhang}}, \bibinfo
  {author} {\bibfnamefont {Y.}~\bibnamefont {Chai}},\ and\ \bibinfo {author}
  {\bibfnamefont {T.}~\bibnamefont {Nan}},\ }\bibfield  {title} {\bibinfo
  {title} {{Recent progress on controlling spin-orbit torques by materials
  design}},\ }\href {https://doi.org/10.1038/s44306-024-00054-z} {\bibfield
  {journal} {\bibinfo  {journal} {npj Spintronics}\ }\textbf {\bibinfo {volume}
  {2}},\ \bibinfo {pages} {56} (\bibinfo {year} {2024})}\BibitemShut {NoStop}%
\bibitem [{\citenamefont {Bychkov}\ and\ \citenamefont
  {Rashba}(1984)}]{bychkov1984oscillatory}%
  \BibitemOpen
  \bibfield  {author} {\bibinfo {author} {\bibfnamefont {Y.~A.}\ \bibnamefont
  {Bychkov}}\ and\ \bibinfo {author} {\bibfnamefont {E.~I.}\ \bibnamefont
  {Rashba}},\ }\bibfield  {title} {\bibinfo {title} {{Oscillatory effects and
  the magnetic susceptibility of carriers in inversion layers}},\ }\href
  {https://doi.org/10.1088/0022-3719/17/33/015} {\bibfield  {journal} {\bibinfo
   {journal} {Journal of Physics C: Solid State Physics}\ }\textbf {\bibinfo
  {volume} {17}},\ \bibinfo {pages} {6039} (\bibinfo {year}
  {1984})}\BibitemShut {NoStop}%
\bibitem [{\citenamefont {Manchon}\ \emph {et~al.}(2015)\citenamefont
  {Manchon}, \citenamefont {Koo}, \citenamefont {Nitta}, \citenamefont
  {Frolov},\ and\ \citenamefont {Jungwirth}}]{manchon2015new}%
  \BibitemOpen
  \bibfield  {author} {\bibinfo {author} {\bibfnamefont {A.}~\bibnamefont
  {Manchon}}, \bibinfo {author} {\bibfnamefont {H.~C.}\ \bibnamefont {Koo}},
  \bibinfo {author} {\bibfnamefont {J.}~\bibnamefont {Nitta}}, \bibinfo
  {author} {\bibfnamefont {S.~M.}\ \bibnamefont {Frolov}},\ and\ \bibinfo
  {author} {\bibfnamefont {T.}~\bibnamefont {Jungwirth}},\ }\bibfield  {title}
  {\bibinfo {title} {{New perspectives for Rashba spin--orbit coupling}},\
  }\href {https://doi.org/10.1038/nmat4360} {\bibfield  {journal} {\bibinfo
  {journal} {Nature Materials}\ }\textbf {\bibinfo {volume} {14}},\ \bibinfo
  {pages} {871} (\bibinfo {year} {2015})}\BibitemShut {NoStop}%
\bibitem [{\citenamefont {Zollner}\ \emph {et~al.}(2025)\citenamefont
  {Zollner}, \citenamefont {Kurpas}, \citenamefont {Gmitra},\ and\
  \citenamefont {Fabian}}]{zollner2025soc}%
  \BibitemOpen
  \bibfield  {author} {\bibinfo {author} {\bibfnamefont {K.}~\bibnamefont
  {Zollner}}, \bibinfo {author} {\bibfnamefont {M.}~\bibnamefont {Kurpas}},
  \bibinfo {author} {\bibfnamefont {M.}~\bibnamefont {Gmitra}},\ and\ \bibinfo
  {author} {\bibfnamefont {J.}~\bibnamefont {Fabian}},\ }\bibfield  {title}
  {\bibinfo {title} {First-principles determination of spin--orbit coupling
  parameters in two-dimensional materials},\ }\href
  {https://doi.org/10.1038/s42254-025-00818-4} {\bibfield  {journal} {\bibinfo
  {journal} {Nature Reviews Physics}\ }\textbf {\bibinfo {volume} {7}},\
  \bibinfo {pages} {255} (\bibinfo {year} {2025})}\BibitemShut {NoStop}%
\bibitem [{\citenamefont {Hayami}\ \emph {et~al.}(2019)\citenamefont {Hayami},
  \citenamefont {Yanagi},\ and\ \citenamefont {Kusunose}}]{hayami2019momentum}%
  \BibitemOpen
  \bibfield  {author} {\bibinfo {author} {\bibfnamefont {S.}~\bibnamefont
  {Hayami}}, \bibinfo {author} {\bibfnamefont {Y.}~\bibnamefont {Yanagi}},\
  and\ \bibinfo {author} {\bibfnamefont {H.}~\bibnamefont {Kusunose}},\
  }\bibfield  {title} {\bibinfo {title} {{Momentum-Dependent Spin Splitting by
  Collinear Antiferromagnetic Ordering}},\ }\href
  {https://doi.org/10.7566/JPSJ.88.123702} {\bibfield  {journal} {\bibinfo
  {journal} {Journal of the Physical Society of Japan}\ }\textbf {\bibinfo
  {volume} {88}},\ \bibinfo {pages} {123702} (\bibinfo {year}
  {2019})}\BibitemShut {NoStop}%
\bibitem [{\citenamefont {Bai}\ \emph {et~al.}(2024)\citenamefont {Bai},
  \citenamefont {Feng}, \citenamefont {Liu}, \citenamefont {{\v{S}}mejkal},
  \citenamefont {Mokrousov},\ and\ \citenamefont
  {Yao}}]{bai2024altermagnetism}%
  \BibitemOpen
  \bibfield  {author} {\bibinfo {author} {\bibfnamefont {L.}~\bibnamefont
  {Bai}}, \bibinfo {author} {\bibfnamefont {W.}~\bibnamefont {Feng}}, \bibinfo
  {author} {\bibfnamefont {S.}~\bibnamefont {Liu}}, \bibinfo {author}
  {\bibfnamefont {L.}~\bibnamefont {{\v{S}}mejkal}}, \bibinfo {author}
  {\bibfnamefont {Y.}~\bibnamefont {Mokrousov}},\ and\ \bibinfo {author}
  {\bibfnamefont {Y.}~\bibnamefont {Yao}},\ }\bibfield  {title} {\bibinfo
  {title} {{Altermagnetism: Exploring New Frontiers in Magnetism and
  Spintronics}},\ }\href {https://doi.org/10.1002/adfm.202409327} {\bibfield
  {journal} {\bibinfo  {journal} {Advanced Functional Materials}\ }\textbf
  {\bibinfo {volume} {34}},\ \bibinfo {pages} {2409327} (\bibinfo {year}
  {2024})}\BibitemShut {NoStop}%
\bibitem [{\citenamefont {Ezawa}(2026{\natexlab{a}})}]{xl2d-88z7}%
  \BibitemOpen
  \bibfield  {author} {\bibinfo {author} {\bibfnamefont {M.}~\bibnamefont
  {Ezawa}},\ }\bibfield  {title} {\bibinfo {title} {Magnetization induced by
  nonlinear response to temperature gradient in ${d}^{\ensuremath{'}},
  {g}^{\ensuremath{'}}$, and ${i}^{\ensuremath{'}}$ altermagnets},\ }\href
  {https://doi.org/10.1103/xl2d-88z7} {\bibfield  {journal} {\bibinfo
  {journal} {Phys. Rev. B}\ }\textbf {\bibinfo {volume} {114}},\ \bibinfo
  {pages} {175304} (\bibinfo {year} {2026}{\natexlab{a}})}\BibitemShut
  {NoStop}%
\bibitem [{\citenamefont {Ezawa}(2026{\natexlab{b}})}]{5ys6-pq68}%
  \BibitemOpen
  \bibfield  {author} {\bibinfo {author} {\bibfnamefont {M.}~\bibnamefont
  {Ezawa}},\ }\bibfield  {title} {\bibinfo {title} {Tunneling magnetoresistance
  in a junction made of {$X$}-wave magnets with ${X}=p,$ $d,$ $f,$ $g,$ $i$},\
  }\href {https://doi.org/10.1103/5ys6-pq68} {\bibfield  {journal} {\bibinfo
  {journal} {Phys. Rev. B}\ }\textbf {\bibinfo {volume} {113}},\ \bibinfo
  {pages} {155303} (\bibinfo {year} {2026}{\natexlab{b}})}\BibitemShut
  {NoStop}%
\bibitem [{\citenamefont {Ezawa}(2026{\natexlab{c}})}]{zxcv-xcyc}%
  \BibitemOpen
  \bibfield  {author} {\bibinfo {author} {\bibfnamefont {M.}~\bibnamefont
  {Ezawa}},\ }\bibfield  {title} {\bibinfo {title} {Fourth- and sixth-order
  nonlinear spin current rectifier in three-dimensional $h$-wave and $j$-wave
  odd-parity magnets},\ }\href {https://doi.org/10.1103/zxcv-xcyc} {\bibfield
  {journal} {\bibinfo  {journal} {Phys. Rev. B}\ }\textbf {\bibinfo {volume}
  {114}},\ \bibinfo {pages} {115404} (\bibinfo {year}
  {2026}{\natexlab{c}})}\BibitemShut {NoStop}%
\bibitem [{\citenamefont {Ezawa}(2025)}]{zt4l-y18j}%
  \BibitemOpen
  \bibfield  {author} {\bibinfo {author} {\bibfnamefont {M.}~\bibnamefont
  {Ezawa}},\ }\bibfield  {title} {\bibinfo {title} {Almost half-quantized
  planar {H}all effects in {$X$}-wave magnets with ${X}=p,d,f,g,i$},\ }\href
  {https://doi.org/10.1103/zt4l-y18j} {\bibfield  {journal} {\bibinfo
  {journal} {Phys. Rev. B}\ }\textbf {\bibinfo {volume} {112}},\ \bibinfo
  {pages} {235307} (\bibinfo {year} {2025})}\BibitemShut {NoStop}%
\bibitem [{\citenamefont {Khodas}\ \emph {et~al.}(2026)\citenamefont {Khodas},
  \citenamefont {{\ifmmode \check{S}\else \v{S}\fi{}}mejkal},\ and\
  \citenamefont {Mazin}}]{p21c-rtv7}%
  \BibitemOpen
  \bibfield  {author} {\bibinfo {author} {\bibfnamefont {M.}~\bibnamefont
  {Khodas}}, \bibinfo {author} {\bibfnamefont {L.}~\bibnamefont {{\ifmmode
  \check{S}\else \v{S}\fi{}}mejkal}},\ and\ \bibinfo {author} {\bibfnamefont
  {I.~I.}\ \bibnamefont {Mazin}},\ }\bibfield  {title} {\bibinfo {title}
  {Nonrelativistic {I}sing superconductivity in $p$-wave magnets},\ }\href
  {https://doi.org/10.1103/p21c-rtv7} {\bibfield  {journal} {\bibinfo
  {journal} {Phys. Rev. B}\ }\textbf {\bibinfo {volume} {114}},\ \bibinfo
  {pages} {L080505} (\bibinfo {year} {2026})}\BibitemShut {NoStop}%
\bibitem [{\citenamefont {Brekke}\ \emph {et~al.}(2024)\citenamefont {Brekke},
  \citenamefont {Sukhachov}, \citenamefont {Giil}, \citenamefont {Brataas},\
  and\ \citenamefont {Linder}}]{PhysRevLett.133.236703}%
  \BibitemOpen
  \bibfield  {author} {\bibinfo {author} {\bibfnamefont {B.}~\bibnamefont
  {Brekke}}, \bibinfo {author} {\bibfnamefont {P.}~\bibnamefont {Sukhachov}},
  \bibinfo {author} {\bibfnamefont {H.~G.}\ \bibnamefont {Giil}}, \bibinfo
  {author} {\bibfnamefont {A.}~\bibnamefont {Brataas}},\ and\ \bibinfo {author}
  {\bibfnamefont {J.}~\bibnamefont {Linder}},\ }\bibfield  {title} {\bibinfo
  {title} {{Minimal Models and Transport Properties of Unconventional $p$-Wave
  Magnets}},\ }\href {https://doi.org/10.1103/PhysRevLett.133.236703}
  {\bibfield  {journal} {\bibinfo  {journal} {Phys. Rev. Lett.}\ }\textbf
  {\bibinfo {volume} {133}},\ \bibinfo {pages} {236703} (\bibinfo {year}
  {2024})}\BibitemShut {NoStop}%
\bibitem [{\citenamefont {Sukhachov}\ \emph {et~al.}(2025)\citenamefont
  {Sukhachov}, \citenamefont {Giil}, \citenamefont {Brekke},\ and\
  \citenamefont {Linder}}]{PhysRevB.111.L220403}%
  \BibitemOpen
  \bibfield  {author} {\bibinfo {author} {\bibfnamefont {P.}~\bibnamefont
  {Sukhachov}}, \bibinfo {author} {\bibfnamefont {H.~G.}\ \bibnamefont {Giil}},
  \bibinfo {author} {\bibfnamefont {B.}~\bibnamefont {Brekke}},\ and\ \bibinfo
  {author} {\bibfnamefont {J.}~\bibnamefont {Linder}},\ }\bibfield  {title}
  {\bibinfo {title} {Coexistence of $p$-wave magnetism and superconductivity},\
  }\href {https://doi.org/10.1103/PhysRevB.111.L220403} {\bibfield  {journal}
  {\bibinfo  {journal} {Phys. Rev. B}\ }\textbf {\bibinfo {volume} {111}},\
  \bibinfo {pages} {L220403} (\bibinfo {year} {2025})}\BibitemShut {NoStop}%
\bibitem [{\citenamefont {Ogawa}\ \emph {et~al.}(2024)\citenamefont {Ogawa},
  \citenamefont {Kanda}, \citenamefont {Murotani} \emph
  {et~al.}}]{ogawa2024bicircular}%
  \BibitemOpen
  \bibfield  {author} {\bibinfo {author} {\bibfnamefont {K.}~\bibnamefont
  {Ogawa}}, \bibinfo {author} {\bibfnamefont {N.}~\bibnamefont {Kanda}},
  \bibinfo {author} {\bibfnamefont {Y.}~\bibnamefont {Murotani}}, \emph
  {et~al.},\ }\bibfield  {title} {\bibinfo {title} {{Programmable generation of
  counterrotating bicircular light pulses in the multi-terahertz frequency
  range}},\ }\href {https://doi.org/10.1038/s41467-024-50186-3} {\bibfield
  {journal} {\bibinfo  {journal} {Nature Communications}\ }\textbf {\bibinfo
  {volume} {15}},\ \bibinfo {pages} {6310} (\bibinfo {year}
  {2024})}\BibitemShut {NoStop}%
\bibitem [{\citenamefont {Hellmann}(1937)}]{hellmann1937einfuhrung}%
  \BibitemOpen
  \bibfield  {author} {\bibinfo {author} {\bibfnamefont {H.}~\bibnamefont
  {Hellmann}},\ }\href@noop {} {\emph {\bibinfo {title} {Einf{\"u}hrung in die
  Quantenchemie}}}\ (\bibinfo  {publisher} {Franz Deuticke},\ \bibinfo
  {address} {Leipzig},\ \bibinfo {year} {1937})\BibitemShut {NoStop}%
\bibitem [{\citenamefont {Feynman}(1939)}]{PhysRev.56.340}%
  \BibitemOpen
  \bibfield  {author} {\bibinfo {author} {\bibfnamefont {R.~P.}\ \bibnamefont
  {Feynman}},\ }\bibfield  {title} {\bibinfo {title} {Forces in molecules},\
  }\href {https://doi.org/10.1103/PhysRev.56.340} {\bibfield  {journal}
  {\bibinfo  {journal} {Phys. Rev.}\ }\textbf {\bibinfo {volume} {56}},\
  \bibinfo {pages} {340} (\bibinfo {year} {1939})}\BibitemShut {NoStop}%
\bibitem [{\citenamefont {Mikami}\ \emph {et~al.}(2016)\citenamefont {Mikami},
  \citenamefont {Kitamura}, \citenamefont {Yasuda}, \citenamefont {Tsuji},
  \citenamefont {Oka},\ and\ \citenamefont {Aoki}}]{mikami2016brillouin}%
  \BibitemOpen
  \bibfield  {author} {\bibinfo {author} {\bibfnamefont {T.}~\bibnamefont
  {Mikami}}, \bibinfo {author} {\bibfnamefont {S.}~\bibnamefont {Kitamura}},
  \bibinfo {author} {\bibfnamefont {K.}~\bibnamefont {Yasuda}}, \bibinfo
  {author} {\bibfnamefont {N.}~\bibnamefont {Tsuji}}, \bibinfo {author}
  {\bibfnamefont {T.}~\bibnamefont {Oka}},\ and\ \bibinfo {author}
  {\bibfnamefont {H.}~\bibnamefont {Aoki}},\ }\bibfield  {title} {\bibinfo
  {title} {{Brillouin-Wigner theory for high-frequency expansion in
  periodically driven systems: Application to Floquet topological
  insulators}},\ }\href {https://doi.org/10.1103/PhysRevB.93.144307} {\bibfield
   {journal} {\bibinfo  {journal} {Physical Review B}\ }\textbf {\bibinfo
  {volume} {93}},\ \bibinfo {pages} {144307} (\bibinfo {year}
  {2016})}\BibitemShut {NoStop}%
\bibitem [{\citenamefont {F{\"u}l{\"o}p}\ \emph {et~al.}(2020)\citenamefont
  {F{\"u}l{\"o}p}, \citenamefont {Tzortzakis},\ and\ \citenamefont
  {Kampfrath}}]{fulop2020strongfield}%
  \BibitemOpen
  \bibfield  {author} {\bibinfo {author} {\bibfnamefont {J.~A.}\ \bibnamefont
  {F{\"u}l{\"o}p}}, \bibinfo {author} {\bibfnamefont {S.}~\bibnamefont
  {Tzortzakis}},\ and\ \bibinfo {author} {\bibfnamefont {T.}~\bibnamefont
  {Kampfrath}},\ }\bibfield  {title} {\bibinfo {title} {{Laser-Driven
  Strong-Field Terahertz Sources}},\ }\href
  {https://doi.org/10.1002/adom.201900681} {\bibfield  {journal} {\bibinfo
  {journal} {Advanced Optical Materials}\ }\textbf {\bibinfo {volume} {8}},\
  \bibinfo {pages} {1900681} (\bibinfo {year} {2020})}\BibitemShut {NoStop}%
\bibitem [{\citenamefont {Kim}\ \emph {et~al.}(2024)\citenamefont {Kim},
  \citenamefont {Kang}, \citenamefont {Jang}, \citenamefont {Roh},
  \citenamefont {Lee}, \citenamefont {Lee}, \citenamefont {Sung}, \citenamefont
  {Lee},\ and\ \citenamefont {Kim}}]{kim2024ionizing}%
  \BibitemOpen
  \bibfield  {author} {\bibinfo {author} {\bibfnamefont {H.}~\bibnamefont
  {Kim}}, \bibinfo {author} {\bibfnamefont {C.}~\bibnamefont {Kang}}, \bibinfo
  {author} {\bibfnamefont {D.}~\bibnamefont {Jang}}, \bibinfo {author}
  {\bibfnamefont {Y.}~\bibnamefont {Roh}}, \bibinfo {author} {\bibfnamefont
  {S.~H.}\ \bibnamefont {Lee}}, \bibinfo {author} {\bibfnamefont {J.~W.}\
  \bibnamefont {Lee}}, \bibinfo {author} {\bibfnamefont {J.~H.}\ \bibnamefont
  {Sung}}, \bibinfo {author} {\bibfnamefont {S.~K.}\ \bibnamefont {Lee}},\ and\
  \bibinfo {author} {\bibfnamefont {K.-Y.}\ \bibnamefont {Kim}},\ }\bibfield
  {title} {\bibinfo {title} {Ionizing terahertz waves with 260 {MV}/cm from
  scalable optical rectification},\ }\href
  {https://doi.org/10.1038/s41377-024-01462-w} {\bibfield  {journal} {\bibinfo
  {journal} {Light: Science \& Applications}\ }\textbf {\bibinfo {volume}
  {13}},\ \bibinfo {pages} {118} (\bibinfo {year} {2024})}\BibitemShut
  {NoStop}%
\bibitem [{\citenamefont {Rudner}\ and\ \citenamefont
  {Lindner}(2020)}]{rudner2020band}%
  \BibitemOpen
  \bibfield  {author} {\bibinfo {author} {\bibfnamefont {M.~S.}\ \bibnamefont
  {Rudner}}\ and\ \bibinfo {author} {\bibfnamefont {N.~H.}\ \bibnamefont
  {Lindner}},\ }\bibfield  {title} {\bibinfo {title} {Band structure
  engineering and non-equilibrium dynamics in {F}loquet topological
  insulators},\ }\href {https://doi.org/10.1038/s42254-020-0170-z} {\bibfield
  {journal} {\bibinfo  {journal} {Nature Reviews Physics}\ }\textbf {\bibinfo
  {volume} {2}},\ \bibinfo {pages} {229} (\bibinfo {year} {2020})}\BibitemShut
  {NoStop}%
\bibitem [{\citenamefont {Chen}\ \emph {et~al.}(2021)\citenamefont {Chen},
  \citenamefont {Wu}, \citenamefont {Hu},\ and\ \citenamefont
  {Yang}}]{chen2021soc2d}%
  \BibitemOpen
  \bibfield  {author} {\bibinfo {author} {\bibfnamefont {J.}~\bibnamefont
  {Chen}}, \bibinfo {author} {\bibfnamefont {K.}~\bibnamefont {Wu}}, \bibinfo
  {author} {\bibfnamefont {W.}~\bibnamefont {Hu}},\ and\ \bibinfo {author}
  {\bibfnamefont {J.}~\bibnamefont {Yang}},\ }\bibfield  {title} {\bibinfo
  {title} {{Spin-Orbit Coupling in 2D Semiconductors: A Theoretical
  Perspective}},\ }\href {https://doi.org/10.1021/acs.jpclett.1c03662}
  {\bibfield  {journal} {\bibinfo  {journal} {The Journal of Physical Chemistry
  Letters}\ }\textbf {\bibinfo {volume} {12}},\ \bibinfo {pages} {12256}
  (\bibinfo {year} {2021})}\BibitemShut {NoStop}%
\bibitem [{\citenamefont {Lin}\ \emph {et~al.}(2024)\citenamefont {Lin},
  \citenamefont {Chen}, \citenamefont {Lu}, \citenamefont {Liang},
  \citenamefont {Feng}, \citenamefont {Yamagami}, \citenamefont {Osiecki},
  \citenamefont {Leandersson}, \citenamefont {Thiagarajan}, \citenamefont {Liu}
  \emph {et~al.}}]{lin2024ruo2}%
  \BibitemOpen
  \bibfield  {author} {\bibinfo {author} {\bibfnamefont {Z.}~\bibnamefont
  {Lin}}, \bibinfo {author} {\bibfnamefont {D.}~\bibnamefont {Chen}}, \bibinfo
  {author} {\bibfnamefont {W.}~\bibnamefont {Lu}}, \bibinfo {author}
  {\bibfnamefont {X.}~\bibnamefont {Liang}}, \bibinfo {author} {\bibfnamefont
  {S.}~\bibnamefont {Feng}}, \bibinfo {author} {\bibfnamefont {K.}~\bibnamefont
  {Yamagami}}, \bibinfo {author} {\bibfnamefont {J.}~\bibnamefont {Osiecki}},
  \bibinfo {author} {\bibfnamefont {M.}~\bibnamefont {Leandersson}}, \bibinfo
  {author} {\bibfnamefont {B.}~\bibnamefont {Thiagarajan}}, \bibinfo {author}
  {\bibfnamefont {J.}~\bibnamefont {Liu}}, \emph {et~al.},\ }\bibfield  {title}
  {\bibinfo {title} {{Observation of Giant Spin Splitting and $d$-wave Spin
  Texture in Room Temperature Altermagnet RuO$_2$}},\ }\href
  {https://arxiv.org/abs/2402.04995} {\bibfield  {journal} {\bibinfo  {journal}
  {arXiv preprint arXiv:2402.04995}\ } (\bibinfo {year} {2024})}\BibitemShut
  {NoStop}%
\bibitem [{\citenamefont {Krempask{\'y}}\ \emph {et~al.}(2024)\citenamefont
  {Krempask{\'y}}, \citenamefont {{\v{S}}mejkal}, \citenamefont {D'Souza},
  \citenamefont {Hajlaoui}, \citenamefont {Springholz}, \citenamefont
  {Uhl{\'i}{\v{r}}ov{\'a}}, \citenamefont {Alarab}, \citenamefont
  {Constantinou}, \citenamefont {Strocov}, \citenamefont {Usanov} \emph
  {et~al.}}]{Krempasky2024}%
  \BibitemOpen
  \bibfield  {author} {\bibinfo {author} {\bibfnamefont {J.}~\bibnamefont
  {Krempask{\'y}}}, \bibinfo {author} {\bibfnamefont {L.}~\bibnamefont
  {{\v{S}}mejkal}}, \bibinfo {author} {\bibfnamefont {S.~W.}\ \bibnamefont
  {D'Souza}}, \bibinfo {author} {\bibfnamefont {M.}~\bibnamefont {Hajlaoui}},
  \bibinfo {author} {\bibfnamefont {G.}~\bibnamefont {Springholz}}, \bibinfo
  {author} {\bibfnamefont {K.}~\bibnamefont {Uhl{\'i}{\v{r}}ov{\'a}}}, \bibinfo
  {author} {\bibfnamefont {F.}~\bibnamefont {Alarab}}, \bibinfo {author}
  {\bibfnamefont {P.~C.}\ \bibnamefont {Constantinou}}, \bibinfo {author}
  {\bibfnamefont {V.}~\bibnamefont {Strocov}}, \bibinfo {author} {\bibfnamefont
  {D.}~\bibnamefont {Usanov}}, \emph {et~al.},\ }\bibfield  {title} {\bibinfo
  {title} {{Altermagnetic lifting of {K}ramers spin degeneracy}},\ }\href
  {https://doi.org/10.1038/s41586-023-06907-7} {\bibfield  {journal} {\bibinfo
  {journal} {Nature}\ }\textbf {\bibinfo {volume} {626}},\ \bibinfo {pages}
  {517} (\bibinfo {year} {2024})}\BibitemShut {NoStop}%
\bibitem [{\citenamefont {Long}\ \emph {et~al.}(2026)\citenamefont {Long},
  \citenamefont {Weinberger}, \citenamefont {Wu}, \citenamefont {Hansen},
  \citenamefont {Tao}, \citenamefont {Shrestha}, \citenamefont {Graf},
  \citenamefont {Skourski}, \citenamefont {Grosche},\ and\ \citenamefont
  {Eaton}}]{Long2026}%
  \BibitemOpen
  \bibfield  {author} {\bibinfo {author} {\bibfnamefont {M.}~\bibnamefont
  {Long}}, \bibinfo {author} {\bibfnamefont {T.~I.}\ \bibnamefont
  {Weinberger}}, \bibinfo {author} {\bibfnamefont {Z.}~\bibnamefont {Wu}},
  \bibinfo {author} {\bibfnamefont {M.~F.}\ \bibnamefont {Hansen}}, \bibinfo
  {author} {\bibfnamefont {R.}~\bibnamefont {Tao}}, \bibinfo {author}
  {\bibfnamefont {M.}~\bibnamefont {Shrestha}}, \bibinfo {author}
  {\bibfnamefont {D.}~\bibnamefont {Graf}}, \bibinfo {author} {\bibfnamefont
  {Y.}~\bibnamefont {Skourski}}, \bibinfo {author} {\bibfnamefont {F.~M.}\
  \bibnamefont {Grosche}},\ and\ \bibinfo {author} {\bibfnamefont {A.~G.}\
  \bibnamefont {Eaton}},\ }\bibfield  {title} {\bibinfo {title} {{3D}
  bulk-resolved $g$-wave altermagnetic order parameter in {C}r{S}b},\ }\href
  {https://doi.org/10.1038/s41586-026-10902-z} {\bibfield  {journal} {\bibinfo
  {journal} {Nature}\ }\textbf {\bibinfo {volume} {656}},\ \bibinfo {pages}
  {854} (\bibinfo {year} {2026})}\BibitemShut {NoStop}%
\bibitem [{\citenamefont {Reimers}\ \emph {et~al.}(2024)\citenamefont
  {Reimers}, \citenamefont {Odenbreit}, \citenamefont {{\v{S}}mejkal},
  \citenamefont {Strocov}, \citenamefont {Constantinou}, \citenamefont
  {Hellenes}, \citenamefont {Jaeschke~Ubiergo}, \citenamefont {Campos},
  \citenamefont {Bharadwaj}, \citenamefont {Chakraborty} \emph
  {et~al.}}]{Reimers2024}%
  \BibitemOpen
  \bibfield  {author} {\bibinfo {author} {\bibfnamefont {S.}~\bibnamefont
  {Reimers}}, \bibinfo {author} {\bibfnamefont {L.}~\bibnamefont {Odenbreit}},
  \bibinfo {author} {\bibfnamefont {L.}~\bibnamefont {{\v{S}}mejkal}}, \bibinfo
  {author} {\bibfnamefont {V.~N.}\ \bibnamefont {Strocov}}, \bibinfo {author}
  {\bibfnamefont {P.}~\bibnamefont {Constantinou}}, \bibinfo {author}
  {\bibfnamefont {A.~B.}\ \bibnamefont {Hellenes}}, \bibinfo {author}
  {\bibfnamefont {R.}~\bibnamefont {Jaeschke~Ubiergo}}, \bibinfo {author}
  {\bibfnamefont {W.~H.}\ \bibnamefont {Campos}}, \bibinfo {author}
  {\bibfnamefont {V.~K.}\ \bibnamefont {Bharadwaj}}, \bibinfo {author}
  {\bibfnamefont {A.}~\bibnamefont {Chakraborty}}, \emph {et~al.},\ }\bibfield
  {title} {\bibinfo {title} {{Direct observation of altermagnetic band
  splitting in {C}r{S}b thin films}},\ }\href
  {https://doi.org/10.1038/s41467-024-46476-5} {\bibfield  {journal} {\bibinfo
  {journal} {Nature Communications}\ }\textbf {\bibinfo {volume} {15}},\
  \bibinfo {pages} {2116} (\bibinfo {year} {2024})}\BibitemShut {NoStop}%
\bibitem [{\citenamefont {Yu}\ \emph {et~al.}(2025)\citenamefont {Yu},
  \citenamefont {Lyngby}, \citenamefont {Shishidou}, \citenamefont {Roig},
  \citenamefont {Kreisel}, \citenamefont {Weinert}, \citenamefont {Andersen},\
  and\ \citenamefont {Agterberg}}]{zk69-k6b2}%
  \BibitemOpen
  \bibfield  {author} {\bibinfo {author} {\bibfnamefont {Y.}~\bibnamefont
  {Yu}}, \bibinfo {author} {\bibfnamefont {M.~B.}\ \bibnamefont {Lyngby}},
  \bibinfo {author} {\bibfnamefont {T.}~\bibnamefont {Shishidou}}, \bibinfo
  {author} {\bibfnamefont {M.}~\bibnamefont {Roig}}, \bibinfo {author}
  {\bibfnamefont {A.}~\bibnamefont {Kreisel}}, \bibinfo {author} {\bibfnamefont
  {M.}~\bibnamefont {Weinert}}, \bibinfo {author} {\bibfnamefont {B.~M.}\
  \bibnamefont {Andersen}},\ and\ \bibinfo {author} {\bibfnamefont {D.~F.}\
  \bibnamefont {Agterberg}},\ }\bibfield  {title} {\bibinfo {title}
  {{Odd-Parity Magnetism Driven by Antiferromagnetic Exchange}},\ }\href
  {https://doi.org/10.1103/zk69-k6b2} {\bibfield  {journal} {\bibinfo
  {journal} {Phys. Rev. Lett.}\ }\textbf {\bibinfo {volume} {135}},\ \bibinfo
  {pages} {046701} (\bibinfo {year} {2025})}\BibitemShut {NoStop}%
\bibitem [{\citenamefont {Dsouza}\ \emph {et~al.}(2026)\citenamefont {Dsouza},
  \citenamefont {Kreisel}, \citenamefont {Andersen}, \citenamefont
  {Agterberg},\ and\ \citenamefont {Christensen}}]{8hq8-38qs}%
  \BibitemOpen
  \bibfield  {author} {\bibinfo {author} {\bibfnamefont {R.}~\bibnamefont
  {Dsouza}}, \bibinfo {author} {\bibfnamefont {A.}~\bibnamefont {Kreisel}},
  \bibinfo {author} {\bibfnamefont {B.~M.}\ \bibnamefont {Andersen}}, \bibinfo
  {author} {\bibfnamefont {D.~F.}\ \bibnamefont {Agterberg}},\ and\ \bibinfo
  {author} {\bibfnamefont {M.~H.}\ \bibnamefont {Christensen}},\ }\bibfield
  {title} {\bibinfo {title} {Odd-parity magnetism in {F}e-based superconductors
  with coplanar magnetic order},\ }\href {https://doi.org/10.1103/8hq8-38qs}
  {\bibfield  {journal} {\bibinfo  {journal} {Phys. Rev. B}\ }\textbf {\bibinfo
  {volume} {113}},\ \bibinfo {pages} {144509} (\bibinfo {year}
  {2026})}\BibitemShut {NoStop}%
\bibitem [{\citenamefont {Razdolski}\ \emph {et~al.}(2017)\citenamefont
  {Razdolski}, \citenamefont {Alekhin}, \citenamefont {Martens}, \citenamefont
  {B{\"u}rstel}, \citenamefont {Diesing}, \citenamefont {M{\"u}nzenberg},
  \citenamefont {Bovensiepen},\ and\ \citenamefont
  {Melnikov}}]{razdolski2017trmoke}%
  \BibitemOpen
  \bibfield  {author} {\bibinfo {author} {\bibfnamefont {I.}~\bibnamefont
  {Razdolski}}, \bibinfo {author} {\bibfnamefont {A.}~\bibnamefont {Alekhin}},
  \bibinfo {author} {\bibfnamefont {U.}~\bibnamefont {Martens}}, \bibinfo
  {author} {\bibfnamefont {D.}~\bibnamefont {B{\"u}rstel}}, \bibinfo {author}
  {\bibfnamefont {D.}~\bibnamefont {Diesing}}, \bibinfo {author} {\bibfnamefont
  {M.}~\bibnamefont {M{\"u}nzenberg}}, \bibinfo {author} {\bibfnamefont
  {U.}~\bibnamefont {Bovensiepen}},\ and\ \bibinfo {author} {\bibfnamefont
  {A.}~\bibnamefont {Melnikov}},\ }\bibfield  {title} {\bibinfo {title}
  {{Analysis of the time-resolved magneto-optical Kerr effect for ultrafast
  magnetization dynamics in ferromagnetic thin films}},\ }\href
  {https://doi.org/10.1088/1361-648X/aa63c6} {\bibfield  {journal} {\bibinfo
  {journal} {Journal of Physics: Condensed Matter}\ }\textbf {\bibinfo {volume}
  {29}},\ \bibinfo {pages} {174002} (\bibinfo {year} {2017})}\BibitemShut
  {NoStop}%
\bibitem [{\citenamefont {Huang}\ \emph {et~al.}(2021)\citenamefont {Huang},
  \citenamefont {Lattery},\ and\ \citenamefont {Wang}}]{huang2021trmoke}%
  \BibitemOpen
  \bibfield  {author} {\bibinfo {author} {\bibfnamefont {D.}~\bibnamefont
  {Huang}}, \bibinfo {author} {\bibfnamefont {D.}~\bibnamefont {Lattery}},\
  and\ \bibinfo {author} {\bibfnamefont {X.}~\bibnamefont {Wang}},\ }\bibfield
  {title} {\bibinfo {title} {{Materials Engineering Enabled by Time-Resolved
  Magneto-Optical Kerr Effect for Spintronic Applications}},\ }\href
  {https://doi.org/10.1021/acsaelm.0c00961} {\bibfield  {journal} {\bibinfo
  {journal} {ACS Applied Electronic Materials}\ }\textbf {\bibinfo {volume}
  {3}},\ \bibinfo {pages} {119} (\bibinfo {year} {2021})}\BibitemShut {NoStop}%
\end{thebibliography}%
\end{document}